\documentclass{aa}  
\usepackage{graphicx}
\usepackage[english]{babel}
\usepackage[utf8]{inputenc}
\usepackage[varg]{txfonts}
\usepackage{amssymb}
\usepackage{amsmath}
\usepackage{mathtools}
\usepackage{xspace}
\usepackage{totcount}
\usepackage{color}
\usepackage[colorinlistoftodos]{todonotes}
\usepackage{verbatim}
\usepackage{graphbox}  
\usepackage{subcaption}
\usepackage{float}
\usepackage{enumerate}
\usepackage{enumitem}
\usepackage{ulem}
\usepackage{hyperref} 
\begin{document} 

\title{Quantitative polarimetry of the disk around HD~169142}
\author{C.~Tschudi\inst{\ref{inst1}} \and  H.~M.~Schmid\inst{\ref{inst1}}}
\institute{Institute for Particle Physics and Astrophysics, ETH Zurich, Wolfgang-Pauli-Strasse 27, 8093 Zurich, Switzerland. \email{chtschud@phys.ethz.ch}\label{inst1}} 

\date{Received 8 April 2021 ;
accepted 10 July 2021} 

 \abstract{Many scattered light images of protoplanetary disks have been obtained with the new generation of adaptive optics (AO) systems at large telescopes. The measured scattered radiation can be used to constrain the dust that forms planets in these disks.} 
{We want to constrain the dust particle properties for the bright, pole-on
transition disk around HD~169142 with accurate measurements and a
quantitative analysis for the polarization and intensity of the scattered
radiation.}
{We investigate high resolution imaging polarimetry of HD~169142 taken in the R$'$ and I$'$ bands with the SPHERE/ZIMPOL AO instrument. The geometry of this pole-on disk is close to rotational symmetry, and we can use azimuthally averaged radial profiles for our analysis. We describe the dependence of the disk polarimetry on the atmospheric turbulence, which strongly impacts  the AO point spread function (PSF). With non-coronagraphic data we can analyze the polarimetric signal of the disk simultaneously with the stellar PSF and determine the polarization of the disk based on simulations of the PSF convolution. We also extract the disk intensity signal and derive the fractional polarization for the R$'$ and I$'$ bands. We compare the scattered flux from the inner and outer disk rings with the corresponding thermal dust emissions measured in the IR and estimate the ratio between scattered and absorbed radiation.} 
{We find for the inner and outer disk rings of HD~169142 mean radii of $170\pm 3$~mas and $522\pm 20$~mas, respectively, and the same small deviations from a perfect ring geometry as previous studies. The AO performance shows strong temporal variation because of the mediocre seeing
of about $1.1''$; this produces PSF peak variations of up to a factor of four and strongly correlated changes for the measured disk polarization of about a factor of two for the inner disk ring and about 1.2 for the more extended outer disk.
This variable PSF convolution effect can be simulated and accurately corrected, and we obtain ratios between the integrated disk polarization flux and total system flux ($\widehat{Q}_\phi/I_{\rm tot}$) of $0.43\pm 0.01~\%$ for the R$'$ band and $0.55\pm 0.01~\%$ for the I$'$ band. This indicates a reddish color for the light reflection by the dust. The inner disk ring contributes about $75~\%$ and the outer disk about $25~\%$ to the total disk flux. The extraction of the scattered intensity of the disk is only possible for the bright, narrow, inner disk ring, and the obtained fractional polarization $\widehat{p}$ for the scattered radiation is $23.6\pm 3.5$~$\%$ for the I$'$ band and $22.0\pm 5.9$~$\%$ for the R$'$ band. The ratio between scattered disk flux and star flux ($\widehat{I}_{\rm disk}/I_{\star}$) is about $2.3\pm 0.3~\%$. This is much smaller than the derived IR excess $F_{\rm fIR}/F_\star=17.6~\%$ for the disk components observed in scattered light. This indicates that only a small fraction of the radiation illuminating the disk is scattered; most is absorbed and reemitted in the IR.}
{We demonstrate the feasibility of accurate quantitative photo-polarimetry of a circumstellar disk  with a radius of less than 0.2$''$, observed with ground-based AO systems, if the PSF convolution effects can be properly taken into account. Accurate measurements are a pre-requisite for finding differences in the dust properties for different disks. The derived fractional polarization of about $23~\%$ in the R and I bands for the compact (20~AU) inner disk of HD~169142 is
lower than the measurement for the more extended disk HD~142527 for the
same wavelength range and significantly lower than the estimates for near-IR data of other
extended protoplanetary disks.}


\maketitle
%

\section{Introduction}\label{sec:intro}
Planetary systems are formed in circumstellar disks made of gas and dust around pre-main-sequence stars. Much has been learned about these systems in recent years thanks to improved observational techniques. Important information about disks, in particular about the dust in the disks, is gained from the analysis of the spectral energy distribution \citep[SED; e.g.,][]{Woitke16}, from detailed Atacama Large Millimeter/submillimeter Array (ALMA) images of the cold, large dust particles in the disk midplane \citep[e.g.,][]{vanderMarel17}, from IR interferometry of the hot dust near the star \citep{Kluska20} and from the imaging in the near-IR and visual of the light scattered by the dust at the disk surface \citep[e.g.,][]{Avenhaus18}. Each of these techniques provides complementary information and improves our understanding of the role of the dust in the planet formation process in circumstellar disks.  
This work describes quantitative measurements of the polarization and intensity of the light scattered by the dust in the disk around HD~169142 based on archival data from the Very Large Telescope (VLT) adaptive optics (AO) instrument Spectro-Polarimetric High-contrast Exoplanet REsearch (SPHERE)/Zurich Imaging Polarimeter (ZIMPOL). A quantitative analysis is a prerequisite for constraining with the model simulation the scattering albedo, the size distribution, and the compactness of the scattering particles, which can be compared with expectations for different dust types, such as icy high albedo grains, carbon-rich dark particles, or highly porous aggregates \citep[e.g.,][]{Kolokolova10,Min16,Tazaki19}. Such investigations will complement studies on the thermal emission of the dust and help us better understand the evolution of the dust particles in these disks and the resulting composition of the forming planets.

The disk integrated intensity (or spectral flux density) is low when compared to the central star, on the order $I_{\rm disk}/I_{\rm star}\la 3~\%$, and depends on the disk geometry. The surface brightness is relatively high for strongly illuminated regions, such as the inner disk wall of transition disks, and faint for disk regions, which are far from the star and illuminated only under gracing incidence, or regions located in the shadow cast by disk structures farther in \citep{Garufi17b}. The measurement of scattered light from a protoplanetary disk requires an instrument with high spatial resolution and high contrast capabilities to resolve the strongly illuminated inner disk regions.  

Early detections of the scattered light from protoplanetary disks  were achieved with the first AO systems at large telescopes and with Hubble Space Telescope (HST) \citep[e.g.,][]{Roddier96,Silber00}. Initially, only extended targets, $r>1''$, could be observed, for instance the circumbinary disk of GG Tau.
Differential polarimetry turned out to be a very useful high contrast technique for separating the scattered, and therefore polarized, light of the disk from the direct, unpolarized, bright starlight \citep{Gledhill89,Kuhn01}. This technique was combined in many high resolution instruments that use AO systems \citep[see][for a review]{Schmid21}, such as VLT/NACO \citep{Lenzen03}, Subaru/CIAO \citep{Murakawa04}, HICIAO \citep{Hodapp08}, Gemini/GPI \citep{Perrin15}, and VLT/SPHERE \citep{Beuzit19,Schmid18,deBoer20}.
Detections of the polarized intensity from about 100 protoplanetary disks have meanwhile been achieved with these systems, and some examples from different instruments are described in \citet{Apai04}, \citet{Quanz11}, \citet{Hashimoto11}, \citet{Muto12}, \citet{Benisty15}, \citet{Monnier17}, and \citet{Garufi17b}.
These data reveal a rich variety of disk morphologies and provide details about the geometric structure, disk dynamics, and possible interactions between the disk and newly formed planets. 

However, almost no accurate photo-polarimetric measurements of the scattered light from protoplanetary disks exist despite the many successful and very clear detections of polarization signals. Typically, rough values for the maximum fractional polarization are given, for example $p \approx 50~\%$ for GG~Tau for $\lambda=1~\mu$m \citep{Silber00}  and $\approx 55~\%$ for AB~Aur at 2~$\mu$m \citep{Perrin09},  based on HST polarimetry, or a maximum polarization of $p \approx 60~\%$ in the H band for HD~34700A \citet{Monnier19} based on ground-based AO observations. There are also strongly  controversial estimates for a given disk, for example for the eastern side of HD~142527 for which \citet{Canovas13} reported a value of $p_{\rm disk}<20$~\% in the H band while \citet{Avenhaus14} guessed a much larger value of $\approx 45~\%$ for the same wavelength. HD~142527 has a very bright and extended disk for which a determination of the fractional polarization is relatively easy. Therefore, \citet{Hunziker21} was able to obtain a first accurate photo-polarimetric measurement for this system for a protoplanetary disk, including a detailed assessment of bias effects and measurements of uncertainties. Among other parameters, the controversial H-band fractional polarizations of the eastern side were finally found to be $p=(35.1\pm 2.0)~\%$. In addition, the corresponding value of $(28.0~\pm 0.8)~\%$ for a visual broad band (VBB $\lambda_c=735$~nm) was also obtained, and this yields a well-defined wavelength dependence of the fractional polarization and, therefore, strong constraints on the scattering dust.  

Quantitative measurements such as those for HD~142527 should be obtained for many circumstellar disks, but this also requires the investigation of measuring procedures for systems with other geometries taken with different instruments and observational conditions. Important problems to be solved are the polarimetric calibration of the AO instrument, the accurate extraction of the disk intensity signal, and the correction of measuring results for the variable intensity smearing and polarization cancelation effects. It will be of great interest to achieve a high accuracy for disks with different geometries, ages, and host stars and for different wavelengths in order to constrain the evolution of the scattering dust particles.
All these challenges are easier or harder to solve depending on the properties of the investigated disk, the instrument used, and the quality of the observations. In this work we carry out a case study for accurate photometric and polarimetric measurements for the bright, pole-on, and almost centro-symmetric transition disk around HD~169142. 

The next section summarizes relevant properties of the HD~169142 system from the literature, and Sect.~3 describes the used SPHERE/ZIMPOL data from the European Southern Observatory (ESO) archive and the data reduction. The data analysis in Sect.~4 characterizes the geometric structure of disk rings, measures and models the radial polarization profile, extracts the disk intensity signal, and compares the scattered light intensities with the thermal dust emission taken from the literature. In Sect.~5 we discuss our measurements for the disk around HD~169142, deduce
scattering parameters for the dust, and provide constraints on the properties of the scattering dust, before concluding in Sect.\ 6.

\section{The HD 169142 disk}\label{11}
HD~169142 is a well-studied young star with a protoplanetary disk seen close to pole-on (see Fig.~\ref{hd169142img}). It was first identified as a ``Vega-like'' star with IR excess emission based on photometry from the IRAS satellite \citep{Walker88}. 
The system is located at a distance of 114~pc \citep{Gaia18} and it cannot be associated
with a well-known star forming region. \citet{Grady07} estimate an age in the range $3-12$~Myr from a close-by, co-moving T~Tauri star binary. Spectral classifications for HD~169142 range from A5~Ve ($T_{\rm eff} \approx 8400$~K) to A8~Ve and it could be a rapidly rotating A-star with a latitudinal temperature gradient \citep{Dunkin97,Grady07}. No clear signatures of gas accretion have been detected to a limit of about $< 2\cdot 10^{-9}\,{\rm M}_\odot{\rm yr}^{-1}$ based on far-UV spectroscopy with IUE satellite or X-ray observations with ROSAT and Chandra space telescopes \citep[see][]{Grady07}. The interstellar extinction is low $E(B-V)\approx 0.1$~mag and the stellar luminosity is estimated to be about 10 ${\rm L}_\odot$ \citep{Sylvester96,Woitke19}.

The transition disk IR excess is dominated by a cold dust component peaking around $25~\mu$m \citep{Sylvester96,Woitke19}. Mid-IR spectroscopy shows strong polycyclic aromatic hydrocarbon (PAH) emission around 6, 8 and 11~$\mu$m, but no typical silicate features \citep{Sylvester95,Meeus01,Acke04,Sloan05, Keller08}. There exists also hot dust with $T\approx 1500$~K at a separation of about 0.1~AU to the star as inferred from near-IR photometry and interferometry \citep[e.g.,][]{Chen18}. 

Scattered light from the disk was first detected for separations $r\approx 0.5''-1.2''$ with seeing limited polarimetric imaging \citep{Kuhn01,Hales06}. \citet{Grady07} measured with HST the disk brightness for  $r=0.57''-1.3"$ and found an azimuthally symmetric disk structure with a radial dependence of $\propto r^{-3.0}$. 
The combination of imaging polarimetry and AO with VLT/NACO by \citet{Quanz13} revealed a central cavity surrounded by a very prominent ring at $r\approx 0.18''$ then a circular gap extending to about $r\approx 0.50''$ surrounded by the weaker, outer disk seen previously. Subsequent imaging polarimetry confirmed this structure but also found deviations from perfect axis-symmetry, which could be caused by interactions with newly formed planets \citep{Momose15,Monnier17,Pohl17,Ligi18,Bertrang18,Bertrang20,Gratton19}. 

Some polarimetric measurements for the disk around HD~169142 are given in, for example, \citet{Quanz13}, \citet{Momose15}, and \citet{Monnier17} as surface brightness profiles, which are at least qualitatively in good agreement. Also, maps, azimuthal profiles, or parameters of a well-fitting disk model are presented \citep{Monnier17}. However, the extraction and comparison of quantitative polarimetric parameters are difficult, because the presented results are very heterogeneous and the involved point spread function (PSF) convolution effects are not described. A quite simple and well-defined polarimetric value is the ratio between the integrated polarized intensity of the disk and the total intensity of the system $Q_\phi/I_{\rm tot}=0.41~\%$ observed for the H band by \citet{Quanz13}, but this value is not corrected for PSF smearing and polarimetric cancelation effects. 

Radio-observations in the millimeter and submillimeter range provide complementary information about the disk structure. \citet{Raman06} derived from CO line maps a disk inclination of 13$^\circ$ and an orientation of the major axis of the projected disk of 5$^\circ$.  High resolution observations in the millimeter range with ALMA show clear circular structures for the thermal emission from large dust grains in the disk midplane and for the line emission from molecular species and both can be associated with the rings seen in scattered light \citep{Osorio14,Carney18,Macias19,Perez19}. The prominent inner ring and weaker features have also been imaged in the IR, in particular in the PAH emission bands as well as for the dust continuum $\lambda = 8.8 - 25~\mu$m \citep[e.g.,][]{Habart06,Honda12,Okamoto17}. 

Important for our study of the disk of HD~169142 is the location of the bright, narrow ring at an ideal separation for our AO observations, and the pole-on orientation providing an almost perfect azimuthal symmetry enabling the analysis of averaged radial profiles for the analysis of the disk polarization and intensity. Also useful is the well-established SED.

\section{Data and data reduction}
\subsection{SPHERE/ZIMPOL data}\label{13}
The data of HD~169142 used in this work were retrieved from the ESO data archive \footnote{http://archive.eso.org/} and are listed in Table~\ref{obsSHOW}. The observations were taken with the VLT instrument SPHERE/ZIMPOL on July~9-10, 2015, by H.~Avenhaus (ESO program 095.C-0404) and they are described in \citet{Bertrang18} including a detailed analysis of the geometric and hydrodynamic structure of the disk. Our study is complementary, because we investigate the photometric and polarimetric properties of the scattering dust in the disk and therefore our data description highlights the aspects relevant to this topic. Technical details about the SPHERE instrument and the ZIMPOL subsystem can be found  in \citet{Beuzit19} and \citet{Schmid18}, respectively, or in the ESO/VLT SPHERE user manual.

All data listed in Table~\ref{obsSHOW} are obtained with ZIMPOL imaging polarimetry in non-coronagraphic mode, with a pixel scale of $3.6~{\rm mas} \times 3.6~{\rm mas}$ and a field of view of about $3.6''\times 3.6''$ centered on HD~169142. Different instrument configurations were used on July 9 and July 10, and the observing conditions also changed.

Our study is mainly based on the data from July 9, 2015, for which the filters $R'$ ($\lambda_c=626~{\rm nm},\,\Delta\lambda=149~{\rm nm}$) and I$'$ ($\lambda_c=790~{\rm nm},\,\Delta\lambda=153~{\rm nm}$) were used simultaneously in ZIMPOL cam1 and cam2 or vice versa \citep{Schmid18}. Part of these data were taken in the high-gain (10.5~e$^-$/ADU) fast modulation polarimetry mode, allowing for observations without saturating the detector at the position of the bright central star. This observing strategy turned out to be most useful for our photo-polarimetry, because it provides the disk intensity and polarization signal simultaneously with the unsaturated stellar PSF in two wavelength bands. In addition, also slow modulation polarimetry was taken, which uses a low-gain (1.5~e$^-$/ADU) and low-read-out noise detector mode, which is optimized for the detection of fainter polarization signals.  The peak of the central star is saturated for these data but they trace better the structure of the outer disk because of the lower read-out noise. 

\begin{figure}[t]
    \includegraphics[trim=0 60 20 70, clip=true,width=0.5\textwidth]{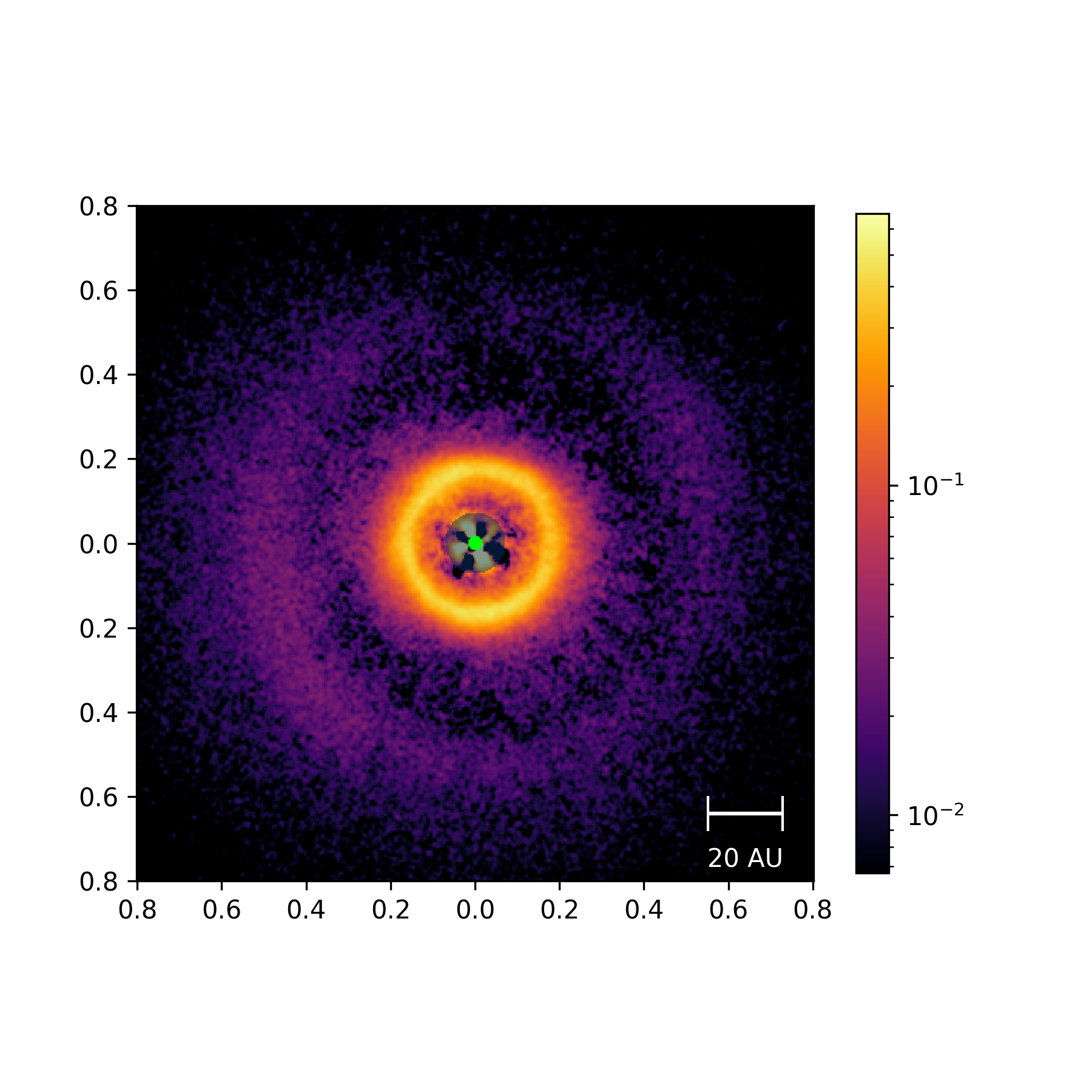}
    \centering
\caption{High signal-to-noise $Q_{\phi}$ azimuthal polarization image of
the HD~169142 protoplanetary disk obtained by stacking all R$'$, I$'$, and
VBB-filter frames from 2015 given in Table~\ref{avdataSPHERE}.
The position of the star is marked with a green circle, and the axes give the
separation in arcseconds. North is up and east to the left.} \label{hd169142img}
\end{figure}

\begin{table}[t]
\caption{\label{avdataSPHERE}Parameters for the used SPHERE/ZIMPOL
polarimetric observation cycles of HD~169142.} 
\label{obsSHOW}
\begin{tabular}{lccccccl}
\hline
\hline
$n_{\rm cyc}$ &  cam1 & cam2 &  det  & nDIT &  DIT & $t_{\rm tot}$ &  note \\
       & filt  & filt &  mode &      &  [s] &   [s]       & \\
\hline
\noalign{\smallskip}
\noalign{2015-07-09, seeing=1.1'', $\tau_0=1.5$ }
$1-4$  & I$'$  & R$'$ & fast & 6 & 6    &  576 \\
$5-8$  & I$'$  & R$'$ & slow & 6 & 10 &  960   & sat. \\     
$9-12$  & I$'$  & R$'$ & fast & 4 & 6    &  384 \\
$13-18$  & R$'$  & I$'$ & slow & 4 & 10 &  960   & sat. \\    
$19-22$  & R$'$  & I$'$ & fast & 4 & 6    &  384 \\
$23-26$  & I$'$  & R$'$ & slow & 4 & 10 &  640   & sat. \\   
$27-30$  & I$'$  & R$'$ & fast & 4 & 6    &  384 \\
\noalign{\smallskip}
\noalign{2015-07-10, seeing=0.75'', $\tau_0=2.1$}
$1-14$ & $VBB$  & $VBB$ & fast & 6  &   10 &  3360 &   sat. \\
\noalign{\smallskip}
\noalign{2018-07-15, seeing=0.51'', $\tau_0=7.5$, low wind effect}
$1-15$ & $VBB$  & $VBB$ & fast & 12  &  4.38 &  3153.6 &   sat. \\
\hline
\end{tabular}
\tablefoot{Seeing and atmospheric coherence time $\tau_0$ are median values for the HD~169142 data of that night; det.mode: detector mode; $t_{\rm tot}$ is
$n_{\rm cyc}\times 4 \times {\rm nDIT}\times {\rm DIT}$; sat: star saturated.}
\end{table}

A standard polarimetric measuring cycle consists of four polarimetric exposures, $Q^+$, $Q^-$, and $U^+$, $U^-$, which in turn consist of nDIT detector integration time (DIT) sub-integrations. The Stokes parameters $Q$ and $U$ for one cycle are obtained with the appropriate combination of exposures $Q=(Q^+-Q^-)/2$ and $U=(U^+-U^-)/2$ \citep{Schmid18}. For HD~169142 a single cycle took a few minutes of telescope time and adding up many cycles yields then a polarimetric observation with higher sensitivity. The polarimetric measurements depend on the PSF quality and the measured polarization is different from cycle to cycle depending on the variable atmospheric conditions and the related AO performance. For this reason Table~1 lists the sequence of polarimetric cycles for the main data for our analysis from July 9, 2015, in detail. The atmospheric conditions for July 9 were for VLT standards less than average, with a seeing around 1.0 arcsec and an atmospheric coherence time of about $\tau_0=1.5$~ms. 

HD~169142 was observed again on July 10, but in the very broad VBB filter ($\lambda_c=735~{\rm nm},\,\Delta\lambda=291~{\rm nm}$) in both ZIMPOL arms and in fast polarimetry mode. The wide VBB filter doubles roughly the flux throughput when compared to the R$'$ or I$'$-filters and therefore this data set provides a higher sensitivity, but the central star is saturated. These data are not included in our photo-polarimetric analysis because the central star cannot be used as simultaneous flux and PSF reference source.
The atmospheric conditions were good on July 10 with  a seeing of about 0.75~arcsec and these data were used for the study of the geometric structure of the disk by \citet{Bertrang18}.

\subsection{Data reduction}\label{datared}
The data were reduced with the IDL-based sz (SPHERE-ZIMPOL) software developed at the Eidgenössische Technische Hochschule (ETH), Zurich.
Basic steps include image extraction, bad pixel cleaning, bias subtraction, flat-fielding, a subtraction of the frame transfer smearing, and a calibration of the polarimetric modulation efficiency. Important for our analysis are the corrections for the differential polarimetric beam shifts, which can be easily determined from the PSF peak of the non-saturated, fast-polarization data. The star is centered in all images and the images are rotated so that north is up and east is to the left. 

In addition we consider the telescope polarization and polarization position angle offset $\delta_{\rm sz}$ of the ZIMPOL instrument as described in \citet{Schmid18}. The telescope polarization adds a fractional polarization $p_{\rm tel}$ to the data with a position angle $\theta_{\rm para}+\delta_{\rm tel}$, where $\theta_{\rm para}$ rotates with the parallactic angle of the observations so that the uncorrected, flux averaged fractional polarization of HD~169142 rotates along a circle in the $Q/I$-$U/I$-plane. The telescope polarization effects and the correction thereof are described in more details in Appendix~\ref{A1_telpol}. We find that the sum of interstellar and intrinsic polarization of HD~169142 is $p\lessapprox0.1~\%$.  This limit is more constraining than previous upper limits obtained with aperture polarimetry of HD~169142 by \citet{Yudin98} and \citet{Chavero06}. According to \cite{Schmid18} we also consider a
polarization position angle offset introduced by the half-wave plate HWP2 and apply for the final $Q$ and $U$ images an extra correction (formulas for $Q_{\rm c4},U_{\rm c4}$) with $\delta_{\rm sz}=+4.5^\circ$ and $+5.8^\circ$ for the R$'$ and I$'$-filters, respectively.

The primary data product of the reduction yields for each polarization cycle four images: the total intensity for the $Q$-measurement $I_{\rm Q}=I_0+I_{90}$, Stokes $Q=I_0-I_{90}$ and the equivalent parameters $I_{\rm U}=I_{45}+I_{135}$, $U=I_{45}-I_{135}$ for Stokes $U$. In principle it should hold that: $I_{\rm Q} = I_{\rm U} = I$.

We use for the polarization images of the disk the azimuthal Stokes parameters $Q_\phi$ and $U_\phi$ with respect to the central star. This approach for describing the polarization of centro-symmetric scattering geometries was initially introduced by \citet{Schmid06} with radial Stokes parameters $Q_r,U_r$, which are related to azimuthal parameters by $Q_\phi=-Q_{\rm r}$ and $U_\phi=-U_{\rm r}$. For circumstellar scattering as for disks the induced polarization is essentially azimuthal or in
$Q_\phi$ direction, while the $U_\phi$ signal is very small $|U_\phi|\ll|Q_\phi|$ in particular for axisymmetric disks seen close to pole-on. The polarized flux $p\cdot I=(Q_\phi^2+U_\phi^2)^{1/2}$ is therefore essentially equal to $\approx Q_\phi$, so that $Q_\phi$ can be used as polarized flux, with the advantage of avoiding bias effects from noisy observational data in the ``traditional'' formula for $p\cdot I$ using the square-root of the sum of squares \citep[see][]{Simmons85}. For HD~169142 the difference between $p\cdot I$ and $Q_\phi$ is small for the inner bright ring but for the faint, noisy outer disk the upward bias for the integrated polarized flux $p\cdot I$ is on the order of a factor of two as illustrated in \cite{Schmid21} and therefore not useful without a correction procedure that takes this strong bias effects into account. The azimuthal polarization is related to $Q$ and $U$ according to $Q_{\phi} = - Q \cos(2 \phi) - U \sin(2 \phi)$ and $U_{\phi} = - Q \sin(2 \phi) + U \cos(2 \phi)$, where $\phi$ is the position angle with respect
to the central star measured from north over east \citep[see also][]{Monnier19}.

For Fig.~\ref{hd169142img} all observations from 2015 listed in Table~\ref{avdataSPHERE} were normalized considering the different DITs, pixel gain values (slow versus fast polarization mode), and filter throughputs and then median-combined. The 2018 data are not included because significant disk movements occurred in the three years \citep{Bertrang20}.

\section{Analysis}\label{sec:analysisHD169142}

\subsection{Geometry of HD 169142}\label{21q}

\begin{figure}[ht]
  \includegraphics[width=0.5\textwidth]{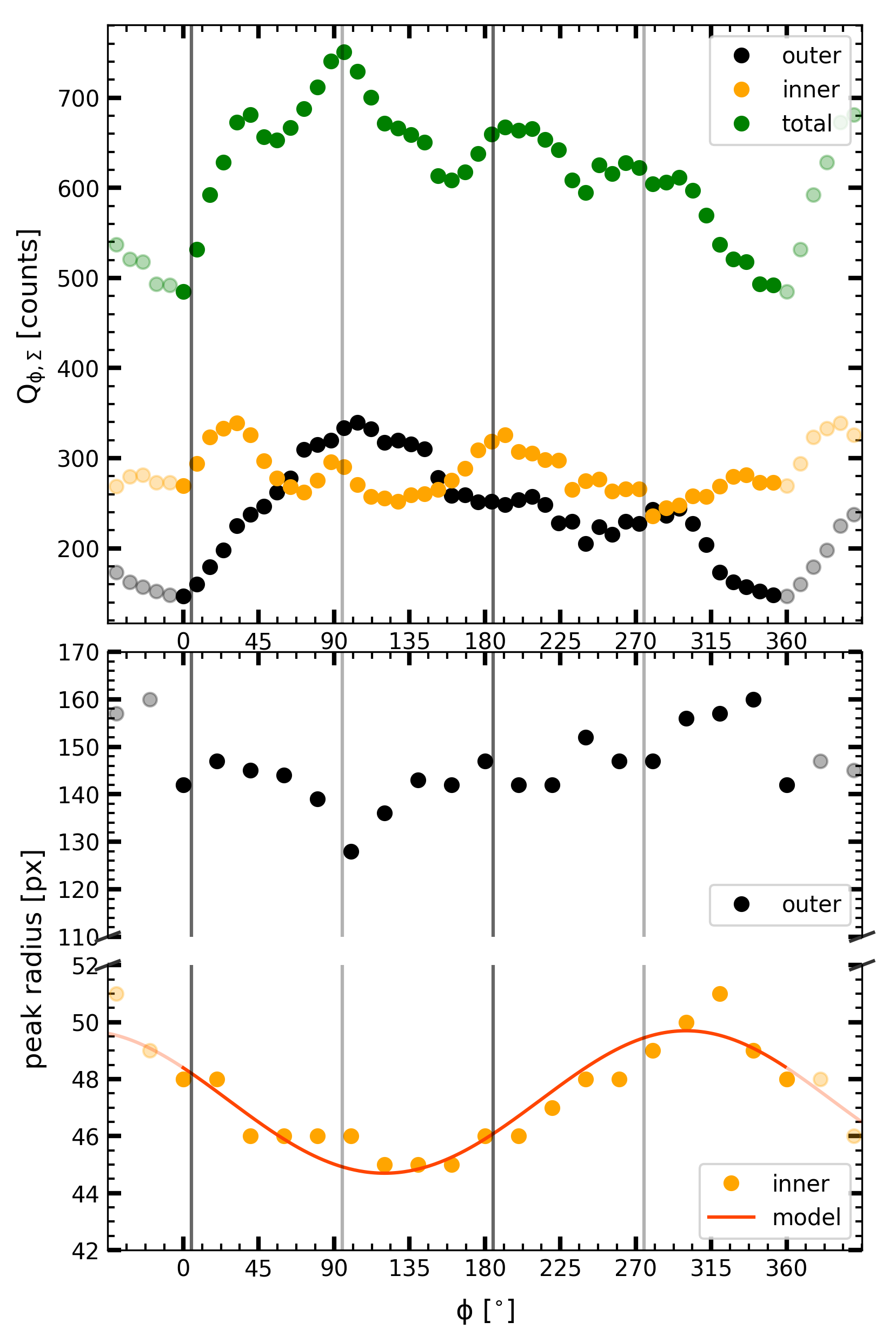}
     \centering
     \caption{ Azimuthal dependence of the polarized flux from the dust disk around HD~169142. Top: Radially integrated polarized flux $Q_{\phi}(\phi)$ for $8^{\circ}$ wide angles for the inner ($30-60$~px) and outer ($110-300$~px) ring and the total disk ($30-300$~px) as a function of azimuthal angle $\phi$.  Bottom: Separation $r_{\rm max}(\phi)$ of the $Q_\phi$ peak flux $r_{\rm max}={\rm max}_r(Q_\phi(r))$ in $20^{\circ}$ angle intervals for the inner and outer dust ring. The red curve is a simple fit of a circle with a 2.5~px center offset (1~px = 3.6~mas).
The major (darker) and minor axis position are marked with vertical gray lines.
     }
\label{geomOV}
\end{figure}

The median $Q_\phi$ polarized flux image of HD~169142 in Fig.~\ref{obsSHOW} is used for the analysis of the disk geometry. We focus our analysis on deviations from axisymmetry for the inner and outer ring. Figure~\ref{geomOV} shows the azimuthal dependence of the radially integrated polarization flux $Q_{\phi,\Sigma}(\phi)$ and the separation of the $Q_\phi$ flux peak $r_{\rm max}(\phi)$ for the inner and outer ring. 

Radio observations of molecular lines for HD~169142 provide kinematic disk maps that yield a major axis orientation of $5^\circ$ and an inclination of $12.5^\circ$ \citep{Raman06}. Brightness peaks in the disk move on Keplerian orbits in clockwise direction around the star \citep{Ligi18,Gratton19} indicating in combination with CO maps that the northern side is moving away from us and the eastern side is the near side of the disk.  We adopt these geometric parameters and the positions of the major and minor axes are also indicated in Fig.~\ref{geomOV}. 

The polarized flux (or azimuthal polarization) $Q_\phi$ as a function of azimuthal angle is calculated from radial integrations of all counts
\begin{displaymath}
  Q_{\phi,\Sigma}(\phi)=\int_{\phi 1}^{\phi 2}\int_a^b Q_\phi(r,\phi)\, r\, {\rm d}r\,{\rm d}\phi
\end{displaymath}
in azimuthal angle intervals of $\Delta\phi = \phi_2-\phi_1 = 8^\circ$. The radial integration goes from $30$~px to $60$~px for the inner ring, from 110~px to 300~px for the outer ring, and from 30~px and 300~px for the entire disk. The resulting $Q_{\phi,\Sigma}(\phi)$ curves are shown in Fig.~\ref{geomOV}. We measure roughly the same amount of polarized flux for the two rings. However, the intrinsic $Q_\phi$-flux of the inner ring is significantly higher after the consideration of observational bias effects as will be described in Sect.~\ref{diskmodchap}.

The shape of the $Q_{\phi,\Sigma}(\phi)$-curves is essentially identical to the J-band ($\lambda=1.26~\mu$m) measurements of HD~169142 presented in \citet{Pohl17} and also described in \cite{Bertrang18} for the VBB filter.  For the inner ring we find deviations $\sigma \approx 9~\%$ from the mean $\overline{Q}_{\phi,\Sigma}(\phi)=281$ count units with maxima at $\phi \approx 30^{\circ}$, $\phi \approx 90^{\circ}$, $\phi \approx 190^{\circ}$ and $\phi \approx 330^{\circ}$. The motion of these maxima is described in \cite{Ligi18} and \cite{Gratton19}.

The outer ring shows stronger azimuthal variations in $Q_{\phi,\Sigma}(\phi)$ with a $\sigma \approx 22\% $, in particular a bright eastern side of the disk, which is about 1.35 times brighter than the western side and with the peak value more than twice as bright than the strong minimum in the north.  We notice an anti-correlation between the brightness of the inner ring and the outer disk, which could be a result of shadowing of the outer disk by the inner ring. 

As second parameter, we derive the separation of the polarization flux peak $r_{\rm max}(\phi)={\rm max}_r(Q_\phi(r,\phi))$ for the inner and outer ring from the central star as a function of azimuthal angle $\phi$. The values are shown in Fig.~\ref{geomOV}, which were derived from smoothed mean radial profiles $Q_\phi(r)$ for $\Delta\phi$ angle intervals with widths of $20^\circ$.

The mean separation for the inner dust ring is $47.2\pm 1.7$ px ($0.170''\pm 0.006''$) and for the outer ring $145 \pm 7$ px ($0.522'' \pm 0.026''$). The inner ring shows for $r_{\rm max}(\phi)$ a systematic deviation from the mean, which looks like a sine-wave with an amplitude of $2.5\pm 0.3$~px. This is essentially equivalent to a circle with a center offset of 2.5~px in direction $\phi=300^\circ\pm 10$ from the star as illustrated in Fig.~\ref{geomOV} by the sine fit curve. The outer ring shows also a smaller separation for the eastern side with $r_{\rm max}\approx 135~px$ and a larger separation $\approx 150~px$ for the western side. These values are not accurate because the determination of the location of the flux peak is rather uncertain for the outer ring. Our determination of the ring radii for HD~169142 agrees with previous measurements of \citet{Bertrang18}.

The smaller ring separation toward the east and the enhanced brightness of the eastern side are expected features of a tilted disk with the eastern side closer to us. The enhanced brightness can be attributed to enhanced forward scattering and the smaller separation is a tilt projection effect for the illuminated surfaces above the disk midplane. However, the two effects are small because the disk inclination is only $i=12.5^\circ$ and they are partly masked by small deviations from axisymmetry of the intrinsic disk structure.

Circumstellar dust rings around HD~169142 are also seen in continuum observation in the millimeter range, which represent the distribution of the large dust grains in the disk midplane. For example the maps of \cite{Perez19} or \cite{Macias19} show rings with only small deviations from circles. The radius of the inner ring is in millimeter wavelengths around $0.22''$, which is substantially larger than the radius measured for the inner ring in scattered light. This radius difference between scattered light and thermal light in the millimeter range is often seen in transition disks \citep[e.g.,][]{Villenave19}, and this is explained by the accumulation of large grains in the gas pressure maximum of the ring, which is, in transition disks, farther out than the scattered light emission from the inner wall. For the outer ring of HD~169142, the millimeter-wave ring has a radius of about $0.53''$, which coincides well with the peak flux radii derived for the scattered light. 

\subsection{Radial polarization profiles}
\subsubsection{Observed mean radial profiles for the polarized flux} 
\label{21b}
Because the disk in HD~169142 is very close to rotational symmetric we can use azimuthally averaged radial polarization profiles $Q_\phi(r)$ for our analysis. This yields a much improved sensitivity, simplifies significantly the analysis, and still provides measurements that represent the properties of the disk well at all azimuthal angles. 
\begin{figure}[ht!]
    \includegraphics[width=0.46\textwidth]{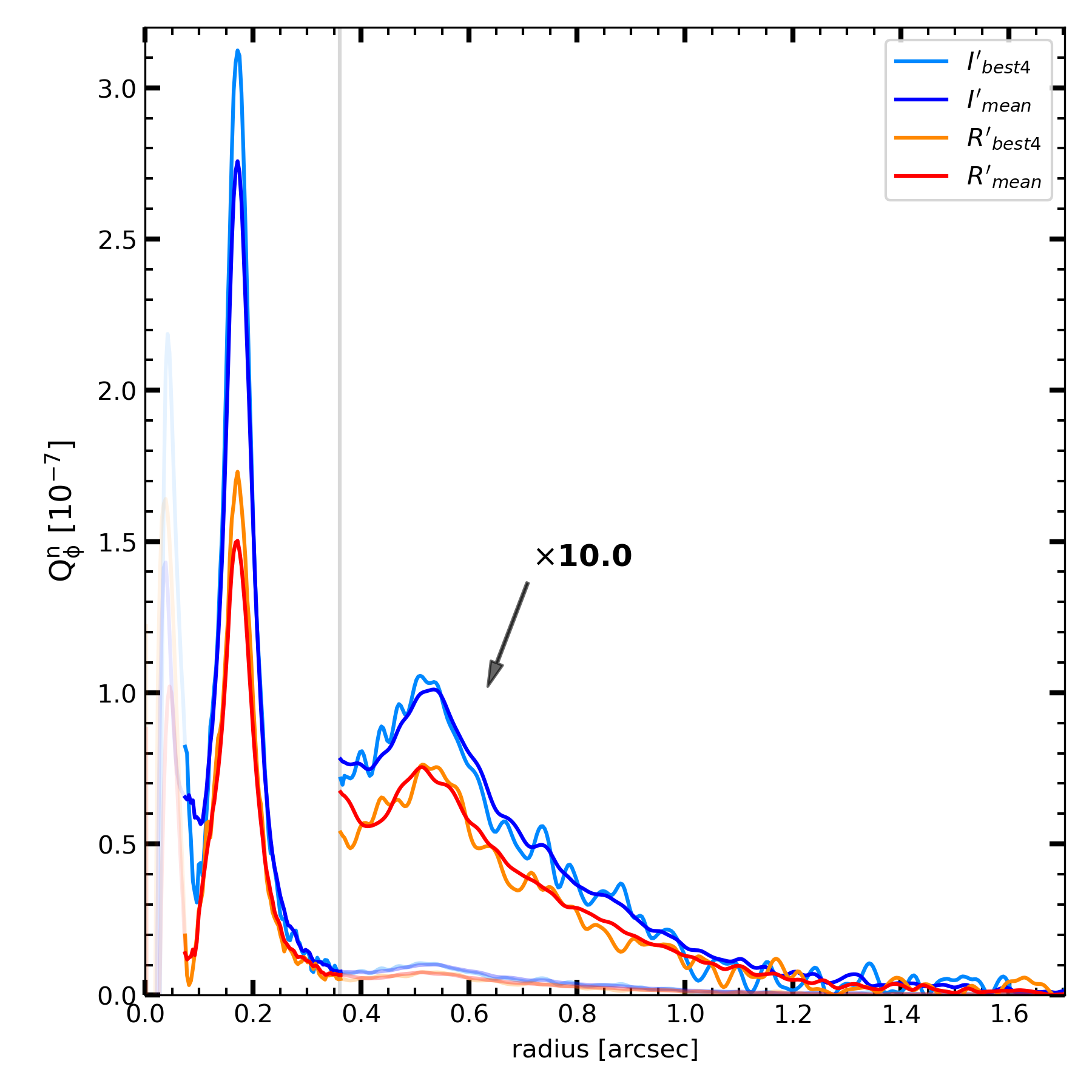}
     \centering
     \caption{Mean radial profiles for the azimuthal polarization $Q_{\phi}^{\rm n}(r)$ for all R$'$(red)- and I$'$(blue)-band cycles from July 9, 2015, for the disk around HD~169142, plotted together with the mean of the four best cycles, shown in lighter colors. For $r>0.36"$ the profiles are smoothed, and the same profiles multiplied by ten are plotted again for better visibility. For $r<0.07"$ the data are dominated by residual noise from the PSF peak. }
\label{finalIRall}
\end{figure}

We compile in Fig.~\ref{finalIRall} the observed radial distribution of the polarized flux $Q_\phi(r)$ for R$'$- and I$'$ bands out to $r=1.7"$. These averaged (red and blue) profiles include all data from July 9, 2015, also the saturated slow polarimetry data, considering the correct weighting of frames taken with different DITs and polarimetric detector mode (gain factors 1.5 and 10.5 e$^-$/ADU). The curves $Q_\phi^{\rm n}(r)=Q_\phi(r)/I_{\rm tot}$ are normalized for each band to the total intensity of the star as measured for the fast polarization data in an aperture with a diameter of $3''$ corresponding to a radius of 416~pixels. The slow polarimetry data reduce significantly the noise further out and therefore this ``total'' average is ideal to measure $Q_\phi^{\rm n}(r)$ for the weak, outer disk $r> 0.36''$. Because of the saturation of the star in the slow polarization data the scale of these images is adjusted to the normalized, and non-saturated fast polarimetry data.
The final radial curves in Fig.~\ref{finalIRall} were smoothed over $0.018"$ (5~px) for the outer disk because of the low signal and over $0.07''$ inside because of the noise from the stellar PSF peak. The observed peak surface brightness in the mean radial profiles differs by more than a factor of ten between inner ring and outer disk with $Q_{\phi,{\rm max}}^{\rm n}=2.7$ and 0.098 for the I$'$ band and 1.5 and 0.073 for the R$'$ band in units of $10^{-7}$ (relative to the total flux of the star), respectively.

We notice strong temporal changes in $Q_\phi^{\rm n}(r)$ for the inner ring caused by PSF variations. Figure~\ref{finalIRall} includes as illustration for these variations averaged profiles for a series of four consecutive cycles 19 to 22 (Table~\ref{obsSHOW}) taken under better atmospheric conditions and therefore with better PSFs. For the inner ring the difference in $Q_\phi$ is clearly visible, while the extended outer disk is much less affected. 

\subsubsection{Impact of the PSF variability}\label{ipsfv}

We investigate the dependence of $Q_\phi$ on the temporal PSF variability with the fast polarimetry data, because they provide the disk profile and the instrumental PSF simultaneously. The observed $Q^{\rm n}_\phi(r)=Q_\phi(r)/I_{\rm tot}$-profiles of the inner disk vary strongly from cycle to cycle.

\begin{figure}[ht!]
    \includegraphics[width=0.46\textwidth]{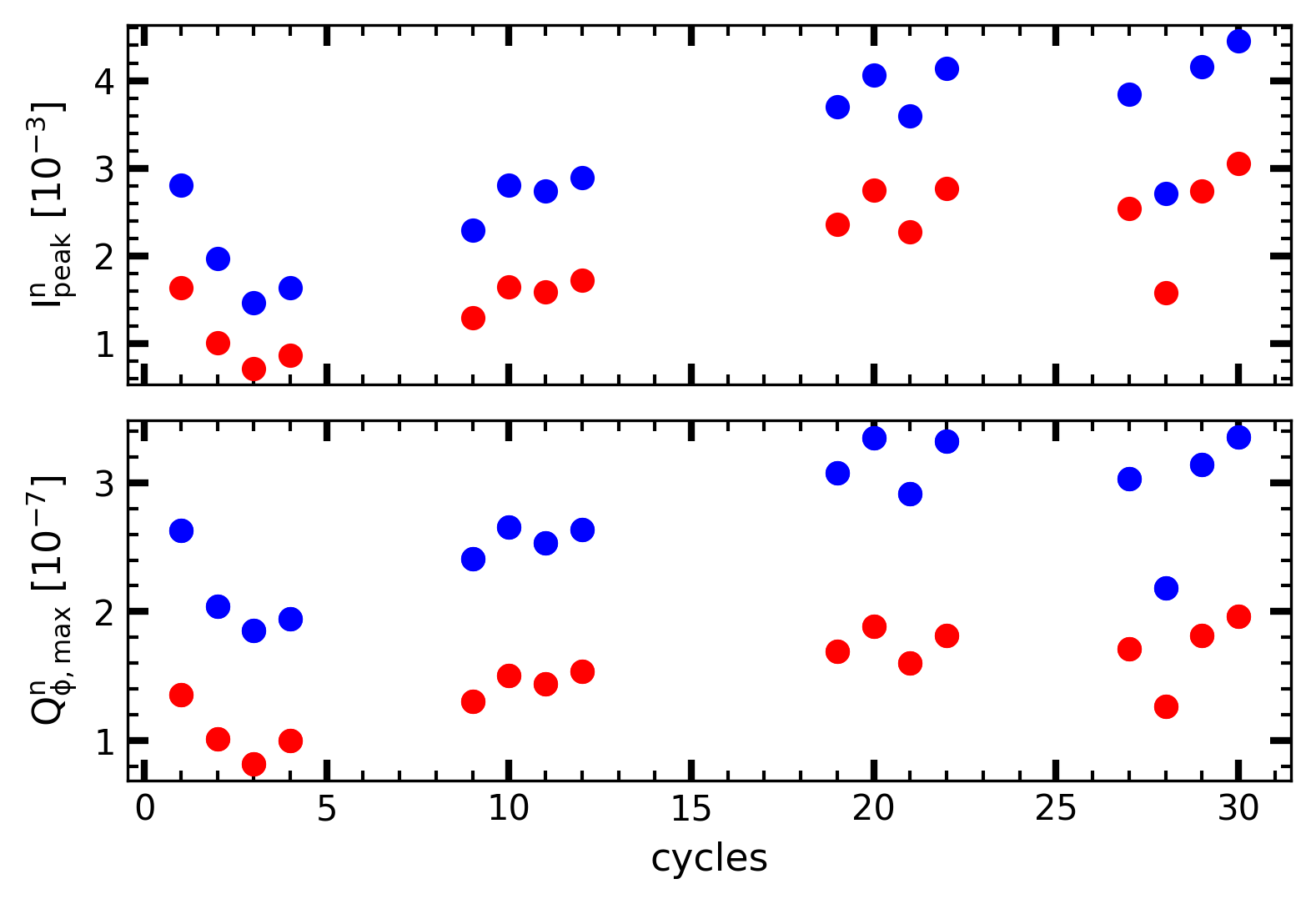}
    \centering
    \caption{Cycle-to-cycle variations in the normalized PSF peak intensity
      $I_{\rm peak}^{\rm n}$ and the peak polarization $Q_{\phi,max}^{\rm n}$ for the
      inner ring of HD~169142 for the R$'$ band in red and the I$'$ band in
      blue (unsaturated fast polarization data only).}
\label{atmVARmaxiqphi}
\end{figure}
\begin{figure}[ht!]
\centering
 \includegraphics[width=0.46\textwidth]{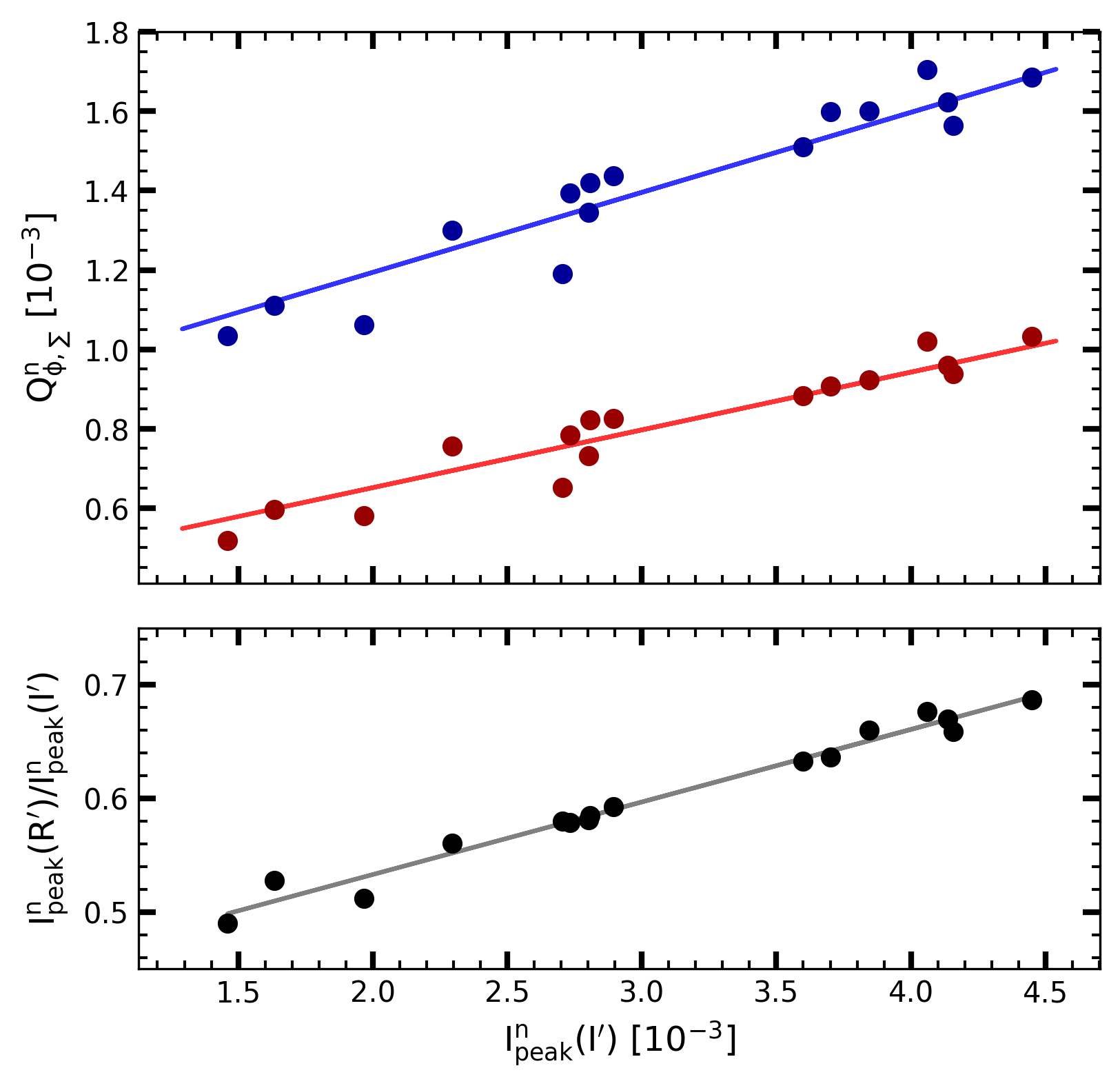}
 \caption{ Correlations with the PSF peak quality $I_{\rm peak}^{\rm n}({\rm I}')$. Top: Correlation between the polarized flux $Q_{\phi,{\sum}}^{\rm n}$ for the inner dust ring and the PSF peak $I_{\rm peak}^{\rm n}$ for the R$'$(red) and I$'$(blue) band. 
Bottom: Correlation of the ratio of peak intensities between the R$'$ and I$'$ band and the $I_{\rm peak}^{\rm n}({\rm I}')$ as an illustration of the stronger dependence of the PSF peak intensity in the R$'$ band on atmospheric conditions.}\label{231irpeakcor}
\end{figure}

Figure~\ref{atmVARmaxiqphi} shows, for each fast polarimetry cycle and both color bands, the variations in the peak flux in the stellar PSF $I_{\rm peak}^{\rm n}$ and the  peak polarization $Q_{\phi,max}^{\rm n}$ polarization for the inner disk ring around $r=0.18''$ for the R$'$ and the I$'$ band. All these values are again normalized to the total star intensity $I_{\rm tot}$ (HD~169142 flux) in a large aperture that is not affected by atmospheric seeing variations. The normalized peak flux $I_{\rm peak}^{\rm n}$ is a simple measure of the Strehl ratio of the data \citep[see][]{Schmid18}.  
As the HD~169142 observations were taken with quite strong atmospheric turbulence, the PSF $I_{\rm peak}^{\rm n}$ shows temporal variations of up to a factor of about four (max/min). Obviously the observed disk peak polarization $Q_{\phi,max}^{\rm n}$ is strongly correlated with variation of up to a factor of about two. The R$'$- and I$'$-band data behave very similar, because they were taken strictly simultaneously with the two arms of ZIMPOL. In Appendix~\ref{A2_psf} we show two-dimensional PSFs and variations and compare the achieved quality.

The variability is investigated in more detail in Fig.~\ref{231irpeakcor}, which shows very well-defined correlations between the integrated polarization $Q_{\phi,\sum}^{\rm n}$ for the inner dust ring and the PSF peak flux. This shows how well the polarization flux measurements are correlated with the instantaneous AO performance.

The color ratio of the normalized PSF peak flux $I_{\rm peak}^{\rm n}({\rm R}')/I_{\rm peak}^{\rm n}({\rm I}')$ in Fig.~\ref{231irpeakcor} shows a positive correlation with PSF quality ($I_{\rm peak}^{\rm n}({\rm I}')$). This indicates that the R$'$-band PSF depends stronger on the observing conditions than the I$'$ band because the atmospheric turbulence affects shorter wavelengths more. 

Of course, measuring such a tight correlation in two color bands for the variations in the stellar PSFs or for the disk polarization is only possible because the data are taken strictly simultaneous in the ZIMPOL double beam imaging polarimeter. This is very useful for an accurate measurement of the color of the disk.

\subsection{Disk polarization model}\label{diskmodchap}

An accurate determination of the intrinsic disk polarization requires an appropriate consideration of the smearing and polarization cancelation effects for our data. For this, we construct a model simulating the intrinsic disk polarization for HD~169142 that reproduces the observed polarimetric signal quite well if convolved with the corresponding PSF available from the simultaneous, unsaturated intensity frame. Our model is an azimuthally averaged radial profile for which we selected after various tests a two-component model $\widehat{Q}_\phi(r)=\widehat{Q}_{\phi,1}(r) + \widehat{Q}_{\phi,2}(r)$.
The inner component is described by a narrow Gaussian disk ring 
\begin{equation*}\label{1g}
\widehat{Q}_{\phi,1}(r) = A_1 \cdot exp \left( - \frac{(r-r_1)^2}{2 \cdot \sigma_1^2} \right),
\end{equation*} 
with peak radius $r_1$, peak amplitude $A_1$, and width $\sigma_1$. The outer component is more extended and can be described by two power laws 
\begin{equation*}\label{1gsl}
\widehat{Q}_{\phi,2}(r) =  A_2 \cdot \left(  \left( \frac{r}{r_2}\right)^{-2 \cdot \alpha_{in}} + \left( \frac{r}{r_2}\right)^{-2 \cdot \alpha_{out}}\right)^{-\frac{1}{2}},
\end{equation*}
with amplitude~$A_2$, approximate peak radius~$r_2$, and inner power law slope $\propto r^{\alpha_{\rm in}}$ with $\alpha_{\rm in} > 0$ and outer slope $r^{\alpha_{\rm out}}$ with $\alpha_{\rm out} < 0 $. We then searched for the $\widehat{Q}_\phi(r)$ model parameters that best fit the observations after convolution with the PSF using a forward modeling method based on approximate Bayesian computations \citep[e.g.,][]{Sisson18}. This is achieved in two steps, first determining the best parameters for the inner disk ring and then for the outer disk, because these two polarization components are essentially detached.

\begin{figure}[ht!]
    \includegraphics[trim=1 1 1 1, clip=true, width=0.48\textwidth]{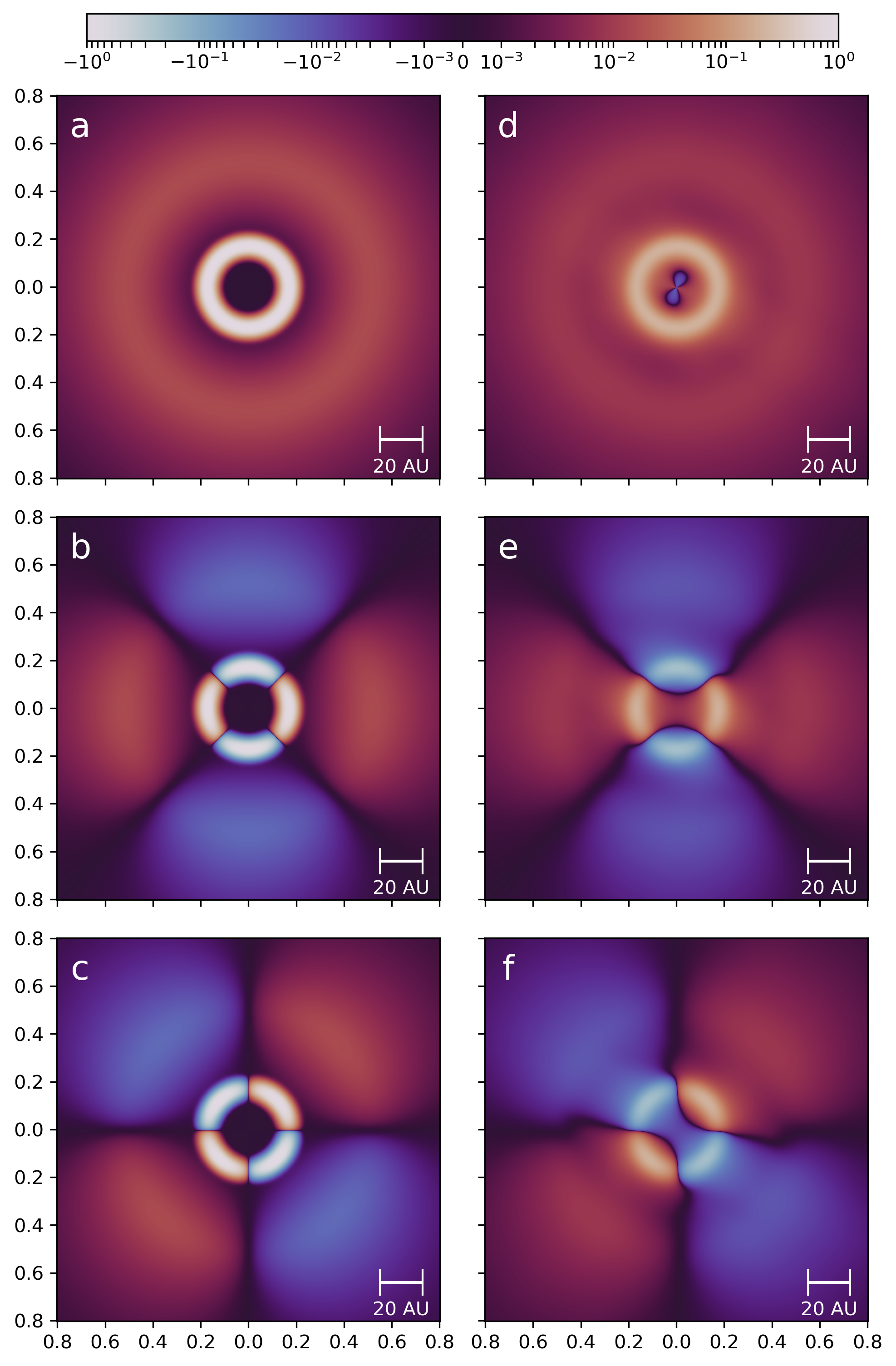}
     \centering
     \caption{Comparison of the Stokes $\widehat{Q}_{\phi}$ (a), $\widehat{Q}$ (b), $\widehat{U}$ (c) disk model and the $Q_{\phi}$ (d), $Q$ (e), $U$ (f) simulation for the PSF-convolved model (PSFs from the four best cycles) for the R$'$ band.  All images are normalized to a model maximum 1. }
\label{hd24comp_n}
\end{figure}

In the forward modeling a large range flat prior was used. Two-dimensional images for Stokes $\widehat{Q}$ and Stokes $\widehat{U}$ were generated for each set of geometric parameters assuming a scattering polarization in strictly azimuthal direction. Then the $\widehat{Q}$ and $\widehat{U}$ frames were convolved with $I$ of the specific observation to simulate the same observing conditions. The $I$ intensity image is a good approximation for the PSF, as more than 97~\% of the intensity is from the star and less than 3~\% from the disk. From the convolved Stokes $Q$ and $U$ images a convolved azimuthal polarization image $Q_{\phi}$ and the corresponding, azimuthally averaged radial profile $Q_{\phi}(r)$ was calculated. The model simulations for two-dimensional, non-convolved $\widehat{Q_\phi}$, $\widehat{Q}$, $\widehat{U}$ model and convolved $Q_\phi$, $Q$, $U$ simulation are shown in Fig.~\ref{hd24comp_n} for the R$'$ band. The small deviations from axisymmetry in the simulations are introduced by the observed PSF used for the convolution. 

 The quality of the calculated profile $Q_{\phi}(r)$ simulation was then evaluated with a comparison with the observed profile $Q_{\phi}(r)$ using as distance metric the sum of the squared residuals and the best solution was adopted. For the optimization we selected the R$'$ and I$'$ data from the fast polarimetry cycles 19-22 (illustrated in lighter color in Fig.~\ref{finalIRall}), which were obtained under the best atmospheric conditions. These data provide the highest spatial resolution and therefore constrain best the ring width $\sigma_1 \approx 5$~px (FWHM $ \approx 11.8$~px or 43~mas) and amplitude $A_1^{\rm n} \approx 10^{-6}$, while the best ring radius $r_1=47.5~$px (171~mas) depends not so much on the data quality.

The derived flux maxima $A_1^{\rm n}$ should be considered as lower limit and the obtained $\sigma_1$ values as upper limits for the inner disk ring. Assuming a narrower ring with a higher peak leads to convolved inner ring images that are essentially indistinguishable. In addition, the ring width of the observed azimuthally averaged profile $Q_\phi(r)$ is slightly larger than the width of the ring for a particular section in the two-dimensional image, because the ring is not exactly centered on the star and slightly distorted so that the azimuthal averaging increases the width.
Therefore, the indicated FWHM of 43~mas for the width of the inner ring is an upper limit for the intrinsic model. Of course, a model with a narrower ring would have an enhanced $A_1^{\rm n}$ parameter so that the total ring flux $\widehat{Q}_{\phi,\Sigma}^{\rm n}$ is preserved. The $\widehat{Q}_{\phi}^{\rm n}(r)=\widehat{Q}_{\phi}(r)/I_{\rm tot}$ model parameter results for R$'$ and I$'$ band are listed in Table~\ref{diskmodtab}.

\begin{table}
  \caption{\label{diskmodtab} Disk model parameters for the normalized
    polarized
    flux $\widehat{Q}_{\phi}^{\rm n}$ for the R$'$ and I$'$ band, where fluxes are
    expressed relative to the total R$'$ and I$'$ intensity of the
    HD 169142 system. }
\centering
\begin{tabular}{lll}
\hline\hline
         & R' & I'  \\
\hline
\noalign{\smallskip}
\noalign{model parameters} 
$A_1^{\rm n}$  [px$^{-2}$] &     $(0.8 \pm 0.04) \cdot 10^{-6}$  &  $(1.1 \pm 0.03) \cdot 10^{-6}$\\
$r_1$  [px]   &    $47.5 \pm  0.3$&  $47.6 \pm 0.2$\\
$\sigma_1$  [px]  &    $5.2 \pm 0.4$&  $4.7 \pm 0.3$ \\
$A_2^{\rm n}$  [px$^{-2}$] &     $(1.8 \pm 0.1) \cdot 10^{-8}$  &  $(2.4 \pm 0.2) \cdot 10^{-8}$\\
$r_2$  [px] &    $146 \pm 6$ &  $135 \pm 6$ \\
$\alpha_{in}$    &   $3.5 \pm 0.7 $& $5.4 \pm 0.9$\\
$\alpha_{out}$   &    $-(4.3 \pm 0.4)$ & $-(3.6 \pm 0.3)$ \\
\noalign{\smallskip}
\noalign{integrated polarized flux $\widehat{Q}_{\phi,\Sigma}^{\rm n}$} 
inner ring & $(3.09 \pm 0.06) \cdot 10^{-3}$ & $(3.94 \pm 0.05) \cdot 10^{-3}$ \\
outer disk & $(1.08 \pm 0.06) \cdot 10^{-3}$ & $(1.44 \pm 0.05) \cdot 10^{-3}$ \\
total disk & $(4.30 \pm 0.10) \cdot 10^{-3}$ & $(5.48 \pm 0.10) \cdot 10^{-3}$\\
\hline    
\end{tabular}
\tablefoot{Inner: 30-60 px, outer: 110-300 px, total: 30-300 px; one pixel corresponds to 3.6~mas.}
\end{table}

The convolution of the bright inner ring with a SPHERE/ZIMPOL PSF yields in the radial profile a weak but significant maximum around $r\approx 150$~px (inner ring radius plus $\approx 100$~px). This artifact is caused by the PSF speckle ring, which is related to the AO control radius of about 100~px for the R band and slightly larger for the I band. The observed peak surface brightness of the outer disk $Q_{\phi,2,{\rm max}}^{\rm n} $ is of order 4.5 $\%$ of the inner ring. However, the observed integrated polarization $Q_{\phi,2,\Sigma}^{\rm n}$ is around 75 $\%$ of the inner disk. About 1/6 of the observed $Q_\phi^{\rm n}$ signal at the position of the outer ring is caused by this PSF convolution artifact from the inner ring. This effect is taken into account in our analysis.

The forward modeling search of the intrinsic parameters of the outer disk is then carried out in a second step keeping the above derived parameters for the inner disk fixed. The found model parameters for the outer disk are also given in Table~\ref{diskmodtab}. One should keep in mind that the outer disk deviates significantly from rotational symmetry and therefore the obtained results represent the azimuthal average. The parameters also depend on each other. A slightly smaller model radius $r_2$ leads consequently to a steeper $\alpha_{\rm in}$ value and vice versa. In general the uncertainty for the fitting parameters of the outer disk is larger. 

\begin{figure}[ht!]
    \includegraphics[width=0.46\textwidth]{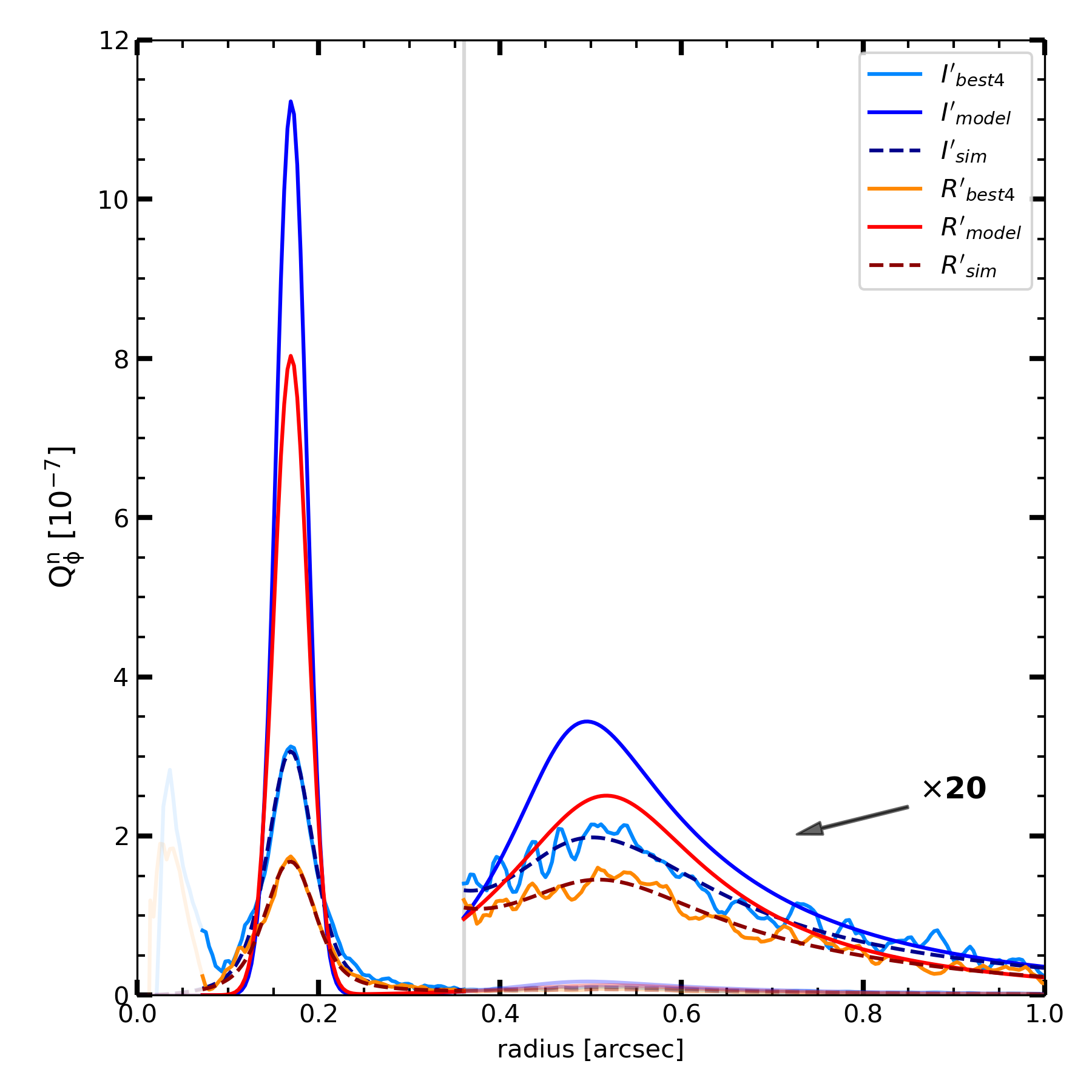}
     \centering
     \caption{R$'$-band (red) and I$'$-band (blue) radial profiles for
       the azimuthal polarization $Q_{\phi}^{\rm n}(r)$ of the disk around HD~169142
       for the derived model ($\rm R_{\rm model}', \rm I_{\rm model}'$), the simulated
       PSF-convolved model ($\rm R_{\rm sim}', \rm I_{\rm sim}'$), and the observation
       of the four best cycles ($\rm R_{\rm best4}', \rm I_{\rm best4}')$. For $r>0.36"$ the profiles are smoothed and all the profiles are plotted again, multiplied by 20, for better visibility. 
}
\label{finalIRunconvnew}
\end{figure}

The obtained intrinsic radial profiles $\widehat{Q}_{\phi}^{\rm n}(r)$ from the disk polarization model is compared in Fig.~\ref{finalIRunconvnew} with the observed profiles ${Q}_{\phi}^{\rm n}(r)$ for the R$'$ and I$'$ bands. The differences between the curves are caused by  the PSF smearing and cancelation effects, which are obviously very strong. The observed peak polarization surface brightness  for the inner disk for cycles 19-22 is reduced by factors of $Q_{\phi,{\rm max}}^{\rm n}/\widehat{Q}_{\phi,{\rm max}}^{\rm n}= 0.22$ and $0.28$ for the R$'$ and I$'$ bands, respectively. The effect is less dramatic for the outer disk, where these ratios are $0.7$ for both filters.

An important quantity for the characterization of disks is the integrated polarized fluxes of the scattered light normalized to the total system intensity flux. For this we integrate the $\widehat{Q}^{\rm n}_\phi(r)$ profiles
\begin{equation}
\widehat{Q}_{\phi,\Sigma}^{\rm n} =\int \widehat{Q}^{\rm n}_\phi(r)\,2\pi r\,{\rm d}r
\end{equation}  
for the entire disk, but also for the inner disk ring $\widehat{Q}_{\phi,1,\Sigma}^{\rm n}$ and the outer, more extended disk $\widehat{Q}_{\phi,2,\Sigma}^{\rm n}$. The resulting values are given in Table~\ref{diskmodtab}. Key numbers are the polarized reflectivity, the ratio between polarized flux and total system flux, which is $\widehat{Q}_{\phi,\Sigma}^{\rm n}=0.0043$ for the R$'$ band and $0.0055$ for the I$'$ band. Thus, the dust in the disk appears ``reddish'' in polarized flux, because it reflects more efficiently in the longer wavelength I$'$ band when compared to the R$'$ band. The same color effect is also obtained for the bright inner ring and the outer extended disk. 

The integrations for $\widehat{Q}_{\phi,1,\Sigma}^{\rm n}$ and $\widehat{Q}_{\phi,2,\Sigma}^{\rm n}$ yield also the relative contributions of >70~\% for the inner, narrow ring and $\approx $ 25~\% of the outer extended disk to the total polarized flux of the disk. This splitting is significantly different from the observations shown in Fig.~\ref{geomOV}, which yield about $Q_{\phi,1,\Sigma}^{\rm n}\approx 1.3 \cdot Q_{\phi,2,\Sigma}^{\rm n}$. The reason is the smearing and polarimetric cancelation, which reduces the integrated polarization signal of the compact ring by a factor of about 2.5 (I$'$) to 3.3 (R$'$), while the reduction for the extended outer disk is less dramatic with a factor of 1.3 (I$'$) or 1.4 (R$'$). Not considering the convolution effects would in our case, for the HD~169142 disk observed with SPHERE/ZIMPOL, underestimate the total polarized flux by about a factor of 1.9. 

\subsection{Simulation of observational variation }\label{23} 
\begin{figure}[ht!]
\centering
  \includegraphics[width=0.46\textwidth]{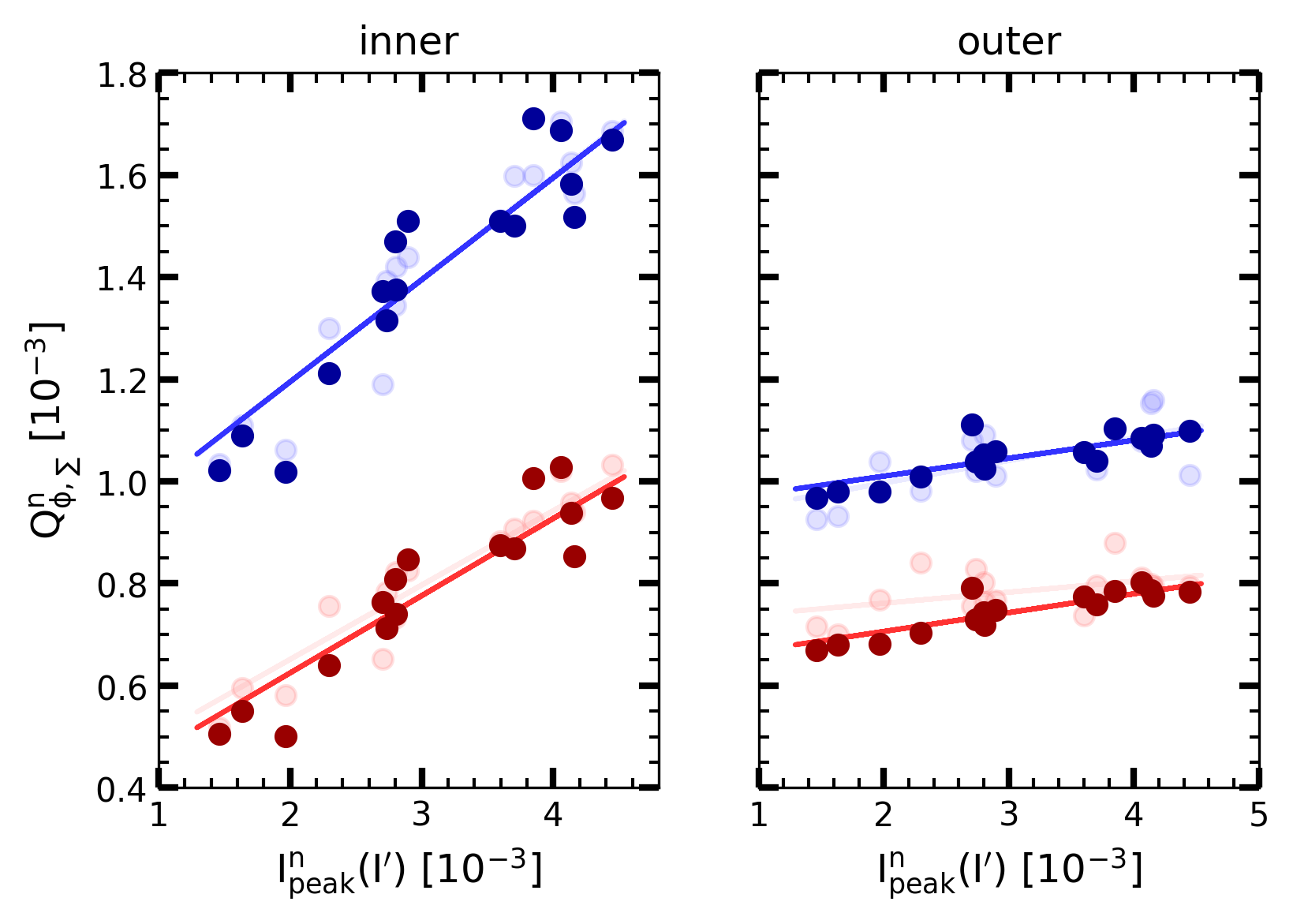}
 \caption{PSF-convolved polarization model flux $Q_{\phi,\sum}^{\rm n}$ and
   corresponding observed flux (light color) for the individual cycles
   for the inner ring (left) and outer disk (right) for R$'$band (red)
   and I$'$ band (blue). The points are plotted versus the I$'$-band
   PSF peak intensity $I_{peak}^{\rm n}(\rm I')$ as a measure of the
   observing conditions.
}\label{simvnew3}
\end{figure}

As a check of our analysis for cycles 19-22, we simulate the cycle-to-cycle variations for the observed polarized flux of the inner ring described in Sect.~\ref{ipsfv}. For this we convolve the derived disk polarization model with the PSF of each cycle and construct the same plot as for the observations (\ref{231irpeakcor}), but now for the simulated $Q_{\phi,\Sigma}^{\rm n}$ polarized disk signal variations in Fig~\ref{simvnew3}. The agreement with the observations is excellent for the relative level of the R$'$ and I$'$ band data, but also for the slope of the correlation between $Q_{\phi,\Sigma}^{\rm n}$ and the PSF peak flux $I_{\rm peak}^{\rm n}$. The mean of the residuals between the convolved model and the observations is essentially zero with a standard deviation of about $\sigma\approx 7~\%$.

The same analysis for the outer disk gives also a very good agreement. The slope of the $Q_{\phi,\Sigma}^{\rm n}$ -- $I_{\rm peak}^{\rm n}$ correlation is much shallower, because the PSF variations have a much smaller impact on extended structures. The observed $Q_{\phi,\Sigma}^{\rm n}$ varies only by about a factor of 1.2 between the cycles with the highest and lowest quality PSF. For the inner disk, the corresponding ratio is about 2. 

\subsection{Extracting the disk intensity signal}\label{diskint}
\begin{figure*}[hbt!]
\hspace*{0.4cm}
\begin{subfigure}{0.46\textwidth}
    \includegraphics[width=1\textwidth]{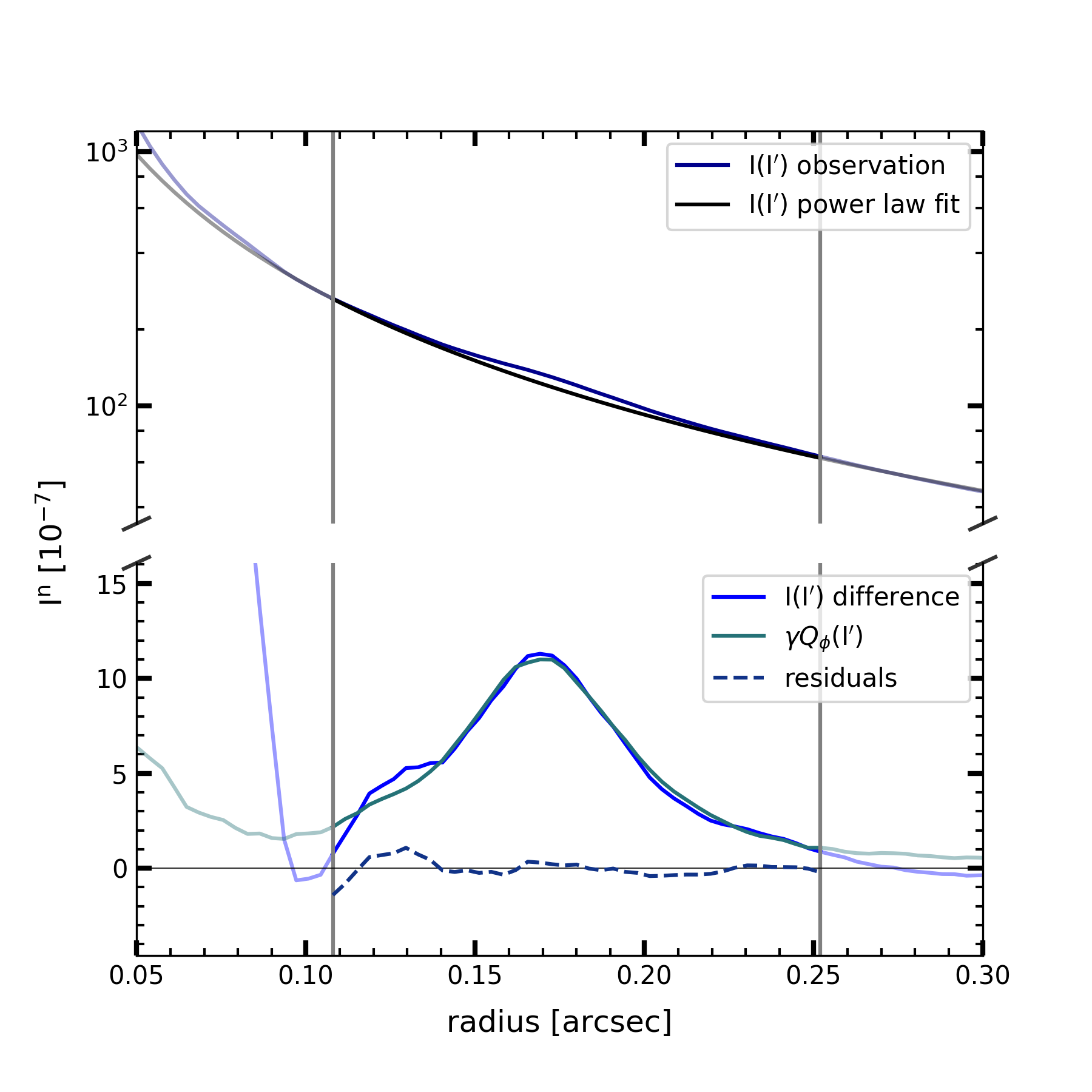}
     \centering
       \caption{I$'$ band.}\label{mcmccurve1i}
\end{subfigure}
\hfill
\begin{subfigure}{0.46\textwidth}
    \includegraphics[width=1\textwidth]{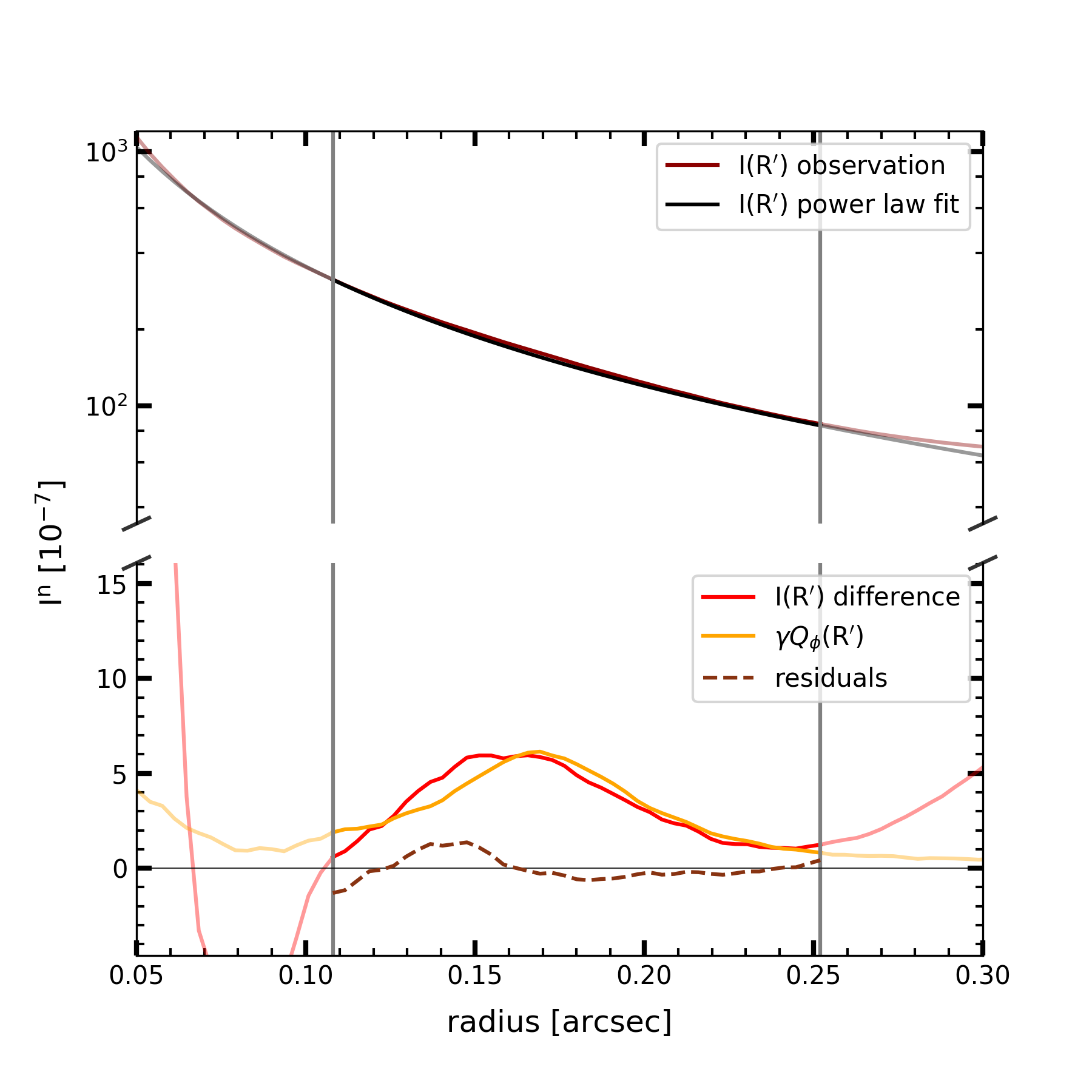}
     \centering
       \caption{R$'$ band.}\label{mcmccurveri}
\end{subfigure}
\hspace*{0.4cm}
\\
\caption{ Azimuthally averaged intensity profiles of
  HD~169142 showing the difference between the stellar PSF and the disk. Upper panels: Radial profiles for the I$'$ band (left) and the R$'$ band (right) for the
  observed intensity $I_{\rm tot}(r)$ and the best power-law fit
  for the stellar PSF $I_{\star}(r)$. Bottom panels: Profiles for
  the best fitting disk intensity model $I_{\rm d}(r)=\gamma Q_{\phi}(r)$,
  together with the difference $I_{\rm tot}(r)-I_{\star}(r)$ and the
  fit residuals (dashed lines). The MCMC fit optimization range is
  between 30 px (0.108'') and 70 px (0.252''), as indicated by the vertical
  lines.}
\label{mcmccurve}
\end{figure*}

The intensity images are strongly dominated by the variable PSF of the star, including an extended speckle halo and narrow features from the instrument as described in \citet{Schmid18}. The disk intensity $I_{\rm d}$ is much weaker and cannot be recognized by eye in the image, but we know accurately the position of the disk from the polarization signal $Q_\phi(r),$ which defines the expected intensity distribution $I_{\rm d}(r)$ quite well.

HD~169142 is a favorable target to extract the disk intensity with SPHERE/ZIMPOL observations, because this disk has a bright, narrow, ring located at $r=47$~px, far enough from the star to be well separated, but also clearly inside the strong speckle ring at $r\approx 100$~px (control radius of AO system). For the outer, extended disk there is not much hope that the disk intensity can be detected in our data. The observed peak polarization at $r=47$~px is $Q_{\phi,{\rm max}}^{\rm n}\approx 2\cdot 10^{-7}$, where the intensity of the stellar PSF is about 100~times  stronger. If we assume a fractional polarization in the range $p=Q_\phi/I\approx 10-50~\%$ for the scattered light, then the expected disk intensity is only a few to several percent of the stellar PSF intensity at this location. The extraction of such a weak disk intensity signal is difficult because of the described PSF variations.

For the extraction, we assume that the radial profile of the disk intensity is just a factor $\gamma = 1/p$ stronger than the polarized intensity $I_{\rm d}(r,\gamma) = \gamma Q_\phi(r)$. This assumption disregards a possible radial dependence $p(r)$ for the inner ring what is well justified because the radial structure of the inner ring is just barely resolved. 

\begin{figure}[ht!]
     \includegraphics[trim=30 1 30 1, clip=true, width=0.5\textwidth]{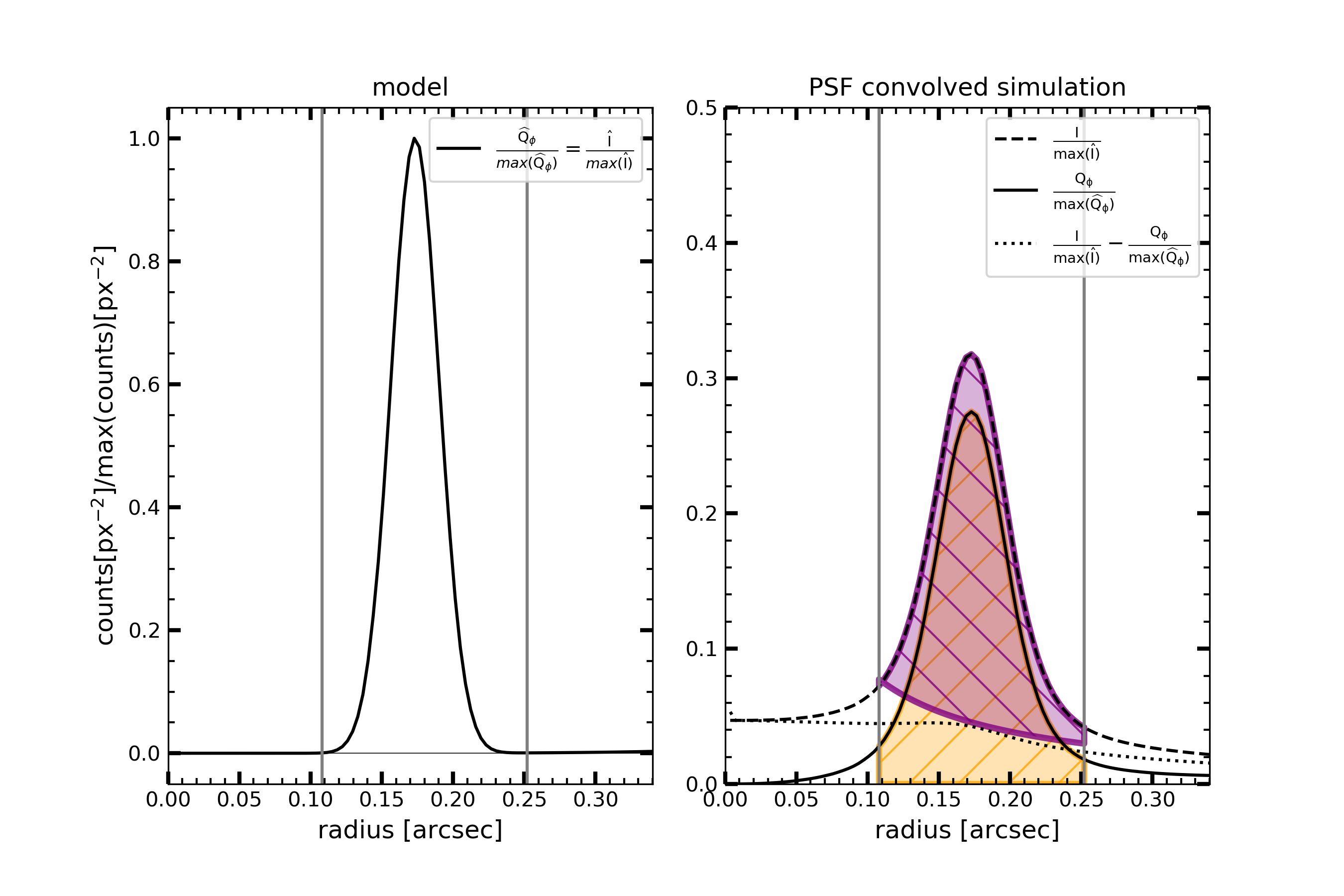}
     \centering
     \caption{Relative PSF convolution differences for the radial profiles of the disk intensity $I_{\rm d}/{\rm max}(\widehat{I}_{\rm d})$ and the  polarized flux $Q_\phi/{\rm max}(\widehat{Q}_\phi)$ for the inner disk ring model of HD~169142. Illustrated are the findings for the I$'$ band model and the effect of the convolution with the mean I$'$ PSF of the four best cycles. The remaining disk intensity $I_{\rm d}$ is shown in purple and the disk polarized intensity $Q_\phi$ in yellow. The correction $C=\widehat{\gamma}/\gamma$ is given by the yellow $\int{Q_\phi}$ divided by the purple intensity $\int{I_{\rm d}}$.
    }
\label{mody1}
\end{figure}
We fit the observed total intensity profile $I_{\rm tot}(r)$ with a model for the stellar PSF profile and the disk intensity $I_{\rm d}(r,\gamma)$ and determine the optimal scaling factor $\gamma$:
\begin{displaymath}
I_{\rm tot, fit}(r) = I_{\star}(r) + I_{\rm d}(r,\gamma) =I_0r^\alpha  + \gamma \cdot Q_\phi(r)\,.
\end{displaymath}

The central peak of the stellar PSF profiles show in our data essentially no Airy rings (see Appendix~\ref{A2_psf}) partly because of the wide passbands $\Delta\lambda\approx 150$~nm of the used R$'$- and I$'$-filters, and partly because of the limited AO-performance. Therefore, we can approximate the stellar PSF in the range of the inner ring, from $r=0.108"$ to $0.252"$, by a power law function and determine the best fit parameters $I_0$, $\alpha$ and $\gamma$, which minimize with a Markov chain Monte Carlo (MCMC) method the sum of squared residuals between observed and fitted profile $\sum R^2 = \sum (I_{\rm tot, fit}(r)-I_{\rm tot}(r))^2$. Best parameters $I_0$, $\alpha$ and $\gamma$, were calculated for each cycle, the means for all cycle, or several subsets of cycles, but always separately for the R$'$ and I$'$ bands. All these determinations give very consistent results for the $\gamma$ parameter and we derived a best $\gamma$ parameter based on weighted least squares, giving more weight to the best cycles considering the $\sum R^2$ error. As MCMC method we use a own implementation of the simple and commonly used Metropolis-Hastings algorithm as described in \citet{Mackay03} with a large range flat prior, normal distributed step size and an acceptance of a point if ${\rm exp(-err_{new}+err_{old})}$ is larger than a random number between 0 and 1.

Figure~\ref{mcmccurve} shows the resulting radial profiles for $I_{\rm d} (r)$ for the R$'$ and I$'$ bands, the observed intensity profiles $I_{\rm tot}(r)$ and the PSF power law fits $I_{\star}(r)$. The two profiles for $I_{\rm tot}(r)$ and $I_{\star}(r)$ in the upper panels are almost indistinguishable and the difference caused by the disk intensity $I_{\rm d} (r)$ can hardly be recognized. The intensity differences between $I_{\rm tot}(r)$ and $I_{\star}(r)$ are plotted in the lower panel on a much finer intensity scale, together with the $I_{\rm d}(r,\gamma)=\gamma Q_\phi(r)$ curve and the fit residuals $R$. The obtained $\gamma$ parameters are 3.9 for the I$'$ band and 4.3 for the R$'$ band, but these values are subject to a systematic bias because the PSF convolution effects for the intensity and polarization are different.

As shown in Fig.~\ref{mody1} the convolution preserves the integrated disk intensity but smears the disk signal and produces a ``smearing'' halo with a substantial flux level inside and outside the disk ring. This is different for the $Q_\phi$-profile, which is based on the Stokes $Q$ and $U$ images composed of positive and negative quadrants. The smearing by the PSF convolution leads to polarization cancelation of the ``smeared'' halo signal for $Q$ and $U$ and the disk integrated polarized flux $Q_\phi$ is reduced. This PSF convolution cancelation effect is demonstrated in Fig.~\ref{hd24comp_n} with two-dimensional images. The dotted line in Fig.~\ref{mody1} is the difference between the convolution of the normalized intensity profile $I_{\rm d}(r)/{\rm max}(\widehat{I_{\rm d}}(r))$ and the convolution of the normalized polarization profile $Q_\phi (r)/{\rm max}(\widehat{Q}_\phi(r))$, where we assumed that both normalized profiles are identical. The difference is large, but our disk intensity extraction procedure can only capture the narrow radial disk bump (purple area in Fig.~\ref{mody1}). The radially extended ``smeared intensity halo'' below the solid purple line and outside of the radial fitting range indicated by the vertical lines cannot be disentangled from the much stronger, stellar PSF profile fitted by a power law. We investigated our procedure and find that the fitting of a scaled $\gamma Q_\phi(r)$ profile to the observed disk intensity bump in the $I_{\rm tot}(r)$ profile accounts only for the purple area in the convolved profile for the normalized intensity bump $I_{\rm d}(r)/{\rm max}(\widehat{I}(r))$. The fitting of the stellar PSF by a power law gives the shape of the solid purple line. This underestimates the flux of the disk intensity bump with respect to the bump for the polarized intensity $\widehat{\gamma}=C\gamma$ of $C(\rm I')=1.075$ and $C(\rm R')=1.068$ and and we apply these small corrections $\widehat{\gamma}=C\,\gamma$ and obtain convolution corrected $\gamma$ factors of $\widehat{\gamma}({\rm I}')=4.2$ and $\widehat{\gamma}({\rm R}') =4.6$.

The derived $\widehat{\gamma}$ parameters give the intrinsic fractional polarization $\widehat{p}$ listed in Table~\ref{fcontrasttab}. The uncertainties for these parameters follow from the weighted least square determination of the many different PSF-fits described above. The relative uncertainties for $\widehat{\gamma}$ or $\widehat{p}$ are smaller ($\approx 15~\%$) for the I$'$ band than for the R$'$ band ($\approx 25~\%$) because the PSF quality is better and therefore the intensity extraction more accurate for the I$'$ band.  The resulting fractional polarizations for the inner disk ring are $\widehat{p}({\rm R}')=22.0~\%$ and $\widehat{p}({\rm I}')=23.6~\%$. This is compatible with no color dependence for the fractional polarization $\widehat{p}$. For the relative intensities $\widehat{I}_{{\rm d},\Sigma}/I_\star$ we consider the small difference between the total system intensity $I_{\rm tot}$ and the intensity of the central, unresolved object $I_\star = I_{\rm tot} - I_{\rm d}$ composed of the stellar intensity and the expected scattered intensity of the hot dust.

The resulting intrinsic disk intensity $\widehat{I_{\rm d}}$ yields roughly the same reddish ${\rm R}'-{\rm I}'$ color as for the polarized flux, but the quite large uncertainties from the $\gamma$ parameter determination makes this color determination inconclusive.  However, we find a correlation between the many $\gamma({\rm R}')$ and $\gamma({\rm I}')$ determinations of simultaneous observations (same cycle). If we determine for a given cycle a relatively low or high $\gamma$ value for the R$'$ band, then we also find typically for the I$'$ band a low or high value, respectively, and we derive a Pearson correlation coefficient of $p=0.71$. Because of this correlation the uncertainties for the wavelength dependence or for the color of the fractional polarization $\widehat{p}$ and the relative disk intensity $\widehat{I}_{\rm d,\Sigma}/I_{\star}$ are smaller than what one would expect if the indicated errors of the R$'$ band and I$'$ band are treated as independent measurements. Therefore, we can restrict the wavelength dependence of the fractional polarization for the inner narrow ring to ratios of about $\widehat{p}({\rm I}')/\widehat{p}({\rm R}')\approx 1.1\pm 0.1$ and the intensity color $(\widehat{I}_{\rm d,\Sigma}/I_{\star})({\rm I}')/(\widehat{I}_{\rm d,\Sigma}/I_{\star})({\rm R}') = 1.25 \pm 0.24$ or expressed in magnitudes $(\widehat{I}_{\rm d,\Sigma}/I_{\star})({\rm R}')-(\widehat{I}_{\rm d,\Sigma} /I_{\star})({\rm I}') = 0.24 \pm 0.21$ mag. Thus we also find a red color for the scattered intensity from the inner bright disk ring, relative to the color of the star. 

The scattered intensity for the outer disk or the entire disk can be guessed from $\widehat{Q_\phi}$ adopting just the derived $\widehat{\gamma}$ from the inner disk ring as approximation for the entire disk. This is a questionable assumption that should be verified in future with measurements. The indicated uncertainties for the guessed outer and total disk values disregard this problem.

\begin{table}[t]
  \caption{\label{fcontrasttab} Model parameters for the scattered intensity for
    the HD 169142 disk: $\widehat{\gamma}$ parameter, fractional polarization $\widehat{p}=1/\widehat{\gamma}$, normalized disk intensities
    $\widehat{I}_{\rm d,\Sigma}/I_{\star}$ measured for the inner ring and for the outer disk and the entire disk adopting the $\widehat{p}$ values from the inner ring.}
\centering
\begin{tabular}{llll}
\hline\hline
comp     & para & R$'$ band        & I$'$ band  \\
\hline
\noalign{\smallskip\noindent fractional polarization parameter \smallskip}
inner & $\widehat{\gamma}$  & $4.6\pm 1.2$     & $4.2\pm 0.6$  \\
        & $\widehat{p}$       & $22.0\pm 5.9$~\% & $23.6\pm 3.5$~\% \\ 
\noalign{\smallskip\noindent normalized flux for inner ring \smallskip}
inner   & $\widehat{I_{d,\Sigma}}/I_{\star}$ & $(1.44\pm0.39)\cdot 10^{-2}$  & $(1.77\pm 0.27)\cdot 10^{-2}$   \\
\noalign{\smallskip\noindent adopted normalized flux (using $\widehat{\gamma}$) \smallskip}
outer & $\widehat{I_{d,\Sigma}}/I_{\star}$ & $(0.56\pm 0.15)\cdot 10^{-2}$  & $(0.75\pm 0.11)\cdot 10^{-2}$   \\
\noalign{\smallskip} 
total  & $\widehat{I_{d,\Sigma}}/I_{\star}$ & $(2.02\pm 0.55)\cdot 10^{-2}$  & $(2.53\pm 0.39)\cdot 10^{-2}$   \\
\hline    
\end{tabular}
 \tablefoot{Reflected light flux contrast between the dust intensity $\widehat{I_{\rm d}}$ and the star intensity $I_{\star}$. Inner: 30-70 px, outer: 100-416 px, total: 0-416 px; $I_{\star} = I_{\rm tot} - \widehat{I}_{\rm d,\sum}$.}
\end{table}

\paragraph{Azimuthal variation in the fractional polarization.}\label{243} 
We tried to derive the azimuthal dependence of the scattered intensity of the inner ring and estimate the fractional polarization for different disk regions, using the same intensity extraction method as for the azimuthally averaged profile. For this we selected four azimuthal wedges with $\Delta \phi = 45^{\circ}$ centered on the major ($\phi=5^\circ,185^\circ$) and minor ($\phi=95^\circ,275^\circ$) axes. We analyzed the mean profiles of the fast polarization R$'$- and I$'$-band data, but unfortunately the $\sum R^2$ errors were quite large and as a consequence, only qualitative estimates can be made. For the following we assume that the azimuthal dependence of the scattered polarized flux $Q_\phi(\phi)$ is small because we found in Sect.~\ref{21q}) only small variations ($\sigma = 9~\%$).

For the northern and southern disk sections, along the major axis, the derived intrinsic fractional polarization $\widehat{p}$ is a few percent larger than the weighted mean of about 23~\% (Table~\ref{fcontrasttab}). For the eastern side ($95 \pm 22.5^{\circ}$) the errors were the smallest, as there was significantly more scattered intensity $\widehat{I}$ and the estimated $\widehat{p}$ is clearly lower at a level of about $\widehat{p} \approx 15$~\%. On the western side, the intensity signal is difficult to extract and the uncertainty is larger, but it seems that $p$ is less than the mean but larger than on the eastern side.

The enhanced scattering intensity on the eastern side, which is also the nearer side for the HD 169142 disk, could be explained by a forward peaked dust scattering phase function. But because the inclination of this disk is small the scattering phase function effects are weak and could be masked by geometric deviations from axisymmetry, which are small but present in this disk. SPHERE/ZIMPOL data taken under better observing conditions should allow more accurate extractions of the disk intensity signal and provide more accurately the azimuthal dependence of the fractional polarization.

\subsection{Comparison of the reflected light with the thermal emission}

The reflected intensity and polarization from the disk in HD~169142 can be compared to the measured IR-excess, which is the result of light absorption and thermal reemission by the same circumstellar disk structures. These two emission components are therefore strongly related and \citet{Garufi17} finds for transition disks (Meeus group~I disks) a correlation between far-IR excess and polarized light contrast of $F({\rm fIR})/F_{\rm tot}\approx  35 \times Q_\phi/I_{\rm tot}$. This relation is based on very heterogeneous literature data for $Q_\phi$ indicating that the correlation must be strong, but it is probably also not accurate. In \citet{Garufi17} the uncertainty in the adopted polarization data, as well as the uncertainty in the derived IR excess values are not discussed. For HD~169142 the far-IR excess in \citet{Garufi17} is 18.2~\%, while \cite{Sylvester96} report $F_{fIR}/F_{\star}=8.8~\%$ and \cite{Seok16} estimate $\approx 24~\%$. In addition there exist detailed model fits to the SED of HD~169142 \citep[e.g.,][]{Honda12,Bertrang18,DianaMod19}, but these studies do not provide the radiation budgets for the different dust emission components. We determine the IR excess again in this work because of the large discrepancy in the available values. 

\subsubsection{Spectral energy distribution}

The SED of HD~169142 is fitted by a stellar source, a near-IR excess from hot dust, PAH emission lines in the 6-12~$\mu$m region, and a strong $\lambda>10~\mu$m excess at longer wavelengths, which is modeled as in \cite{Honda12}, \cite{Seok16}, and \cite{Bertrang18} by the combined thermal emission of warm dust from the inner ring and cold dust form the outer disk.

For our determination we took the data from the DIANA database~\citep{Diana19} and their resulting model fit, which matches excellently these data points. We interpolated the points to obtain an SED with constant ${\rm d\,log}\lambda$ steps with a step size of 0.01, which is about the average spacing of available data points. 

For the stellar component we fitted the stellar model and the reddening correction of $A_V = 0.06$ from \cite{DianaMod19}, where the star is a PHOENIX spectrum with $T_{\rm eff}=7800K$ \citep{Brott05}. \cite{DianaMod19} used 145~pc as distance to HD~169142 instead of the newer GAIA DR2 estimate of 114~pc and therefore we introduce the fitting factor $A_0$ to adjust the star spectrum to the rescaled SED. For the reddening correction we use the model from~\cite{Fitzpatrick04}.

The IR emission is fitted by three Planck functions, $B_\lambda(A_{\rm x},T_{\rm x}),\, x=1-3$, for the hot dust, for the inner ring and the outer disk, each with the two free parameters $A_{\rm x}$ for the flux scale and $T_{\rm x}$ for the temperature. For the determination of the seven free model parameters we applied a  Metropolis-Hastings MCMC \citep{Mackay03} procedure with a large range flat prior, normal distributed step size, and an acceptance of a point if ${\rm exp(-err_{new}+err_{old})}$ is larger than a random number between 0 and 1. The error function is the sum of squared residuals with an increased weighting for the low flux values (in units of $W/m^2$) at longer wavelengths. The wavelength range of the PAH features between $6-12~\mu$m was ignored for the fitting. We used the difference between the observed strong spectral features in this wavelength range and the Planck fit as measure for the PAH flux $F_{\rm PAH}$ in HD~169142.

An important additional constraint is included in the fitting procedure for disentangling in the SED the contributions of the inner, narrow ring and the outer extended disk. The resolved mid-IR imaging data of \cite{Honda12} show for the wavelengths of 18.8~$\mu$m and also 24.5~$\mu$m that the inner ring dominates strongly the flux at these wavelengths. Therefore, we included in the fitting a corresponding penalty factor for this wavelength region.

\begin{figure}[ht!]
    \includegraphics[width=0.46\textwidth]{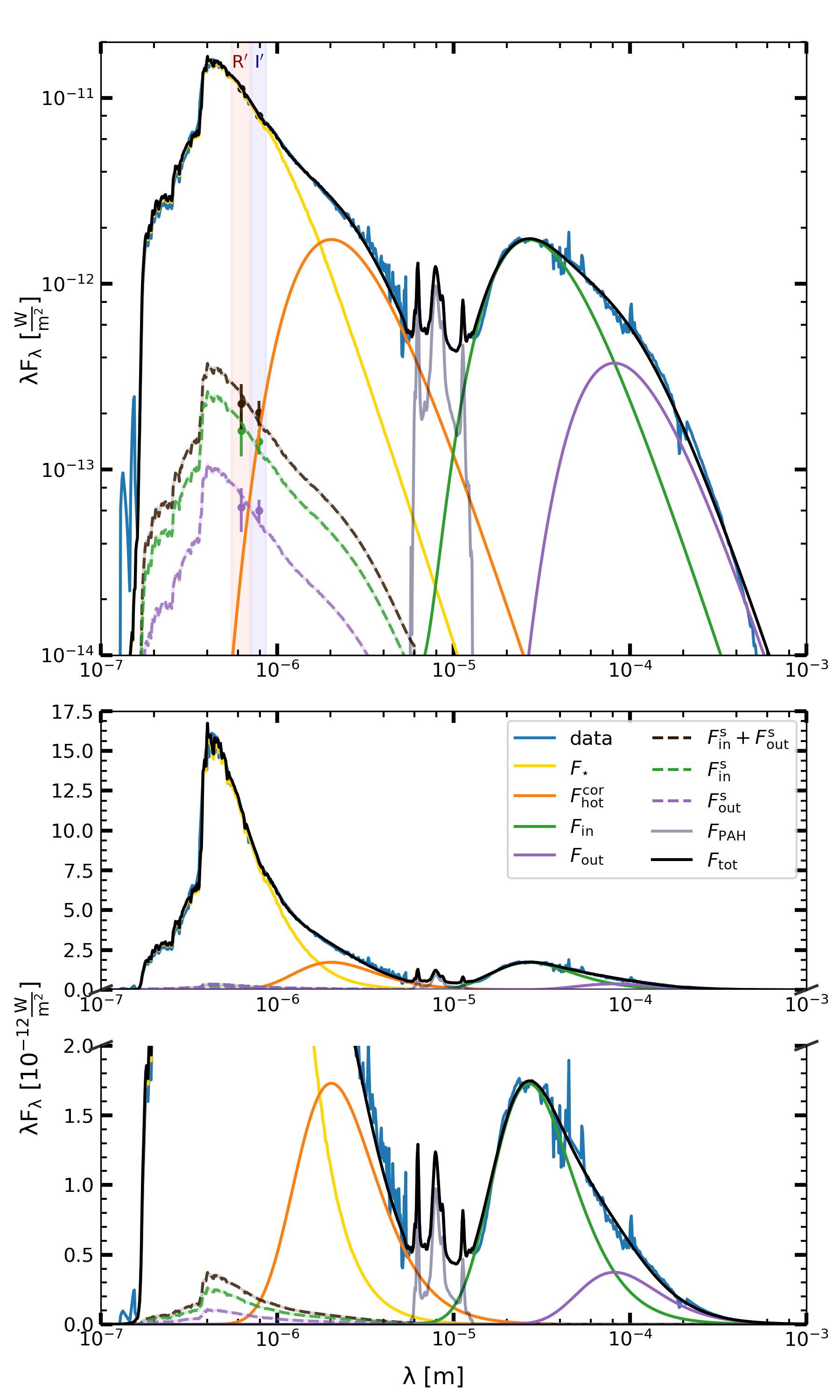}
     \centering
     \caption{Detailed SED model fit for HD~169142. Blue represents our merged data fit, black our total model, yellow the radiation
       of the star, orange the hot dust close to the star, green the
       warm inner ring, purple the cold outer disk, and gray the PAH
       estimate. The dots are our measurements
       of the dust reflection and the dashed lines our estimates for
       the spectral distribution of the scattering.}
\label{SEDhd169142_fit_new}
\end{figure}

The resulting fit to the SED is shown in Fig.~\ref{SEDhd169142_fit_new} and the corresponding integrated flux values and black-body temperatures are listed in Table~\ref{SEDbestfit_new}. The integrated flux of the PAH features in the mid-IR is calculated from the residuals between data and fit in this spectral region (5.5-15 $\mu m$). We use in Table \ref{SEDbestfit_new} and the SED model the symbol  $F_{\rm star}$ for the integrated flux of the stellar SED, which includes the stellar emission and the scattered stellar radiation from the hot dust and the disk. The symbol $F_\star$ is used for the stellar flux corrected for the scattered (stellar) radiation from the disk, or the emission of the star plus only the scattered stellar light from the hot dust. This is equivalent to the stellar intensity $I_{\star} = I_{\rm tot} - \widehat{I}_{\rm d,\sum}$ used in the imaging observations for the unresolved central object that includes also the emission of the star and the scattered emission of the hot dust. 

\begin{table}
  \caption{\label{SEDbestfit_new} Derived integrated fluxes and
    temperatures for the different emission components
    in the SED according to the fit model shown in
    Fig.~\ref{SEDhd169142_fit_new}. }
\centering
\begin{tabular}{llllll}
\hline\hline
comp   & flux $F$    & $F/F_\star$ & temp & remark  \\
            & [$10^{-12}{\rm Wm}^{-2}]$ & [\%] & [K]        & (symbol) \\
\hline
\noalign{\smallskip\noindent SED components \smallskip}
star        &  $16.66 \pm 0.10$      & 102.3    & 7800     & $F_{\rm star}$\\
hot dust     &   $2.40 \pm 0.25$      & 14.8     & $1810 \pm 20$ & $F_{\rm hot}$   \\

inner disk  &   $2.35 \pm 0.29$     & 14.4       & $137 \pm 5$   & $F_{\rm in}$    \\
outer disk  &   $0.51 \pm 0.14$     & 3.1       & $ 45 \pm 3$   & $F_{\rm out}$        \\
PAH         &   $0.27 \pm 0.04$       & 1.6          & n/a     & $F_{\rm PAH}$         \\
\noalign{\smallskip\noindent total \smallskip}
sum         & $22.18  \pm 0.42$      & ~136.2       & --           & $F_{\rm tot}$ \\
\noalign{\smallskip\noindent scattered flux for the hot dust and disk\smallskip}
hot dust$^s$    & n/a         & n/a                  &--        &  $F_{\rm hot}^{\rm s}$  \\
inner disk$^s$  &   $0.30 \pm 0.04$    & 1.8       & --           & $F_{\rm in}^{\rm s}$  \\
outer disk$^s$ &   $0.12 \pm 0.02$   & 0.7       & --            & $F_{\rm out}^{\rm s}$  \\
\noalign{\smallskip\noindent SED components corrected for disk scattering
  \smallskip}
star$^\star$        &  $16.29 \pm 0.10$      & 100        & 7800     & $F_{\star}$ \\
hot dust$^\star$    &   $2.35 \pm 0.25$      & 14.4       & $1810 \pm 20$  & $F_{\rm hot}^{\rm cor}$  \\

\hline    
\end{tabular}
\tablefoot{($F_{\rm star}$+$F_{\rm hot}$) = ($F_{\rm in}^{\rm s}$+$F_{\rm out}^{\rm s}$) + ($F_{\star}$+$F_{\rm hot}^{\rm cor}$). Components illustrated in Fig.~\ref{SEDhd169142_fit_new}.}
\end{table}

Of course, our dust model based on just three single-temperature black-body components is strongly oversimplifying the complex emission properties and temperature distribution of the emitting dust. Nonetheless, we get a very good fit to the SED and can accurately quantify the flux from the individual spectral components. The radiation budget in Table~\ref{SEDbestfit_new} yields a total IR excess of $F_{\rm IR, tot}/F_{\star}=0.339$ where $F_{\rm IR, tot}$ includes the four SED contributions of the hot dust in the near-IR, the inner ring and outer disk in the mid- to far-IR range and the PAH-emission features ($F_{\rm IR, tot}$ is the whole SED minus $F_{\rm star}$).

The extended disk resolved with scattered light imaging is responsible for the thermal emission from about $10~\mu$m to $1$~mm and we obtain for this part of the SED without PAH an excess of $F_{\rm fIR}/F_{\star}=0.175$  in good agreement with the value $0.182$ given in \citet{Garufi17}. The value $F_{\rm fIR}/F_{\rm star}=0.088$ from \cite{Sylvester96} is much lower probably because only the less accurate far-IR data points from the IRAS satellite were available at the time of that work. \cite{Seok16} focused their model mainly on the PAH-features in HD~169142, but they also give an overall SED fit that (slightly) underestimates the data points in the visual and near-IR, which also underestimates $F_{\rm star}$; therefore, the resulting relative IR excess ($F_{\rm fIR}/F_{\rm star}\approx 0.24$) is probably higher than our value and the result from \citet{Garufi17}.

For the individual dust components $F/F_{\star}$ our obtained relative values yield for the hot 14.8~\%, warm 14.4~\%, cold 3.1~\% and PAH 1.6~\% spectral dust components. The corresponding values from \cite{Seok16} for the hot $\approx 12.7~\%$, warm $\approx 16.4~\%$, cold $\approx 7.8~\%$ and PAH $\approx 4.1~\%$ components show a quite good agreement for the strong hot and warm dust components, but they obtained  clearly higher values for the PAH features and the cold dust component. For the PAH emission our simple estimate is not considering spectral regions with low PAH flux and therefore the determination of \cite{Seok16} should be more accurate. For the cold dust, the fit of \cite{Seok16} overestimates clearly the data points in the submillimeter range and this could explain partly the discrepancy with our fit results (see below). Overall, we conclude that our simple dust model fit using three black body components provides good  values for the energy budget for the hot and warm dust and probably also for the cold dust.

Very interesting for the investigation of the light scattering and light absorption by the dust in the disk of HD~169142 is the thermal IR emission from the warm and cold dust components, which correspond to the narrow ring and the outer disk components seen in scattered light. The ``Planck curve fitting procedure'' gives a ratio of $F_{\rm in}/F_{\rm out}\approx 4.6\pm 1.8$ between these two thermal components. However, the formal fitting uncertainties from Table~\ref{SEDbestfit_new} disregard errors introduced by our use of simple Planck-curve models. The flux $F_{\rm in}$ for the inner disk ring seems to be quite reliable, because it must fit the emission peak around $\lambda\approx 25~\mu$m according to the thermal IR imaging of \cite{Honda12}. However, the flux of the ``cold'' component and therefore the ratio $F_{\rm in}/F_{\rm out}$ depends strongly on the adopted spectral shape for the dominant ``warm'' component. The model of \cite{Seok16} gives a much lower ratio of $F_{\rm in}/F_{\rm out}=2.1$ by  calculating more sophisticated spectral shapes for the SEDs of the warm and cold dust, but their fit for the cold dust overestimates the emission in the submillimeter range and underestimates the flux at 7~cm, but all these data points at $\lambda >200~\mu$m are not relevant for the energy budget (see Fig.~\ref{SEDhd169142_fit_new}). One may wonder how much the derived cold dust flux $F_{\rm out}/F_{\rm star}$  would change in a detailed model like the one from \cite{Seok16} if the weights for the data points are optimized for the determination of the energy budget. Considering these points we extend the uncertainty range from the black body fit for the flux ratio between the warm and cold dust components to $F_{\rm in}/F_{\rm out}\approx 2-6.4$. This is similar to the ratio in scattered light between the inner narrow ring and the outer disk, where we derive for the I$'$ band a value of $\widehat{I}_{d,\Sigma}^{\rm in}/\widehat{I}_{d,\Sigma}^{\rm out}=2.4\pm 0.7$ (Table~\ref{fcontrasttab}) for the intrinsic, integrated dust intensities.  This indicates that the ratio between thermal and scattered flux of the dust in the disk is roughly similar for the warm, inner ring and the cold outer disk, but we may suspect from the ratios derived in this work that the scattering with respect to the thermal emission is more ``efficient'' for the outer disk.

Finding significant differences between scattering and absorption could be very important for the characterization of the dust. For example, it is expected that forward scattering dust would produce a larger differences in the reflectivity between a steep inner wall illuminated close to perpendicularly, when compared to a flat disk surface illuminated under gracing incidence. This difference would be smaller for isotropically scattering dust as expected for small dust grains. Alternatively, a reflectivity change between inner and outer disk could also point to a radial evolution of dust particles because of the presence of an ice line. More accurate far-IR SED measurements or spatially resolved observations would be extremely interesting for such detailed investigations of the dust reflectivity.

\subsubsection{Ratios between scattered and thermal radiation components}

The observed scattered and thermal radiation from the disk is essentially produced by the same dust in the surface layers of illuminated disk regions. The relative strength of these two emission components can therefore be used to investigate the single scattering albedo $\omega$ of the dust particles from the normalized scattered flux and the fractional IR excess
\begin{displaymath} 
  \Lambda_I = \frac{\widehat{I}_{d,\Sigma}/{I_\star}}{F_{\rm IR}/F_{\star}}\,
\end{displaymath}
or from the equivalent ratios for the normalized polarized flux
\begin{displaymath} 
  \Lambda_\phi = \frac{\widehat{Q}_{\phi,\Sigma}/{I_\star}}{F_{\rm IR}/F_{\star}}\,.
\end{displaymath}
These $\Lambda$ ratios based on observable quantities can be considered as a measure for the disk reflectivity ${\cal{R}}\approx {I_{d,\Sigma}}/{I_\star}$ or equivalent for the reflected polarization and the normalized disk absorption ${{\cal{K}}\approx {F_{\rm IR}}/F_{\star}}$ or
\begin{displaymath}
\Lambda_I \approx \frac{{\cal{R}}(\lambda)}{\cal{K}}\,\Phi({\rm disk)}\,.  
\end{displaymath}
The disk reflectivity ${\cal{R}}$ is a function of the wavelength-dependent dust scattering albedo $\omega_{\lambda}$  and therefore an estimate of ${\cal{R}}$ is only representative for the observed wavelength.  The absorption ${\cal{K}}$ derived from the normalized thermal emission represents a wavelength integrated mean of the absorbed radiation by the dust particles in the disk.  Existing ${I_{d,\Sigma}}/{I_\star}$ data suggest that protoplanetary disks show often a higher reflectivity for longer wavelengths \citep[e.g.,][]{Mulders13,Stolker16,Hunziker21}.

The $\Lambda$ ratios depend also on the disk geometry, certainly on the disk inclination, the separation and structure of strongly illuminated disk regions, the distribution of unresolved absorbing and scattering hot dust near the star, and probably other factors that are not well defined and which are lumped together in the function $\Phi({\rm disk)}$. With model calculations and measurements of the $\Lambda$ ratios for many disks it should be possible to constrain the $\Phi({\rm disk)}$ term so that the $\Lambda$ ratios provide useful information about the scattering and absorption properties of the dust particles in the disk. In addition, we may gain insights about the impact of the hot dust by comparing $\Lambda$ ratios for disks with strong and weak near-IR emission from hot dust. 

A first study of the ratio $\Lambda_\phi$ between polarized scattered radiation and thermal radiation from circumstellar disk was carried out by \citet{Garufi17} based on literature data. They found for transition disks a well-defined average value of about $2.9$~\%, or ${\rm log}\Lambda_\phi = -1.54~{\rm dex}$ although with a quite large scatter and unclear measuring uncertainties. Our study on HD 169142 provides values of ${\rm log}\Lambda_\phi = -1.60  $ (R$'$) and $-1.49$ (I$'$) with a much improved precision of about $\pm 0.06~{\rm dex}$.

\begin{table}[t]
  \caption{\label{scattherm} Comparison of scattered radiation with thermal radiation. }
\centering
\begin{tabular}{llllll}
\hline\hline
para   & filter &  inner  [\%] & outer [\%] & total  [\%] \\
    &  &  x="in" & x="out" & x="tot" \\
\hline
$F_{\rm IR, x}/F_\star$ &  & $14.4 \pm 1.5 $ & $3.1 \pm 0.9 $ & $17.6 \pm 2.0 $  \\
\noalign{\smallskip\noindent measured $\Lambda$ parameters \smallskip}
\shortstack{$\Lambda_{\phi,x}$ \vspace{1.5mm}} & \shortstack{R' \\ I'} & \shortstack{$2.2 \pm 0.3 $\\ $2.8 \pm 0.3 $} & \shortstack{$3.5 \pm 1.0 $\\ $4.7 \pm 1.3 $} & \shortstack{$2.5 \pm 0.3 $\\ $3.2 \pm 0.4 $} \\
\shortstack{$\Lambda_{\rm I,x}$ \vspace{1.5mm}} & \shortstack{R' \\ I'} & \shortstack{$10.0 \pm 3.0 $\\ $12.3 \pm 2.4 $} &  & \\
\noalign{\smallskip\noindent estimated $\Lambda$ parameters (for $p\approx 22.5 \%$) \smallskip}
\shortstack{$\Lambda_{\rm I,x}$ \vspace{1.5mm}} & \shortstack{R' \\ I'} &  & \shortstack{$\approx 17.9 \pm 6.9 $\\ $\approx 24.0 \pm 7.5 $} & \shortstack{$\approx 11.5 \pm 3.4 $\\ $\approx 14.4 \pm 2.8 $}  \\
\hline    
\end{tabular}
\tablefoot{$\Lambda_{\phi,x}=({\widehat{Q}_{\phi,\Sigma,x}/I_{\rm \star}})({F_{\rm IR,x}/F_{\star}})$, \
$\Lambda_{\rm I,x}=({\widehat{I}_{d,\Sigma,x}/I_{\rm \star}})/{(F_{\rm IR,x}/F_{\star})}$. \\  }
\end{table}

Table~\ref{scattherm} summarizes the $\Lambda$ double ratios following from the result of previous sections (Tables~\ref{diskmodtab},~\ref{fcontrasttab}, and~\ref{SEDbestfit_new}). Separate values are given for the R$'$ and I$'$ band and for the bright, inner ring, the extended outer disk and the total disk. The most accurate parameter are obtained for  the double ratio $\Lambda_\phi$ between polarized flux and the IR excess. This double ratio is larger for the I$'$ band when compared to the R$'$ band, mirroring the derived color dependence of the normalized polarized flux (Tab.~\ref{diskmodtab}). The $\Lambda_\phi$ parameter is about 1.65 times larger for the outer disk when compared to the inner ring. This result depends on the splitting of the SED curve between the warm and the cold dust emission as discussed in the previous section and a more sophisticated model fit to the far-IR SED data of HD~169142 could clarify this issue. A lower $\Lambda$ ratio, or less scattering, for the inner disk can be explained by the incidence angle for the inner ring, where we see the scattering from a steep inner wall illuminated with nearly normal incidence, while for the outer disk the reflection is from the disk surface, which is illuminated under gracing incidence. It is expected for dust with a typical preference for forward scattering that more light is absorbed for the inner disk wall with respect to the disk surface
\citep[e.g.,][]{Mulders13}. 

The double ratio for the intensity $\Lambda_I$ of the inner disk ring yields values of $10.0~\%$ for the R$'$ band and $12.3~\%$ for the I$'$ band. We can also make some guesses on the $\Lambda_I$ parameters of the outer disk and the total disk. However, for this we are adopting the measured fractional polarization $p$ derived for the inner disk, also for the outer disk. Further, we adopt also the not-so-well-determined two-component splitting for the inner and outer disk for the IR SED, where uncertainties have a much larger impact on the weaker cold component. Thus, the estimated parameter are based on several assumptions and should consequently be considered very cautiously. Therefore, we have put these result in extra lines in Table~\ref{scattherm} under the label ``estimated parameters.''

\paragraph{Unresolved issues.}
For the moment, we can use the $\Lambda_I$ double ratio as first guess for the ratio between disk reflection at a given wavelength and the spectrally averaged absorption ${\cal{R}}/{\cal{K}}\approx 0.1$ for the disk in HD~169142. We can also use the more accurately measured parameter $\Lambda_\phi$ for the ratio between polarized reflectivity and absorption $(p\times {\cal{R}})/{\cal{K}} \approx 0.03$ (I$'$ band) for the characterization of the disk. However, for a more detailed analysis of dust properties based on comparisons with other disks, we should investigate the impact of various simplifications made in our analysis and elaborate appropriate corrections to eliminate systematic effects like disk inclination and geometry or the color dependence of $\cal{R}(\lambda)$.

The rough guess on the disk scattering and absorption assumes that the irradiation of the disk is well described by $I_\star$,  the measured intensity of the unresolved central object, and $F_\star$,  the stellar spectral component in the SED corrected for the contribution of the scattered light from the disk. This assumption is problematic for HD~169142 because of the presence of absorbing hot dust near the star as revealed by the strong near-IR excess ($F_{\rm hot}/F_\star=14.4~\%$). This hot dust,  could be distributed in the disk plane and reduce substantially the illumination of the disk imaged in scattered light. Therefore, the observed stellar radiation $I_\star$ and $F_\star$ are overestimating the disk illumination. Possibly, one can assume for this ``disk case'' that both $I_\star$ and $F_\star$ are overestimated by about the same factor, so that the impact of a disk of hot dust on the double ratios $\Lambda_I$ and $\Lambda_\phi$ is rather small. This scenario assumes that scattered and thermal emission from the hot dust disk is emitted predominantly perpendicular to the disk plane and is not adding to the illumination of the resolved disk.

Alternatively, the absorbing hot dust could also be distributed in an optically thin halo around the star. If this halo is close to spherical, then the disk irradiation by stellar light is reduced by about the same amount as the observed stellar radiation $I_\star$ and $F_\star$. However, the near-IR emission of the hot dust halo would then also contribute significantly to the disk irradiation and we should use modified $\Lambda$ ratios where the SED components of the star and the hot dust are taken into account
\begin{displaymath} 
  {\Lambda'}_I = \frac{\widehat{I}_{\rm d,\Sigma}/{I_\star}}
    {F_{\rm IR}/(F_{\star}+F_{\rm hot})}\,.
\end{displaymath}
This  would reduce for the case of HD~169142 the $\Lambda$ ratios by roughly 15~\%. Similar scenarios apply for the light scattering, but the impact of dust scattering on the radiation budget and the $\Lambda$ ratios is expected to be significantly smaller when compared to the effect of the thermal emission of the hot dust.

\section{Discussion}\label{sec:conclusion}
 
This work presents a quantitative analysis of scattered light observations for the transition disk around HD~169142 taken with the AO system SPHERE at VLT. Such studies have the potential to clarify the properties of the scattering dust particles and may provide evidence for dust evolution in protoplanetary disks if the results from systems with different properties are compared. The signatures of different types of dust particles are expected to be small and therefore a high measuring accuracy is required for meaningful interpretations. The data analysis must consider several instrumental effects and there are hardly any studies in the literature with a focus on quantifying the uncertainties, and getting the accurate results for the measurements of the photo-polarimetric parameters of a circumstellar disk. Therefore, we carried out a step by step analysis of HD~169142 based on imaging polarimetry with an AO system. In this section we discuss the key points for the determination of accurate scattered light parameters for the disk around HD~169142 and then we put the derived results into context for the investigation of the dust particle properties in protoplanetary disks.

\subsection{SPHERE/ZIMPOL measurements of HD 169142}

The used SPHERE/ZIMPOL instrument and the selected disk HD~169142 are both special when compared to the many existing polarimetric disk observations.

ZIMPOL is unique, because it works at optical wavelengths ($500-900$~nm), and uses a fast modulation-demodulation technique for the polarimetric measurements.  ZIMPOL observations have, because of the short wavelength, a lower Strehl ratio and therefore more halo light and larger temporal variations in the PSFs when compared to IR-systems. On the other side, ZIMPOL has the advantage of using detectors with a very high dynamic range so that the circumstellar disk can be observed simultaneously in two wavelength bands together with the corresponding stellar PSF \citep{Schmid18}. This helps significantly in the signal calibration and the determination of wavelength dependences. The strong PSF variations induced by the atmospheric turbulence can be calibrated and corrected very well, as described in this work.

HD~169142 is a very special disk target because it is seen essentially pole-on and appears as highly axisymmetric system with a circular inner ring and a broader outer disk. Consequently, also smearing and polarization cancelation effects are to first order ``axisymmetric'' and the scattered light signal can be analyzed in terms of radial profiles with much improved S/Ns because of the azimuthal averaging. Therefore, we can measure exactly the impact of PSF variations for individual polarimetric observation cycles and correct for it. For inclined disks, the simple radial profile analysis carried out for HD~169142 is not applicable, but many effects are qualitatively similar.

SPHERE/ZIMPOL provides as prerequisite for a quantitative analysis of well-calibrated polarimetric data \citep{Schmid18}. AO systems are complex and require dedicated and extensive polarimetric monitoring to quantify cross-talk effects, which depend typically on the telescope pointing direction and which can convert linear polarization into circular polarization and reduce significantly ($\ga 20~\%$) the measurable signal. Fortunately, there exist several other well-calibrated AO polarimeters, such as GPI \citep{Perrin15} or SPHERE/IRDIS \citep{deBoer20,vanHolstein20}, and this enables very important cross-checks of results. HD~169142 is one of the good examples where the same object was observed with Gemini/GPI \citep{Monnier17}, VLT/SPHERE \citep[][and this work]{Pohl17,Bertrang18,Gratton19}, and Subaru/HICIAO \citep{Momose15} and the agreement of the described $Q_\phi$ signal is good (see Sect.~\ref{21q}). Unfortunately, quantitative comparisons of the measurements are difficult because the results are presented in different ways or because the intrinsic disk parameters cannot be separated easily from instrumental effects. 

\subsubsection{Polarimetric correction for the PSF convolution}
The ZIMPOL data of HD~169142 demonstrate the very strong impact of the variable PSF smearing and polarization cancelation effect for the measurement of the azimuthal polarization $Q_\phi$ for the compact ($r\approx 0.17''$) inner disk. The good message is that the observed and highly variable $Q_\phi$ disk signal, with variations of up to a factor of two, is tightly correlated with the measured PSF quality. Therefore, the observed $Q_\phi$ signal can be simulated with a disk model and the simultaneously registered PSF with an accuracy of a few percent. Not correcting $Q_\phi$ for the PSF smearing and cancelation effects would underestimate the signal for the inner ring of HD~169142 by a factor of about three for the best polarimetric cycles taken under good atmospheric conditions and more for average or bad cycles (see, e.g., Figs.~\ref{finalIRunconvnew} and \ref{simvnew3}). For the more extended $r\approx 0.5''$ outer disk of HD~169142 the $Q_\phi$ degradation is ``only'' at a level of about 30~\%.

The wavelength dependence of the polarized flux relative to the star $Q_\phi^{\rm n}(\lambda)$ is important for the characterization of the dust in circumstellar disks, because the disk geometry can be assumed to be the same for different wavelengths and a gradient for $Q_\phi^{\rm n} (\lambda)$ is therefore caused by the wavelength dependence of the particle properties. With ZIMPOL, this color effect has been measured accurately as the R$'$- and I$'$-band data for HD~169142 were taken simultaneously and therefore under identical atmospheric conditions.

\subsubsection{Extracting the disk intensity $I_{\rm d}$}
Measuring the disk intensity $I_{\rm d}$ together with $Q_\phi$ yields the fractional polarization $p=Q_\phi/I_{\rm d}$ of the scattered light, which is a strong function of the scattering angle $\alpha$. For circumstellar disks the angle $\alpha$ can typically be well determined and therefore a $p$ measurement is quite unambiguous and provides well-defined constraints for the reflected light.

Unfortunately, it is difficult with ground-based AO observations to extract the disk intensity $I_{\rm d}$, because of the strongly variable intensity PSF of the bright central star. Polarimetric observations have the advantage that the location of the disk is accurately traced by the $Q_\phi$ signal so that photometric extraction windows can be accurately defined. Nonetheless, favorable conditions are still required for a successful extraction, such as for the inner ring of HD~169142, which is bright, narrow, axisymmetric and for ZIMPOL located far enough from the star but close enough to avoid the strong speckle ring at $r\approx 0.5''$ from the AO system. For these reasons, we were successful in extracting the disk intensity signal.

The PSF smearing needs again to be considered for the derivation of $p$ as $Q_\phi$ is smeared and partially canceled, while $I_{\rm d}$ is only smeared out. This effect reduces for our ZIMPOL observations the measured $p=25.6~\%$ for the inner ring of HD~169142 to a de-biased intrinsic fractional polarization value of $\widehat{p}=23.6~\%$ (I$'$ band).

\subsection{Dust properties for the disk of HD~169142}

\subsubsection{Derived photo-polarimetric results}
For HD~169142 we obtained as most basic photo-polarimetric parameter the intrinsic, normalized, azimuthal polarization $\widehat{Q}_{\phi,\Sigma}^{\rm n}=\widehat{Q}_{\phi,\Sigma}/I_{\rm tot}$ of $(0.430~\pm~0.010)~\%$ for the $R'$ band and $(0.548~\pm~0.010)~\%$ for the $I'$ band for the scattered light from the resolved disk covering the radial range from $r=0.07''$ to $1.6''$. The inner narrow disk ring contributes 72~\% to this total. The derived value for the polarized flux of $\approx 0.5~\%$ for HD~169142 is typical for transition disks with a large central gap \citep{Garufi17} where a substantial fraction of the stellar light is scattered by the inner wall of the outer disk. These result yield also the intrinsic ${\rm R}'-{\rm I}'$ color of the polarized flux relative to the central star of
\begin{displaymath}
m_{{\rm R}'}(\widehat{Q}_\phi^{\rm n})-m_{{\rm I}'}(\widehat{Q}_\phi^{\rm n})=0.264\pm 0.045\, {\rm mag}\,,
\end{displaymath}
indicating a slightly reddish color with respect to the star.  We find no significant color difference between the strong inner ring component ($0.26\pm 0.05$~mag) and the outer disk ($0.31\pm0.10$~mag). The color of the normalized polarization flux can also be expressed as double ratio with respect to the wavelength 
\begin{displaymath}
{\cal{L}}_\phi(\lambda) = \frac{\widehat{Q}_\phi^{\rm n}(\lambda_2)/\widehat{Q}_\phi^{\rm n}(\lambda_1)}{\lambda_2/\lambda_1} = 1.01\,,
\end{displaymath}
or roughly $\widehat{Q}_\phi^{\rm n}\propto \lambda$ for the R$'$ to I$'$ band range. The red color for $\widehat{Q}_\phi^{\rm n}(\lambda)$ is also supported by the measurement of \citet{Quanz13} who obtained  $Q_\phi^{\rm n}=0.41~\%$ for the H band with VLT/NACO, without a correction for the PSF smearing and cancelation effects. The intrinsic value $\widehat{Q}_\phi^{\rm n}$ for the H band should therefore be significantly larger, perhaps by a factor of 1.5 or even more, in agreement with the derived red color above. A red wavelength dependence of the $Q_\phi^{\rm n}(\lambda)$ signal is also measured for the transition disk HD~142527 by \citet{Hunziker21}, who obtained for $Q_\phi^{\rm n}$ for the near-IR H band a value that is a factor of 2.3 higher than for the ``visual'' VBB band. Interestingly, this yields a double ratio of ${\cal{L}}_\phi(\lambda)=1.04$ or the same rough $\widehat{Q}_\phi^{\rm n}\propto \lambda$ relation, but for a much wider wavelength range. Similar red color trends are also seen in other protoplanetary disks as for instance HD~135344B \citep{Stolker16}, but the $Q_\phi^{\rm n}$ color has not been quantified accurately.

The second important result of this study is the determination of the fractional polarization of the scattered light from the disk in HD~169142. We find for the I$'$ band a value of
\begin{displaymath}
\widehat{p} = \widehat{Q}_\phi/\widehat{I}_{\rm d} = 23.6~(\pm 3.5)~\%,
\end{displaymath}  
and for the R$'$ band a similar but less constrained value of $22.0~(\pm 5.9)~\%$. The achieved accuracy is not high enough to claim a positive or negative wavelength trend for $p(\lambda)$.

The fractional polarization is a strong function of the scattering angle $p(\alpha)$ and is expected to have for many types of dust a maximum near $\alpha\approx 90^\circ$, and low values for forward $\alpha\approx 0^\circ$ and backward scattering $\alpha\approx 180^\circ$. For transition disk seen pole-on like HD~169142, the main contribution to the brightness of the inner ring comes most likely from reflections near the upper rim of the inner wall, where the scattering angle is about $\approx 80^\circ$. Therefore, we can assume that the measured $p$ value is close to the maximum fractional polarization produced by the reflection from the disk wall. 

It is again interesting to compare with HD~142527 ($i\approx 20^\circ$), where \citet{Hunziker21} find $p=(28.0\pm 0.8)$~\% for the polarization of the backside wall for the VBB band ($\lambda_{\rm c}$ = 735~nm). In HD~142527 the backside wall is spatially resolved and one can see the change of the fractional polarization as a function of $\alpha$ for the range of scattering angles from $\alpha=90^\circ$ to $130^\circ$. The peak polarization is a bit higher than the average, at a level of $p_{\rm peak}\approx 31~\%$. We conclude from this comparison that the fractional polarization produced by the dust in the inner ring of HD~169142 is lower by more than one sigma when compared to HD~142527.

For wavelength $\lambda<1~\mu$m, no other $p$ values have been measured for the reflecting dust in protoplanetary disks. At near-IR wavelength there are the accurate results for HD~142527 of \citet{Hunziker21} for the H band, who find for the back wall a mean value of $p=(35.1\pm 2.0)~\%$, which is a significantly higher fractional polarization than in the optical.  Other estimates of $p$ for protoplanetary disks are unfortunately rare. \citet{Monnier19} obtained GPI-observations for the disk HD~34700A and report a peak fractional polarization of about $p_{\rm peak}\approx 50~\%$ for the J band and $\approx 60~\%$ for the H band, and \citet{Perrin09} obtained from $\lambda=2~\mu$m HST polarimetry of the disk around AB~Aur a maximum value of $p_{\rm peak}\approx 55~\%$, which is also much higher than for HD~169142. One reason is probably the general trend that dust scattering produces a higher fractional polarization at longer wavelengths but
it also seems that the dust particles in HD~169142 are different when compared to HD~142527, HD~34700B or AB~Aur and produce a lower level of fractional polarization for the scattered light. In Sect.~\ref{53DP}, we speculate that the dust particles  in the inner ring of HD~169142 produce less polarization as they are more compact as a result of the stronger stellar irradiation compared to the other objects.

Another interesting result for HD~169142 follows from the comparison of the scattered light with the thermal emission from the same disk region. This yields for the dust in the disk a rough measure of the ratio between the reflected radiation and the absorbed and reemitted radiation in the IR. 

The comparison of the normalized polarized flux $\widehat{Q}_{\phi,\Sigma}/I_\star$ with the near-IR excess $F_{\rm IR}/F_\star$ for the warm, inner ring and the cold, outer disk yields $\Lambda_\phi\approx 2.2$ to $4.7~\%$, which is, according to the compilation of \citet{Garufi17}, quite typical for transition disks. In HD~169142 we can even split the scattered radiation and the thermal emission for the two imaged disk components and find that the values for $\Lambda_\phi$ are marginally higher for the outer disk, which is illuminated under grazing incidence.

Adopting the fractional polarization $\widehat{p} \approx 22.5~\%$ gives then for the $\Lambda$ ratio between normalized scattered intensity $\widehat{I}_{\rm d,\Sigma}/I_{\star}$ and IR excess $F_{\rm IR}/F_{\star}$ the value $\Lambda_I\approx 12.3~\%$. This value indicates roughly that the disk surface albedo is on the order of $10~\%$ within an uncertainty of probably a factor of two.

\subsubsection{Scattering parameters for the dust}
The photo-polarimetric results described above constrain the averaged scattering parameter of the dust in the disk. A simple description of scattering grains consists of three parameters, the single scattering albedo $\omega$, the asymmetry factor $g$ for the adopted Henyey-Greenstein intensity phase function, and a Rayleigh-scattering-like angle dependence of the induced fractional polarization with $p_{\rm max}$ for the maximum value for  $\theta=90^\circ$ \citep{Graham07,Engler17}.  Protoplanetary disks are optically thick and therefore one has to take multiple scattering effects into account. This complicates the conversion from the photo-polarimetric disk parameters to dust scattering parameters. We use for the interpretation preliminary results of a disk model grid calculating the conversion between disk scattering parameters and dust scattering parameters (Ma \& Schmid, in preparation).

For the low inclination disk HD~169142 the scattering angle $\alpha$ is close to $90^\circ$ and therefore the measured fractional polarization of $\widehat{p}=23~\%$ of the scattered light is close to the peak value $p_{\rm peak}$ produced by the reflection from the disk surface. The observed $p_{\rm peak}$ value requires for the dust scattering parameters a maximum polarization $p_{\rm max}>p_{\rm peak}$, because under multiple scattering conditions the fractional polarization produced by single scatterings is always reduced by the more randomly polarized (the equivalent of unpolarized) contribution from photons undergoing more than one scattering. Thus, $p_{\rm max}$ must be higher than $p_{\rm peak}$, $p_{\rm max}/p_{\rm peak}\ga 2$ if multiple scattering dominate the reflected light, while $p_{\rm max}/p_{\rm peak}\approx 1-2$, if the observed photons undergo on average between 1 and 2 scatterings only.

Thus, the depolarization effect is strong for high albedo particles $\omega\ga 0.7$ producing a lot of multiple scattered light and weak for low albedo $\omega\la 0.3$ particles. Our measurements for $\Lambda_I$ yield a rough guess for the ratio between surface reflectivity and absorption or for the surface albedo of about $10~\%$ for the disk of HD~169142. As described by \citet{Mulders13}  the surface albedo depends also strongly on the scattering asymmetry parameter $g$ of the particles. If forward scattering is strong (e.g., $g=0.75$) then the photons will be scattered into deeper dust layers where the probability for absorption is high, while for little forward scattering ($g=0.25$) many photons are undergoing only one scattering, and escape or stay near the surface and can escape after a few scatterings.

Our measurement $\Lambda_I\approx 10$~\% can be met by high albedo, forward scattering particles, producing a high scattering polarization ($\omega\approx 0.7$, $g\approx 0.75$, and $p_{\max}\approx 50~\%$) or lower albedo, almost isotropically scattering particles with lower scattering polarization ($\omega\approx 0.5$, $g\approx 0.25$ and $p_{\max}\approx 30~\%$), or particles with about intermediate scattering parameters ($\omega\approx 0.6$, $g\approx 0.5$, and $p_{\max}\approx 40~\%$). 

Unfortunately, the pole-on disk HD~169142 is not suited to constrain the scattering asymmetry parameter $g$, which can be better estimated in inclined disks from the front side to back side brightness contrast.  In future, it should be possible to deduce a ``typical'' $g$ parameter for protoplanetary disk from observations of inclined disks and then one can narrow down further the possible parameter range for $\omega$ and $p_{\rm max}$ for the dust in HD~169142.\\

\subsection{Constraints on dust properties}\label{53DP}
The dust scattering parameters derived for HD~169142 and other protoplanetary disks inform us about the physical properties of the scattering grains, such as size distribution, composition, and structure. Small grains with radii $a<\lambda/(2\pi)$ are in the Rayleigh regime and they produce a high fractional polarization $p_{\rm max}$ and only moderate forward scattering $g$, while large spherical grains $a>\lambda/(2\pi)$ have a much smaller $p_{\rm max}$ but larger $\omega$.  Therefore, interstellar dust as described by \citet{Draine03} with particle size distribution found by \citet{Weingartner01} is expected to show a strong wavelength dependence for $p_{\rm max}$, decreasing from more than 50~\% in H band ($a/\lambda$ small), to 40~\% in the I band and only 30~\% in the R band ($a/\lambda$ large), while the asymmetry parameter increases from $g\approx 0.2$ in the H band to about 0.5 in the R band. We cannot constrain the $g$ parameter for the pole-on HD~169142, but inclined protoplanetary disks, for example HD~142527 \citep{Hunziker21}, show a strong forward scattering effect at $\lambda=0.73~\mu$m, but without the expected decrease in $g$ for the near-IR (1.6~$\mu$m) as inferred from the front to back side disk brightness ratio. This indicates that the scattering particles in these disks are larger than for the interstellar dust. However, the Mie theory predicts for a distribution of large spherical particles with a composition such as interstellar dust a low fractional scattering polarization $p_{\rm max} \lessapprox 30~\%$ in the visual and near-IR. Considering multiple  scattering led to a dilution of the observable fractional polarization for a optically thick disk to $p_{\rm max} \lessapprox 20~\%$,  which is still compatible with our observations of HD~169142, but certainly not with the higher fractional disk polarization ${\rm max}(p)\approx 30~\% - 60~\%$ measured for, for example, HD~142527, AB~Aur, or HD34700A \citep{Hunziker21,Perrin09,Monnier19}. This is a well-known general problem of the Mie theory for describing the dust scattering in protoplanetary disks. Therefore, models for more complex dust were introduced, based on aggregates or porous grains, which produce a high scattering asymmetry $g>0.5$ as found for large particles but also a high scattering polarization $p_{\rm max} \gtrapprox 30~\%$ from the (Rayleigh scattering like)  interaction of the radiation with small-scale substructures of the particle.

Models for more complex particles are for example discussed in \citet{Min16} for compact dust aggregates with a porosity of about 25~\% or \citet{Tazaki19} for fluffy aggregates with a porosity $>80~\%$. Large particles with a low porosity as described in \citet{Min16} tend to produce a lower scattering polarization $p_{\rm max} \lessapprox 50~\%$ compatible with the observed fractional disk polarization of $p\approx 23~\%$ for HD~169142, while a disk with $p \approx 50~\%$ requires more porous grains as discussed in \citet{Tazaki19}.

We may speculate that the relatively low fractional polarization found for HD~169142 could be explained by the presence of more compact grains. HD~169142 is special because as of now no fractional polarization measurements exist for dust scattering in protoplanetary disks that is so close to the star ($r=0.17''$ or 19~AU) and as strongly illuminated as the inner disk wall of this object.  Therefore, the corresponding black body temperature for the scattering dust of $\approx 137~$K is much higher than for other disks with $p$ measurements. We therefore speculate that the strong stellar irradiation could  be responsible for the presence of more compact dust particles. This tentative hypothesis should be considered cautiously and should be tested with similar measurements of the dust scattering for other disks with strong irradiation.  

Another important piece of information for testing dust models is the wavelength dependence of the scattering parameters because it seems reasonable to assume that the disk geometry is identical for different colors. We find for HD~169142 a red color for the disk reflectivity in polarized light, similar to other disks. For large particles $a>1~\mu$m the color of the disk reflectivity is expected to depend strongly on the wavelength dependence of the single scattering albedo $\omega(\lambda)$, and much less on $g(\lambda)$ or $p_{\rm max}(\lambda)$. The scattering albedo is certainly a strong function of the particle composition, with high albedos for silicate rich and icy particles and low albedos for carbon rich particles. The wavelength dependences of the albedo for complex grains have not yet been investigated in much detail. For example, \citet{Tazaki19} find for their $a>1~\mu$m aggregate grains predominantly a gray or even slightly blue dependence for $\omega(\lambda)$. The less porous grains in the model of \citet{Min16}, which are composed of  silicates, carbon and iron sulfides typical for solar abundances, show little wavelength dependence for $\omega(\lambda)$ for $\lambda<2~\mu$m and an increase with wavelength for $\lambda \gtrapprox 2~\mu$m. It seems that there are some clear discrepancy between our observations and these dust models. 

It is beyond the scope of this paper to discuss the merits of different dust models. But we like to strongly emphasize that accurate polarimetric measurements of the scattered light from protoplanetary disks, such as the ones presented in this paper for HD~169142, are now possible and more will become available.  These data provide interesting new constraints and will help to establish the models for the scattering dust in disks and clarify its relations to the resulting composition of the planets formed in these disks.

\section{Conclusions}
The new generation of AO polarimeters provides the opportunity to obtain quantitative polarimetric measurements for the light scattered by the dust in circumstellar disks \citep[e.g.,][]{Schmid21}. Such observations are potentially very useful for constraining the properties of the dust that may also become part of newly formed planets. Apart from a few previous polarimetric studies based on HST data \citep[e.g.,][]{Perrin09}, such investigations are rather new for protoplanetary disks, and  one needs to first establish the data analysis procedures and the diagnostics for the interpretation of the polarization results. This paper is one of the first studies with a primary focus on a quantitative polarimetry of protoplanetary disks based on ground-based AO instruments.

We selected HD~169142 as a simple target for our basic study; it is a well-resolved, bright, pole-on, axisymmetric transition disk. Therefore, we were able to carry out an analysis based on azimuthally averaged radial profiles instead of two-dimensional disk maps. We found that the measured disk polarization depends strongly on the variable AO performance, a fact that is often overlooked in disk polarimetry. We show that the corresponding PSF convolution effects can be accurately modeled if the PSFs for the observations are well known. Convolution must be taken into account for accurate measurements of the polarized disk flux. The same type of correction can be applied to inclined disks but requires, of course, an analysis procedure that considers the azimuthal dependences.

Due to the fact that the SPHERE/ZIMPOL instrument simultaneously measures the stellar intensity PSF and the polarized disk flux in two pass-bands, we achieved, after detailed correction for the convolution effects, a very high accuracy for the intrinsic polarized flux relative to the total system intensity of $\widehat{Q}^n_{\phi,\Sigma}=\widehat{Q}_\phi/I_{\rm tot}=(4.30\pm 0.10)\cdot 10^{-3}$ for the R$'$ band and $(5.48\pm 0.10)\cdot 10^{-3}$ for the I$'$ band. This color dependence for the polarized reflectivity of the disk surface  is an easily measurable quantity that seems to be very useful for the characterization of the scattering dust. Thus, for HD~169142 the dust has a red color in polarized scattered light, similar to the previous study for HD~142527 \citep{Hunziker21}.

It was also possible to extract the intensity signal for the bright inner disk ring of HD~169142, and we derived a intrinsic fractional polarization of about 23~\% for the scattered light, considering again  the important convolution and signal extraction effects. In addition, we compared the scattered intensity with the thermal emission components of the dust and estimated a rough value for the disk surface albedo of about 10~\%. 

The imprints of different types of dust particles on the scattering polarization are expected to be subtle, and accurately measuring the polarization parameters for the reflectivity of the dust is important. Only this will allow, from measurements of different types of protoplanetary disks, an assessment of the heterogeneity or homogeneity of dust scattering properties, and this is important for deriving new constraints regarding the dust evolution processes in these systems.

Of course, we would like to ultimately derive the size distribution, structure, and composition of the dust grains in the studied disks. This requires model calculations for protoplanetary disks for the interpretation of the polarization data that consider the detailed geometry of the disk and the multiple scatterings taking place at the surface and which use realistic descriptions for the light scattering by the complex dust particles. Much progress has been made in modeling the thermal emission of disks \citep[e.g.,][]{Chiang97,Dullemond04,Woitke16}, the general hydrodynamic structures \citep[e.g.,][]{Armitage11}, the overall disk appearance in polarized scattered light \citep[e.g.,][]{Murakawa10,Min12,Dong12}, and the light scattering properties of different types of dust \citep{Min16,Tazaki19}. However, these models are still far from a comprehensive radiative transfer model for a given circumstellar disk, and they include ad hoc assumptions and produce unsolved parameter ambiguities for the best fitting solution to disk observations.
 
For example, a larger fraction of disk models in the literature use the Mie theory for homogeneous spherical particles with a composition typical for silicates \citep{Draine84}. This light scattering description is not able to easily explain the scattered light observations of protoplanetary disks; it typically shows the properties of strong forward scattering, as expected for large grains, and a contradictory high fractional scattering polarization of $>20~\%$, as expected for small spherical grains. For this reason, more complex particles, so-called aggregates, are proposed; they are large with respect to the wavelength in order to produce a strong forward scattering diffraction peak but have a lot of small-scale features so that the light can interact with small structures similar to scattering by small particles (like Rayleigh scattering), producing a relatively high fractional scattering polarization \citep{Min16,Tazaki19}.

Model calculations for the expected polarization signal for disks with such complex dust grains are presented by \cite{Tazaki19}. It is interesting to note that for pole-on disks they predict the fractional polarization $\widehat{p}$ at $\lambda=1.6~\mu$m for grains with different porosities. For disks with low porosity (25~percent) compact grains they calculate $\widehat{p}=22~\%$, while disks with highly porous (85~percent) grains produce $\widehat{p}=46~\%$. The measured fractional polarization for HD~169142 and HD~142527 lies between these two cases.  \cite{Tazaki19} also predict blue colors in the near-IR for the polarized reflectivity of grains with high porosity ($\ga$~85~percent), and one may expect that this is not compatible with the red polarized reflectivity colors in this work despite the fact that the color is measured at shorter wavelengths. The red colors for the disk reflectivity of HD~169142 and HD~142527 are more compatible with large, low porosity ($\approx$~25~percent) dust aggregates, as described by \cite{Min16}.

More such model calculations are required to better constrain the dust properties in circumstellar disks based on polarimetric measurements as presented in this work. In particular, the models should provide diagnostic parameters for the determination of dust properties that can easily be compared with observational data and do not depend strongly on often unknown details of the disk geometry. Such a parameter is the polarized reflectivity color derived in this work for HD~169142, where we expect that the measured color value depends predominantly on the dust scattering properties. Also, the fractional polarization of the scattered light $\widehat{p}$ is such a ``dust scattering parameter'' because the scattering angle, the most important geometric parameter for the $\widehat{p}$ of a surface, is typically well constrained for circumstellar disks. The comparison between the scattered intensity and thermal emission of the disk as described in this work by the $\Lambda$ double ratios provide additional information about the dust scattering opacity. In addition, inclined disks will be very useful for constraining the scattering asymmetry parameter. Combining all this observational information from high quality data with sophisticated dust scattering models will advance our knowledge of the scattering dust in protoplanetary disks in the near future.

\begin{acknowledgements}
We would like to thank the anonymous referee for helping to improve this manuscript with valuable comments. We thank Jie Ma for giving access to preliminary results of her model calculations. We acknowledge the financial support by the Swiss National Science Foundation through grant 200020\_162630/1. 
Based on data obtained from the ESO Science Archive Facility.
\end{acknowledgements}

\bibpunct{(}{)}{;}{a}{}{,} 
\bibliographystyle{aa}
\bibliography{bib/project}

\begin{thebibliography}{86}
\expandafter\ifx\csname natexlab\endcsname\relax\def\natexlab#1{#1}\fi

\bibitem[{{Acke} \& {van den Ancker}(2004)}]{Acke04}
{Acke}, B. \& {van den Ancker}, M.~E. 2004, \aap, 426, 151

\bibitem[{{Apai} {et~al.}(2004){Apai}, {Pascucci}, {Brandner}, {Henning},
  {Lenzen}, {Potter}, {Lagrange}, \& {Rousset}}]{Apai04}
{Apai}, D., {Pascucci}, I., {Brandner}, W., {et~al.} 2004, \aap, 415, 671

\bibitem[{Armitage(2011)}]{Armitage11}
Armitage, P.~J. 2011, Annual Review of Astronomy and Astrophysics, 49, 195

\bibitem[{{Avenhaus} {et~al.}(2018){Avenhaus}, {Quanz}, {Garufi}, {Perez},
  {Casassus}, {Pinte}, {Bertrang}, {Caceres}, {Benisty}, \&
  {Dominik}}]{Avenhaus18}
{Avenhaus}, H., {Quanz}, S.~P., {Garufi}, A., {et~al.} 2018, \apj, 863, 44

\bibitem[{{Avenhaus} {et~al.}(2014){Avenhaus}, {Quanz}, {Schmid}, {Meyer},
  {Garufi}, {Wolf}, \& {Dominik}}]{Avenhaus14}
{Avenhaus}, H., {Quanz}, S.~P., {Schmid}, H.~M., {et~al.} 2014, \apj, 781, 87

\bibitem[{{Benisty} {et~al.}(2015){Benisty}, {Juhasz}, {Boccaletti},
  {Avenhaus}, {Milli}, {Thalmann}, {Dominik}, {Pinilla}, {Buenzli}, {Pohl},
  {Beuzit}, {Birnstiel}, {de Boer}, {Bonnefoy}, {Chauvin}, {Christiaens},
  {Garufi}, {Grady}, {Henning}, {Huelamo}, {Isella}, {Langlois}, {M{\'e}nard},
  {Mouillet}, {Olofsson}, {Pantin}, {Pinte}, \& {Pueyo}}]{Benisty15}
{Benisty}, M., {Juhasz}, A., {Boccaletti}, A., {et~al.} 2015, \aap, 578, L6

\bibitem[{{Bertrang} {et~al.}(2018){Bertrang}, {Avenhaus}, {Casassus},
  {Montesinos}, {Kirchschlager}, {Perez}, {Cieza}, \& {Wolf}}]{Bertrang18}
{Bertrang}, G.~H.~M., {Avenhaus}, H., {Casassus}, S., {et~al.} 2018, \mnras,
  474, 5105

\bibitem[{{Bertrang} {et~al.}(2020){Bertrang}, {Flock}, {Keppler}, {Trifonov},
  {Penzlin}, {Avenhaus}, {Henning}, \& {Montesinos}}]{Bertrang20}
{Bertrang}, G. H.~M., {Flock}, M., {Keppler}, M., {et~al.} 2020, arXiv
  e-prints, arXiv:2007.11565

\bibitem[{{Beuzit} {et~al.}(2019){Beuzit}, {Vigan}, {Mouillet}, {Dohlen},
  {Gratton}, {Boccaletti}, {Sauvage}, {Schmid}, {Langlois}, {Petit},
  {Baruffolo}, {Feldt}, {Milli}, {Wahhaj}, {Abe}, {Anselmi}, {Antichi},
  {Barette}, {Baudrand}, {Baudoz}, {Bazzon}, {Bernardi}, {Blanchard}, {Brast},
  {Bruno}, {Buey}, {Carbillet}, {Carle}, {Cascone}, {Chapron}, {Charton},
  {Chauvin}, {Claudi}, {Costille}, {De Caprio}, {de Boer}, {Delboulb{\'e}},
  {Desidera}, {Dominik}, {Downing}, {Dupuis}, {Fabron}, {Fantinel}, {Farisato},
  {Feautrier}, {Fedrigo}, {Fusco}, {Gigan}, {Ginski}, {Girard}, {Giro},
  {Gisler}, {Gluck}, {Gry}, {Henning}, {Hubin}, {Hugot}, {Incorvaia}, {Jaquet},
  {Kasper}, {Lagadec}, {Lagrange}, {Le Coroller}, {Le Mignant}, {Le Ruyet},
  {Lessio}, {Lizon}, {Llored}, {Lundin}, {Madec}, {Magnard}, {Marteaud},
  {Martinez}, {Maurel}, {M{\'e}nard}, {Mesa}, {M{\"o}ller-Nilsson}, {Moulin},
  {Moutou}, {Orign{\'e}}, {Parisot}, {Pavlov}, {Perret}, {Pragt}, {Puget},
  {Rabou}, {Ramos}, {Reess}, {Rigal}, {Rochat}, {Roelfsema}, {Rousset}, {Roux},
  {Saisse}, {Salasnich}, {Santambrogio}, {Scuderi}, {Segransan}, {Sevin},
  {Siebenmorgen}, {Soenke}, {Stadler}, {Suarez}, {Tiph{\`e}ne}, {Turatto},
  {Udry}, {Vakili}, {Waters}, {Weber}, {Wildi}, {Zins}, \& {Zurlo}}]{Beuzit19}
{Beuzit}, J.~L., {Vigan}, A., {Mouillet}, D., {et~al.} 2019, \aap, 631, A155

\bibitem[{{Brott} \& {Hauschildt}(2005)}]{Brott05}
{Brott}, I. \& {Hauschildt}, P.~H. 2005, in ESA Special Publication, Vol. 576,
  The Three-Dimensional Universe with Gaia, 565

\bibitem[{{Canovas} {et~al.}(2013){Canovas}, {M{\'e}nard}, {Hales},
  {Jord{\'a}n}, {Schreiber}, {Casassus}, {Gledhill}, \& {Pinte}}]{Canovas13}
{Canovas}, H., {M{\'e}nard}, F., {Hales}, A., {et~al.} 2013, \aap, 556, A123

\bibitem[{{Carney} {et~al.}(2018){Carney}, {Fedele}, {Hogerheijde}, {Favre},
  {Walsh}, {Bruderer}, {Miotello}, {Murillo}, {Klaassen}, {Henning}, \& {van
  Dishoeck}}]{Carney18}
{Carney}, M.~T., {Fedele}, D., {Hogerheijde}, M.~R., {et~al.} 2018, \aap, 614,
  A106

\bibitem[{{Chavero} {et~al.}(2006){Chavero}, {G{\'o}mez}, {Whitney}, \&
  {Saffe}}]{Chavero06}
{Chavero}, C., {G{\'o}mez}, M., {Whitney}, B.~A., \& {Saffe}, C. 2006, \aap,
  452, 921

\bibitem[{{Chen} {et~al.}(2018){Chen}, {K{\'o}sp{\'a}l}, {{\'A}brah{\'a}m},
  {Kreplin}, {Matter}, \& {Weigelt}}]{Chen18}
{Chen}, L., {K{\'o}sp{\'a}l}, {\'A}., {{\'A}brah{\'a}m}, P., {et~al.} 2018,
  \aap, 609, A45

\bibitem[{{Chiang} \& {Goldreich}(1997)}]{Chiang97}
{Chiang}, E.~I. \& {Goldreich}, P. 1997, \apj, 490, 368

\bibitem[{{de Boer} {et~al.}(2020){de Boer}, {Langlois}, {van Holstein},
  {Girard}, {Mouillet}, {Vigan}, {Dohlen}, {Snik}, {Keller}, {Ginski}, {Stam},
  {Milli}, {Wahhaj}, {Kasper}, {Schmid}, {Rabou}, {Gluck}, {Hugot}, {Perret},
  {Martinez}, {Weber}, {Pragt}, {Sauvage}, {Boccaletti}, {Le Coroller},
  {Dominik}, {Henning}, {Lagadec}, {M{\'e}nard}, {Turatto}, {Udry}, {Chauvin},
  {Feldt}, \& {Beuzit}}]{deBoer20}
{de Boer}, J., {Langlois}, M., {van Holstein}, R.~G., {et~al.} 2020, \aap, 633,
  A63

\bibitem[{{Dionatos} {et~al.}(2019){Dionatos}, {Woitke}, {G{\"u}del},
  {Degroote}, {Liebhart}, {Anthonioz}, {Antonellini}, {Baldovin-Saavedra},
  {Carmona}, {Dominik}, {Greaves}, {Ilee}, {Kamp}, {M{\'e}nard}, {Min},
  {Pinte}, {Rab}, {Rigon}, {Thi}, \& {Waters}}]{Diana19}
{Dionatos}, O., {Woitke}, P., {G{\"u}del}, M., {et~al.} 2019, \aap, 625, A66

\bibitem[{{Dong} {et~al.}(2012){Dong}, {Rafikov}, {Zhu}, {Hartmann}, {Whitney},
  {Brandt}, {Muto}, {Hashimoto}, {Grady}, {Follette}, {Kuzuhara}, {Tanii},
  {Itoh}, {Thalmann}, {Wisniewski}, {Mayama}, {Janson}, {Abe}, {Brandner},
  {Carson}, {Egner}, {Feldt}, {Goto}, {Guyon}, {Hayano}, {Hayashi}, {Hayashi},
  {Henning}, {Hodapp}, {Honda}, {Inutsuka}, {Ishii}, {Iye}, {Kandori}, {Knapp},
  {Kudo}, {Kusakabe}, {Matsuo}, {McElwain}, {Miyama}, {Morino}, {Moro-Martin},
  {Nishimura}, {Pyo}, {Suto}, {Suzuki}, {Takami}, {Takato}, {Terada}, {Tomono},
  {Turner}, {Watanabe}, {Yamada}, {Takami}, {Usuda}, \& {Tamura}}]{Dong12}
{Dong}, R., {Rafikov}, R., {Zhu}, Z., {et~al.} 2012, \apj, 750, 161

\bibitem[{{Draine}(2003)}]{Draine03}
{Draine}, B.~T. 2003, \apj, 598, 1017

\bibitem[{{Draine} \& {Lee}(1984)}]{Draine84}
{Draine}, B.~T. \& {Lee}, H.~M. 1984, \apj, 285, 89

\bibitem[{{Dullemond} \& {Dominik}(2004)}]{Dullemond04}
{Dullemond}, C.~P. \& {Dominik}, C. 2004, \aap, 421, 1075

\bibitem[{{Dunkin} {et~al.}(1997){Dunkin}, {Barlow}, \& {Ryan}}]{Dunkin97}
{Dunkin}, S.~K., {Barlow}, M.~J., \& {Ryan}, S.~G. 1997, \mnras, 286, 604

\bibitem[{{Engler} {et~al.}(2017){Engler}, {Schmid}, {Thalmann}, {Boccaletti},
  {Bazzon}, {Baruffolo}, {Beuzit}, {Claudi}, {Costille}, {Desidera}, {Dohlen},
  {Dominik}, {Feldt}, {Fusco}, {Ginski}, {Gisler}, {Girard}, {Gratton},
  {Henning}, {Hubin}, {Janson}, {Kasper}, {Kral}, {Langlois}, {Lagadec},
  {M{\'e}nard}, {Meyer}, {Milli}, {Mouillet}, {Olofsson}, {Pavlov}, {Pragt},
  {Puget}, {Quanz}, {Roelfsema}, {Salasnich}, {Siebenmorgen}, {Sissa},
  {Suarez}, {Szulagyi}, {Turatto}, {Udry}, \& {Wildi}}]{Engler17}
{Engler}, N., {Schmid}, H.~M., {Thalmann}, C., {et~al.} 2017, \aap, 607, A90

\bibitem[{{Fitzpatrick}(2004)}]{Fitzpatrick04}
{Fitzpatrick}, E.~L. 2004, in ASP, Vol. 309, Astrophysics of Dust, 33

\bibitem[{{Gaia Collaboration}(2018)}]{Gaia18}
{Gaia Collaboration}. 2018, VizieR Online Data Catalog, I/345

\bibitem[{{Garufi} {et~al.}(2017{\natexlab{a}}){Garufi}, {Benisty}, {Stolker},
  {Avenhaus}, {de Boer}, {Pohl}, {Quanz}, {Dominik}, {Ginski}, {Thalmann}, {van
  Boekel}, {Boccaletti}, {Henning}, {Janson}, {Salter}, {Schmid}, {Sissa},
  {Langlois}, {Beuzit}, {Chauvin}, {Mouillet}, {Augereau}, {Bazzon}, {Biller},
  {Bonnefoy}, {Buenzli}, {Cheetham}, {Daemgen}, {Desidera}, {Engler}, {Feldt},
  {Girard}, {Gratton}, {Hagelberg}, {Keller}, {Keppler}, {Kenworthy}, {Kral},
  {Lopez}, {Maire}, {Menard}, {Mesa}, {Messina}, {Meyer}, {Milli}, {Min},
  {Muller}, {Olofsson}, {Pawellek}, {Pinte}, {Szulagyi}, {Vigan}, {Wahhaj},
  {Waters}, \& {Zurlo}}]{Garufi17b}
{Garufi}, A., {Benisty}, M., {Stolker}, T., {et~al.} 2017{\natexlab{a}}, The
  Messenger, 169, 32

\bibitem[{{Garufi} {et~al.}(2017{\natexlab{b}}){Garufi}, {Meeus}, {Benisty},
  {Quanz}, {Banzatti}, {Kama}, {Canovas}, {Eiroa}, {Schmid}, {Stolker}, {Pohl},
  {Rigliaco}, {M{\'e}nard}, {Meyer}, {van Boekel}, \& {Dominik}}]{Garufi17}
{Garufi}, A., {Meeus}, G., {Benisty}, M., {et~al.} 2017{\natexlab{b}}, \aap,
  603, A21

\bibitem[{{Gledhill} \& {Scarrott}(1989)}]{Gledhill89}
{Gledhill}, T.~M. \& {Scarrott}, S.~M. 1989, \mnras, 236, 139

\bibitem[{{Grady} {et~al.}(2007){Grady}, {Schneider}, {Hamaguchi}, {Sitko},
  {Carpenter}, {Hines}, {Collins}, {Williger}, {Woodgate}, {Henning},
  {M{\'e}nard}, {Wilner}, {Petre}, {Palunas}, {Quirrenbach}, {Nuth},
  {Silverstone}, \& {Kim}}]{Grady07}
{Grady}, C.~A., {Schneider}, G., {Hamaguchi}, K., {et~al.} 2007, \apj, 665,
  1391

\bibitem[{{Graham} {et~al.}(2007){Graham}, {Kalas}, \& {Matthews}}]{Graham07}
{Graham}, J.~R., {Kalas}, P.~G., \& {Matthews}, B.~C. 2007, \apj, 654, 595

\bibitem[{{Gratton} {et~al.}(2019){Gratton}, {Ligi}, {Sissa}, {Desidera},
  {Mesa}, {Bonnefoy}, {Chauvin}, {Cheetham}, {Feldt}, {Lagrange}, {Langlois},
  {Meyer}, {Vigan}, {Boccaletti}, {Janson}, {Lazzoni}, {Zurlo}, {De Boer},
  {Henning}, {D'Orazi}, {Gluck}, {Madec}, {Jaquet}, {Baudoz}, {Fantinel},
  {Pavlov}, \& {Wildi}}]{Gratton19}
{Gratton}, R., {Ligi}, R., {Sissa}, E., {et~al.} 2019, \aap, 623, A140

\bibitem[{{Habart} {et~al.}(2006){Habart}, {Natta}, {Testi}, \&
  {Carbillet}}]{Habart06}
{Habart}, E., {Natta}, A., {Testi}, L., \& {Carbillet}, M. 2006, \aap, 449,
  1067

\bibitem[{{Hales} {et~al.}(2006){Hales}, {Gledhill}, {Barlow}, \&
  {Lowe}}]{Hales06}
{Hales}, A.~S., {Gledhill}, T.~M., {Barlow}, M.~J., \& {Lowe}, K.~T.~E. 2006,
  \mnras, 365, 1348

\bibitem[{{Hashimoto} {et~al.}(2011){Hashimoto}, {Tamura}, {Muto}, {Kudo},
  {Fukagawa}, {Fukue}, {Goto}, {Grady}, {Henning}, {Hodapp}, {Honda},
  {Inutsuka}, {Kokubo}, {Knapp}, {McElwain}, {Momose}, {Ohashi}, {Okamoto},
  {Takami}, {Turner}, {Wisniewski}, {Janson}, {Abe}, {Brandner}, {Carson},
  {Egner}, {Feldt}, {Golota}, {Guyon}, {Hayano}, {Hayashi}, {Hayashi}, {Ishii},
  {Kandori}, {Kusakabe}, {Matsuo}, {Mayama}, {Miyama}, {Morino}, {Moro-Martin},
  {Nishimura}, {Pyo}, {Suto}, {Suzuki}, {Takato}, {Terada}, {Thalmann},
  {Tomono}, {Watanabe}, {Yamada}, {Takami}, \& {Usuda}}]{Hashimoto11}
{Hashimoto}, J., {Tamura}, M., {Muto}, T., {et~al.} 2011, \apjl, 729, L17

\bibitem[{Hodapp {et~al.}(2008)Hodapp, Suzuki, Tamura, Abe, Suto, Kandori,
  Morino, Nishimura, Takami, Guyon, Jacobson, Stahlberger, Yamada, Shelton,
  Hashimoto, Tavrov, Nishikawa, Ukita, Izumiura, Hayashi, Nakajima, Yamada, \&
  Usuda}]{Hodapp08}
Hodapp, K.~W., Suzuki, R., Tamura, M., {et~al.} 2008, in Ground-based and
  Airborne Instrumentation for Astronomy II, ed. I.~S. McLean \& M.~M. Casali,
  Vol. 7014, International Society for Optics and Photonics (SPIE), 488 -- 499

\bibitem[{{Honda} {et~al.}(2012){Honda}, {Maaskant}, {Okamoto}, {Kataza},
  {Fukagawa}, {Waters}, {Dominik}, {Tielens}, {Mulders}, {Min}, {Yamashita},
  {Fujiyoshi}, {Miyata}, {Sako}, {Sakon}, {Fujiwara}, \& {Onaka}}]{Honda12}
{Honda}, M., {Maaskant}, K., {Okamoto}, Y.~K., {et~al.} 2012, \apj, 752, 143

\bibitem[{{Hunziker} {et~al.}(2021){Hunziker}, {Schmid}, {Ma}, {Menard},
  {Avenhaus}, {Boccaletti}, {Beuzit}, {Chauvin}, {Dohlen}, {Dominik}, {Engler},
  {Ginski}, {Gratton}, {Henning}, {Langlois}, {Milli}, {Mouillet}, {Tschudi},
  {van Holstein}, \& {Vigan}}]{Hunziker21}
{Hunziker}, S., {Schmid}, H.~M., {Ma}, J., {et~al.} 2021, \aap, 648, A110

\bibitem[{{Keller} {et~al.}(2008){Keller}, {Sloan}, {Forrest}, {Ayala},
  {D'Alessio}, {Shah}, {Calvet}, {Najita}, {Li}, {Hartmann}, {Sargent},
  {Watson}, \& {Chen}}]{Keller08}
{Keller}, L.~D., {Sloan}, G.~C., {Forrest}, W.~J., {et~al.} 2008, \apj, 684,
  411

\bibitem[{{Kluska} {et~al.}(2020){Kluska}, {Berger}, {Malbet}, {Lazareff},
  {Benisty}, {Le Bouquin}, {Absil}, {Baron}, {Delboulb{\'e}}, {Duvert},
  {Isella}, {Jocou}, {Juhasz}, {Kraus}, {Lachaume}, {M{\'e}nard},
  {Millan-Gabet}, {Monnier}, {Moulin}, {Perraut}, {Rochat}, {Pinte}, {Soulez},
  {Tallon}, {Thi}, {Thi{\'e}baut}, {Traub}, \& {Zins}}]{Kluska20}
{Kluska}, J., {Berger}, J.~P., {Malbet}, F., {et~al.} 2020, \aap, 636, A116

\bibitem[{{Kolokolova} \& {Kimura}(2010)}]{Kolokolova10}
{Kolokolova}, L. \& {Kimura}, H. 2010, \aap, 513, A40

\bibitem[{{Kuhn} {et~al.}(2001){Kuhn}, {Potter}, \& {Parise}}]{Kuhn01}
{Kuhn}, J.~R., {Potter}, D., \& {Parise}, B. 2001, \apjl, 553, L189

\bibitem[{{Lenzen} {et~al.}(2003){Lenzen}, {Hartung}, {Brandner}, {Finger},
  {Hubin}, {Lacombe}, {Lagrange}, {Lehnert}, {Moorwood}, \&
  {Mouillet}}]{Lenzen03}
{Lenzen}, R., {Hartung}, M., {Brandner}, W., {et~al.} 2003, SPIE Conf. Ser.,
  Vol. 4841, {NAOS-CONICA first on sky results in a variety of observing
  modes}, 944--952

\bibitem[{{Ligi} {et~al.}(2018){Ligi}, {Vigan}, {Gratton}, {de Boer},
  {Benisty}, {Boccaletti}, {Quanz}, {Meyer}, {Ginski}, {Sissa}, {Gry},
  {Henning}, {Beuzit}, {Biller}, {Bonnefoy}, {Chauvin}, {Cheetham}, {Cudel},
  {Delorme}, {Desidera}, {Feldt}, {Galicher}, {Girard}, {Janson}, {Kasper},
  {Kopytova}, {Lagrange}, {Langlois}, {Lecoroller}, {Maire}, {M{\'e}nard},
  {Mesa}, {Peretti}, {Perrot}, {Pinilla}, {Pohl}, {Rouan}, {Stolker},
  {Samland}, {Wahhaj}, {Wildi}, {Zurlo}, {Buey}, {Fantinel}, {Fusco}, {Jaquet},
  {Moulin}, {Ramos}, {Suarez}, \& {Weber}}]{Ligi18}
{Ligi}, R., {Vigan}, A., {Gratton}, R., {et~al.} 2018, \mnras, 473, 1774

\bibitem[{{Mac{\'\i}as} {et~al.}(2019){Mac{\'\i}as}, {Espaillat}, {Osorio},
  {Anglada}, {Torrelles}, {Carrasco-Gonz{\'a}lez}, {Flock}, {Linz}, {Bertrang},
  {Henning}, {G{\'o}mez}, {Calvet}, \& {Dent}}]{Macias19}
{Mac{\'\i}as}, E., {Espaillat}, C.~C., {Osorio}, M., {et~al.} 2019, \apj, 881,
  159

\bibitem[{{Mackay}(2003)}]{Mackay03}
{Mackay}, D. J.~C. 2003, {Information Theory, Inference and Learning
  Algorithms}

\bibitem[{{Meeus} {et~al.}(2001){Meeus}, {Waters}, {Bouwman}, {van den Ancker},
  {Waelkens}, \& {Malfait}}]{Meeus01}
{Meeus}, G., {Waters}, L.~B.~F.~M., {Bouwman}, J., {et~al.} 2001, \aap, 365,
  476

\bibitem[{{Min} {et~al.}(2012){Min}, {Canovas}, {Mulders}, \& {Keller}}]{Min12}
{Min}, M., {Canovas}, H., {Mulders}, G.~D., \& {Keller}, C.~U. 2012, \aap, 537,
  A75

\bibitem[{{Min} {et~al.}(2016){Min}, {Rab}, {Woitke}, {Dominik}, \&
  {M{\'e}nard}}]{Min16}
{Min}, M., {Rab}, C., {Woitke}, P., {Dominik}, C., \& {M{\'e}nard}, F. 2016,
  \aap, 585, A13

\bibitem[{{Momose} {et~al.}(2015){Momose}, {Morita}, {Fukagawa}, {Muto},
  {Takeuchi}, {Hashimoto}, {Honda}, {Kudo}, {Okamoto}, {Kanagawa}, {Tanaka},
  {Grady}, {Sitko}, {Akiyama}, {Currie}, {Follette}, {Mayama}, {Kusakabe},
  {Abe}, {Brandner}, {Brand t}, {Carson}, {Egner}, {Feldt}, {Goto}, {Guyon},
  {Hayano}, {Hayashi}, {Hayashi}, {Henning}, {Hodapp}, {Ishii}, {Iye},
  {Janson}, {Kandori}, {Knapp}, {Kuzuhara}, {Kwon}, {Matsuo}, {McElwain},
  {Miyama}, {Morino}, {Moro-Martin}, {Nishimura}, {Pyo}, {Serabyn}, {Suenaga},
  {Suto}, {Suzuki}, {Takahashi}, {Takami}, {Takato}, {Terada}, {Thalmann},
  {Tomono}, {Turner}, {Watanabe}, {Wisniewski}, {Yamada}, {Takami}, {Usuda}, \&
  {Tamura}}]{Momose15}
{Momose}, M., {Morita}, A., {Fukagawa}, M., {et~al.} 2015, \pasj, 67, 83

\bibitem[{{Monnier} {et~al.}(2017){Monnier}, {Harries}, {Aarnio}, {Adams},
  {Andrews}, {Calvet}, {Espaillat}, {Hartmann}, {Hinkley}, {Kraus}, {McClure},
  {Oppenheimer}, {Perrin}, \& {Wilner}}]{Monnier17}
{Monnier}, J.~D., {Harries}, T.~J., {Aarnio}, A., {et~al.} 2017, \apj, 838, 20

\bibitem[{{Monnier} {et~al.}(2019){Monnier}, {Harries}, {Bae}, {Setterholm},
  {Laws}, {Aarnio}, {Adams}, {Andrews}, {Calvet}, {Espaillat}, {Hartmann},
  {Kraus}, {McClure}, {Miller}, {Oppenheimer}, {Wilner}, \& {Zhu}}]{Monnier19}
{Monnier}, J.~D., {Harries}, T.~J., {Bae}, J., {et~al.} 2019, \apj, 872, 122

\bibitem[{{Mulders} {et~al.}(2013){Mulders}, {Min}, {Dominik}, {Debes}, \&
  {Schneider}}]{Mulders13}
{Mulders}, G.~D., {Min}, M., {Dominik}, C., {Debes}, J.~H., \& {Schneider}, G.
  2013, \aap, 549, A112

\bibitem[{{Murakawa}(2010)}]{Murakawa10}
{Murakawa}, K. 2010, \aap, 518, A63

\bibitem[{{Murakawa} {et~al.}(2004){Murakawa}, {Suto}, {Tamura}, {Kaifu},
  {Takami}, {Takato}, {Oya}, {Hayano}, {Gaessler}, \& {Kamata}}]{Murakawa04}
{Murakawa}, K., {Suto}, H., {Tamura}, M., {et~al.} 2004, \pasj, 56, 509

\bibitem[{{Muto} {et~al.}(2012){Muto}, {Grady}, {Hashimoto}, {Fukagawa},
  {Hornbeck}, {Sitko}, {Russell}, {Werren}, {Cur{\'e}}, {Currie}, {Ohashi},
  {Okamoto}, {Momose}, {Honda}, {Inutsuka}, {Takeuchi}, {Dong}, {Abe},
  {Brandner}, {Brandt}, {Carson}, {Egner}, {Feldt}, {Fukue}, {Goto}, {Guyon},
  {Hayano}, {Hayashi}, {Hayashi}, {Henning}, {Hodapp}, {Ishii}, {Iye},
  {Janson}, {Kandori}, {Knapp}, {Kudo}, {Kusakabe}, {Kuzuhara}, {Matsuo},
  {Mayama}, {McElwain}, {Miyama}, {Morino}, {Moro-Martin}, {Nishimura}, {Pyo},
  {Serabyn}, {Suto}, {Suzuki}, {Takami}, {Takato}, {Terada}, {Thalmann},
  {Tomono}, {Turner}, {Watanabe}, {Wisniewski}, {Yamada}, {Takami}, {Usuda}, \&
  {Tamura}}]{Muto12}
{Muto}, T., {Grady}, C.~A., {Hashimoto}, J., {et~al.} 2012, \apjl, 748, L22

\bibitem[{{Okamoto} {et~al.}(2017){Okamoto}, {Kataza}, {Honda}, {Yamashita},
  {Fujiyoshi}, {Miyata}, {Sako}, {Fujiwara}, {Sakon}, {Fukagawa}, {Momose}, \&
  {Onaka}}]{Okamoto17}
{Okamoto}, Y.~K., {Kataza}, H., {Honda}, M., {et~al.} 2017, \aj, 154, 16

\bibitem[{{Osorio} {et~al.}(2014){Osorio}, {Anglada}, {Carrasco-Gonz{\'a}lez},
  {Torrelles}, {Mac{\'\i}as}, {Rodr{\'\i}guez}, {G{\'o}mez}, {D'Alessio},
  {Calvet}, {Nagel}, {Dent}, {Quanz}, {Reggiani}, \& {Mayen-Gijon}}]{Osorio14}
{Osorio}, M., {Anglada}, G., {Carrasco-Gonz{\'a}lez}, C., {et~al.} 2014, \apjl,
  791, L36

\bibitem[{{P{\'e}rez} {et~al.}(2019){P{\'e}rez}, {Casassus}, {Baruteau},
  {Dong}, {Hales}, \& {Cieza}}]{Perez19}
{P{\'e}rez}, S., {Casassus}, S., {Baruteau}, C., {et~al.} 2019, \aj, 158, 15

\bibitem[{{Perrin} {et~al.}(2015){Perrin}, {Duchene}, {Millar-Blanchaer},
  {Fitzgerald}, {Graham}, {Wiktorowicz}, {Kalas}, {Macintosh}, {Bauman},
  {Cardwell}, {Chilcote}, {De Rosa}, {Dillon}, {Doyon}, {Dunn}, {Erikson},
  {Gavel}, {Goodsell}, {Hartung}, {Hibon}, {Ingraham}, {Kerley}, {Konapacky},
  {Larkin}, {Maire}, {Marchis}, {Marois}, {Mittal}, {Morzinski}, {Oppenheimer},
  {Palmer}, {Patience}, {Poyneer}, {Pueyo}, {Rantakyr{\"o}}, {Sadakuni},
  {Saddlemyer}, {Savransky}, {Soummer}, {Sivaramakrishnan}, {Song}, {Thomas},
  {Wallace}, {Wang}, \& {Wolff}}]{Perrin15}
{Perrin}, M.~D., {Duchene}, G., {Millar-Blanchaer}, M., {et~al.} 2015, \apj,
  799, 182

\bibitem[{{Perrin} {et~al.}(2009){Perrin}, {Schneider}, {Duchene}, {Pinte},
  {Grady}, {Wisniewski}, \& {Hines}}]{Perrin09}
{Perrin}, M.~D., {Schneider}, G., {Duchene}, G., {et~al.} 2009, \apjl, 707,
  L132

\bibitem[{{Pohl} {et~al.}(2017){Pohl}, {Benisty}, {Pinilla}, {Ginski}, {de
  Boer}, {Avenhaus}, {Henning}, {Zurlo}, {Boccaletti}, {Augereau}, {Birnstiel},
  {Dominik}, {Facchini}, {Fedele}, {Janson}, {Keppler}, {Kral}, {Langlois},
  {Ligi}, {Maire}, {M{\'e}nard}, {Meyer}, {Pinte}, {Quanz}, {Sauvage},
  {Sezestre}, {Stolker}, {Szul{\'a}gyi}, {van Boekel}, {van der Plas},
  {Villenave}, {Baruffolo}, {Baudoz}, {Le Mignant}, {Maurel}, {Ramos}, \&
  {Weber}}]{Pohl17}
{Pohl}, A., {Benisty}, M., {Pinilla}, P., {et~al.} 2017, \apj, 850, 52

\bibitem[{{Quanz} {et~al.}(2013){Quanz}, {Avenhaus}, {Buenzli}, {Garufi},
  {Schmid}, \& {Wolf}}]{Quanz13}
{Quanz}, S.~P., {Avenhaus}, H., {Buenzli}, E., {et~al.} 2013, \apjl, 766, L2

\bibitem[{{Quanz} {et~al.}(2011){Quanz}, {Schmid}, {Geissler}, {Meyer},
  {Henning}, {Brandner}, \& {Wolf}}]{Quanz11}
{Quanz}, S.~P., {Schmid}, H.~M., {Geissler}, K., {et~al.} 2011, \apj, 738, 23

\bibitem[{{Raman} {et~al.}(2006){Raman}, {Lisanti}, {Wilner}, {Qi}, \&
  {Hogerheijde}}]{Raman06}
{Raman}, A., {Lisanti}, M., {Wilner}, D.~J., {Qi}, C., \& {Hogerheijde}, M.
  2006, \aj, 131, 2290

\bibitem[{{Roddier} {et~al.}(1996){Roddier}, {Roddier}, {Northcott}, {Graves},
  \& {Jim}}]{Roddier96}
{Roddier}, C., {Roddier}, F., {Northcott}, M.~J., {Graves}, J.~E., \& {Jim}, K.
  1996, \apj, 463, 326

\bibitem[{{Schmid}(2021)}]{Schmid21}
{Schmid}, H.~M. 2021, IAU Symp. (in press), 360

\bibitem[{{Schmid} {et~al.}(2018){Schmid}, {Bazzon}, {Roelfsema}, {Mouillet},
  {Milli}, {Menard}, {Gisler}, {Hunziker}, {Pragt}, {Dominik}, {Boccaletti},
  {Ginski}, {Abe}, {Antoniucci}, {Avenhaus}, {Baruffolo}, {Baudoz}, {Beuzit},
  {Carbillet}, {Chauvin}, {Claudi}, {Costille}, {Daban}, {de Haan}, {Desidera},
  {Dohlen}, {Downing}, {Elswijk}, {Engler}, {Feldt}, {Fusco}, {Girard},
  {Gratton}, {Hanenburg}, {Henning}, {Hubin}, {Joos}, {Kasper}, {Keller},
  {Langlois}, {Lagadec}, {Martinez}, {Mulder}, {Pavlov}, {Podio}, {Puget},
  {Quanz}, {Rigal}, {Salasnich}, {Sauvage}, {Schuil}, {Siebenmorgen}, {Sissa},
  {Snik}, {Suarez}, {Thalmann}, {Turatto}, {Udry}, {van Duin}, {van Holstein},
  {Vigan}, \& {Wildi}}]{Schmid18}
{Schmid}, H.~M., {Bazzon}, A., {Roelfsema}, R., {et~al.} 2018, \aap, 619, A9

\bibitem[{{Schmid} {et~al.}(2006){Schmid}, {Joos}, \& {Tschan}}]{Schmid06}
{Schmid}, H.~M., {Joos}, F., \& {Tschan}, D. 2006, \aap, 452, 657

\bibitem[{{Seok} \& {Li}(2016)}]{Seok16}
{Seok}, J.~Y. \& {Li}, A. 2016, \apj, 818, 2

\bibitem[{{Silber} {et~al.}(2000){Silber}, {Gledhill}, {Duch{\^e}ne}, \&
  {M{\'e}nard}}]{Silber00}
{Silber}, J., {Gledhill}, T., {Duch{\^e}ne}, G., \& {M{\'e}nard}, F. 2000,
  \apjl, 536, L89

\bibitem[{{Simmons} \& {Stewart}(1985)}]{Simmons85}
{Simmons}, J.~F.~L. \& {Stewart}, B.~G. 1985, \aap, 142, 100

\bibitem[{{Sisson} {et~al.}(2018){Sisson}, {Fan}, \& {Beaumont}}]{Sisson18}
{Sisson}, S.~A., {Fan}, Y., \& {Beaumont}, M.~A. 2018, arXiv e-prints,
  arXiv:1802.09720

\bibitem[{{Sloan} {et~al.}(2005){Sloan}, {Keller}, {Forrest}, {Leibensperger},
  {Sargent}, {Li}, {Najita}, {Watson}, {Brandl}, {Chen}, {Green},
  {Markwick-Kemper}, {Herter}, {D'Alessio}, {Morris}, {Barry}, {Hall}, {Myers},
  \& {Houck}}]{Sloan05}
{Sloan}, G.~C., {Keller}, L.~D., {Forrest}, W.~J., {et~al.} 2005, \apj, 632,
  956

\bibitem[{{Stolker} {et~al.}(2016){Stolker}, {Dominik}, {Avenhaus}, {Min}, {de
  Boer}, {Ginski}, {Schmid}, {Juhasz}, {Bazzon}, {Waters}, {Garufi},
  {Augereau}, {Benisty}, {Boccaletti}, {Henning}, {Langlois}, {Maire},
  {M{\'e}nard}, {Meyer}, {Pinte}, {Quanz}, {Thalmann}, {Beuzit}, {Carbillet},
  {Costille}, {Dohlen}, {Feldt}, {Gisler}, {Mouillet}, {Pavlov}, {Perret},
  {Petit}, {Pragt}, {Rochat}, {Roelfsema}, {Salasnich}, {Soenke}, \&
  {Wildi}}]{Stolker16}
{Stolker}, T., {Dominik}, C., {Avenhaus}, H., {et~al.} 2016, \aap, 595, A113

\bibitem[{{Sylvester} {et~al.}(1995){Sylvester}, {Barlow}, \&
  {Skinner}}]{Sylvester95}
{Sylvester}, R.~J., {Barlow}, M.~J., \& {Skinner}, C.~J. 1995, \apss, 224, 405

\bibitem[{{Sylvester} {et~al.}(1996){Sylvester}, {Skinner}, {Barlow}, \&
  {Mannings}}]{Sylvester96}
{Sylvester}, R.~J., {Skinner}, C.~J., {Barlow}, M.~J., \& {Mannings}, V. 1996,
  \mnras, 279, 915

\bibitem[{{Tazaki} {et~al.}(2019){Tazaki}, {Tanaka}, {Muto}, {Kataoka}, \&
  {Okuzumi}}]{Tazaki19}
{Tazaki}, R., {Tanaka}, H., {Muto}, T., {Kataoka}, A., \& {Okuzumi}, S. 2019,
  \mnras, 485, 4951

\bibitem[{{van der Marel}(2017)}]{vanderMarel17}
{van der Marel}, N. 2017, {The ALMA Revolution: Gas and Dust in Transitional
  Disks}, ed. M.~{Pessah} \& O.~{Gressel}, Vol. 445, 39

\bibitem[{{van Holstein} {et~al.}(2020){van Holstein}, {Girard}, {de Boer},
  {Snik}, {Milli}, {Stam}, {Ginski}, {Mouillet}, {Wahhaj}, {Schmid}, {Keller},
  {Langlois}, {Dohlen}, {Vigan}, {Pohl}, {Carbillet}, {Fantinel}, {Maurel},
  {Orign{\'e}}, {Petit}, {Ramos}, {Rigal}, {Sevin}, {Boccaletti}, {Le
  Coroller}, {Dominik}, {Henning}, {Lagadec}, {M{\'e}nard}, {Turatto}, {Udry},
  {Chauvin}, {Feldt}, \& {Beuzit}}]{vanHolstein20}
{van Holstein}, R.~G., {Girard}, J.~H., {de Boer}, J., {et~al.} 2020, \aap,
  633, A64

\bibitem[{{Villenave} {et~al.}(2019){Villenave}, {Benisty}, {Dent},
  {M{\'e}nard}, {Garufi}, {Ginski}, {Pinilla}, {Pinte}, {Williams}, {de Boer},
  {Morino}, {Fukagawa}, {Dominik}, {Flock}, {Henning}, {Juh{\'a}sz}, {Keppler},
  {Muro-Arena}, {Olofsson}, {P{\'e}rez}, {van der Plas}, {Zurlo}, {Carle},
  {Feautrier}, {Pavlov}, {Pragt}, {Ramos}, {Sauvage}, {Stadler}, \&
  {Weber}}]{Villenave19}
{Villenave}, M., {Benisty}, M., {Dent}, W.~R.~F., {et~al.} 2019, \aap, 624, A7

\bibitem[{Walker \& Wolstencroft(1988)}]{Walker88}
Walker, H.~J. \& Wolstencroft, R.~D. 1988, PASP, 100, 1509

\bibitem[{{Weingartner} \& {Draine}(2001)}]{Weingartner01}
{Weingartner}, J.~C. \& {Draine}, B.~T. 2001, \apj, 548, 296

\bibitem[{{Woitke} {et~al.}(2019){Woitke}, {Kamp}, {Antonellini}, {Anthonioz},
  {Baldovin-Saveedra}, {Carmona}, {Dionatos}, {Dominik}, {Greaves},
  {G{\"u}del}, {Ilee}, {Liebhardt}, {Menard}, {Min}, {Pinte}, {Rab}, {Rigon},
  {Thi}, {Thureau}, \& {Waters}}]{DianaMod19}
{Woitke}, P., {Kamp}, I., {Antonellini}, S., {et~al.} 2019, \pasp, 131, 064301

\bibitem[{Woitke {et~al.}(2019)Woitke, Kamp, Antonellini, Anthonioz,
  Baldovin-Saveedra, Carmona, Dionatos, Dominik, Greaves, Güdel, Ilee,
  Liebhardt, Menard, Min, Pinte, Rab, Rigon, Thi, Thureau, \&
  Waters}]{Woitke19}
Woitke, P., Kamp, I., Antonellini, S., {et~al.} 2019, PASP, 131, 064301

\bibitem[{{Woitke} {et~al.}(2016){Woitke}, {Min}, {Pinte}, {Thi}, {Kamp},
  {Rab}, {Anthonioz}, {Antonellini}, {Baldovin-Saavedra}, {Carmona}, {Dominik},
  {Dionatos}, {Greaves}, {G{\"u}del}, {Ilee}, {Liebhart}, {M{\'e}nard},
  {Rigon}, {Waters}, {Aresu}, {Meijerink}, \& {Spaans}}]{Woitke16}
{Woitke}, P., {Min}, M., {Pinte}, C., {et~al.} 2016, \aap, 586, A103

\bibitem[{{Yudin} \& {Evans}(1998)}]{Yudin98}
{Yudin}, R.~V. \& {Evans}, A. 1998, \aaps, 131, 401

\end{thebibliography}

\begin{appendix} 
\section{Telescope polarization correction}\label{A1_telpol}

The aluminum coating of the M3 mirror in the VLT introduces instrumental polarization, which is partly but not completely compensated for by the following optical components. The details are described in~\cite{Schmid18} with correction formulas and parameters for different filters. The measured polarization for uncorrected cycles are expected to lie on a circle in the $Q/I$-$U/I$ plane because the M3-mirror orientation of a Nasmyth telescope rotates with respect to the sky during the observations.  The radius of the circle is defined by the telescope polarization $p_{\rm tel}$, and the position angle $\theta_{\rm tel}$ depends on the parallactic angle $\theta_{\rm para}(t)$ and a SPHERE-ZIMPOL specific position angle offset $\delta_{\rm tel}$, which depends on the used passband filter. The midpoint ($q_{\rm m},u_{\rm m}$) of the circle can be offset from (0,0) because of an intrinsic polarization of the star or interstellar polarization introduced by dust between us and the target.
\begin{figure}[ht!]
    \includegraphics[width=0.46\textwidth]{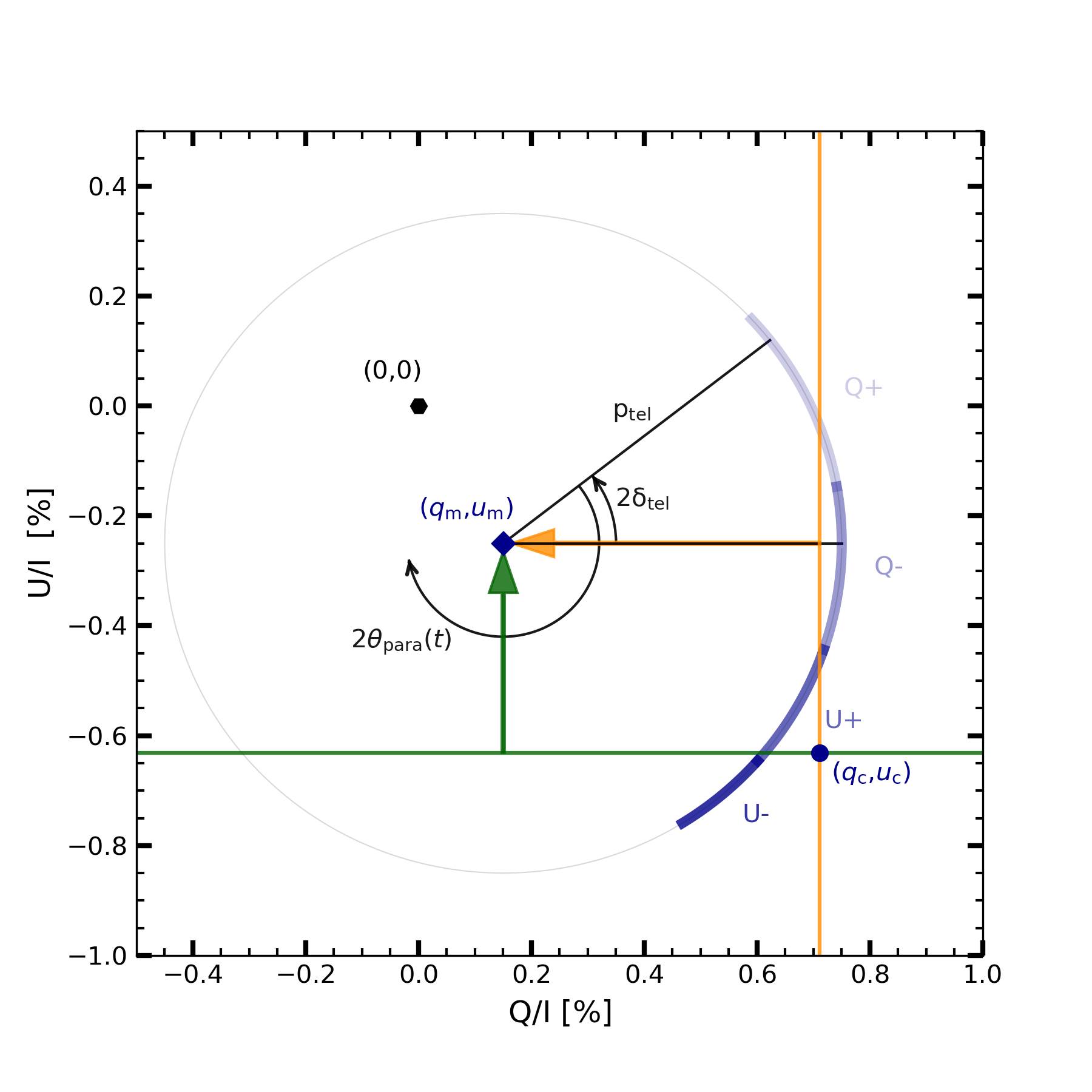}
     \centering
     \caption{Measured $Q/I$ and $U/I$
       parameters and correction for the temporal evolution of the telescope
       polarization position angle due to the changing parallactic angle
       during one polarimetric cycle.  
       The blue circle sections show the effect of the changing
       telescope polarization angle $\theta_{\rm tel}$ during the
       $Q^+$ (lightest blue), $Q^-$, $U^+$, and $U^-$ (darkest blue) integrations.
       The required corrections for Stokes $Q/I$ and $U/I$ are indicated by the
       orange and green arrows, respectively.
}
\label{TelPolAp0}
\end{figure}
\begin{figure}[ht!]
    \includegraphics[width=0.46\textwidth]{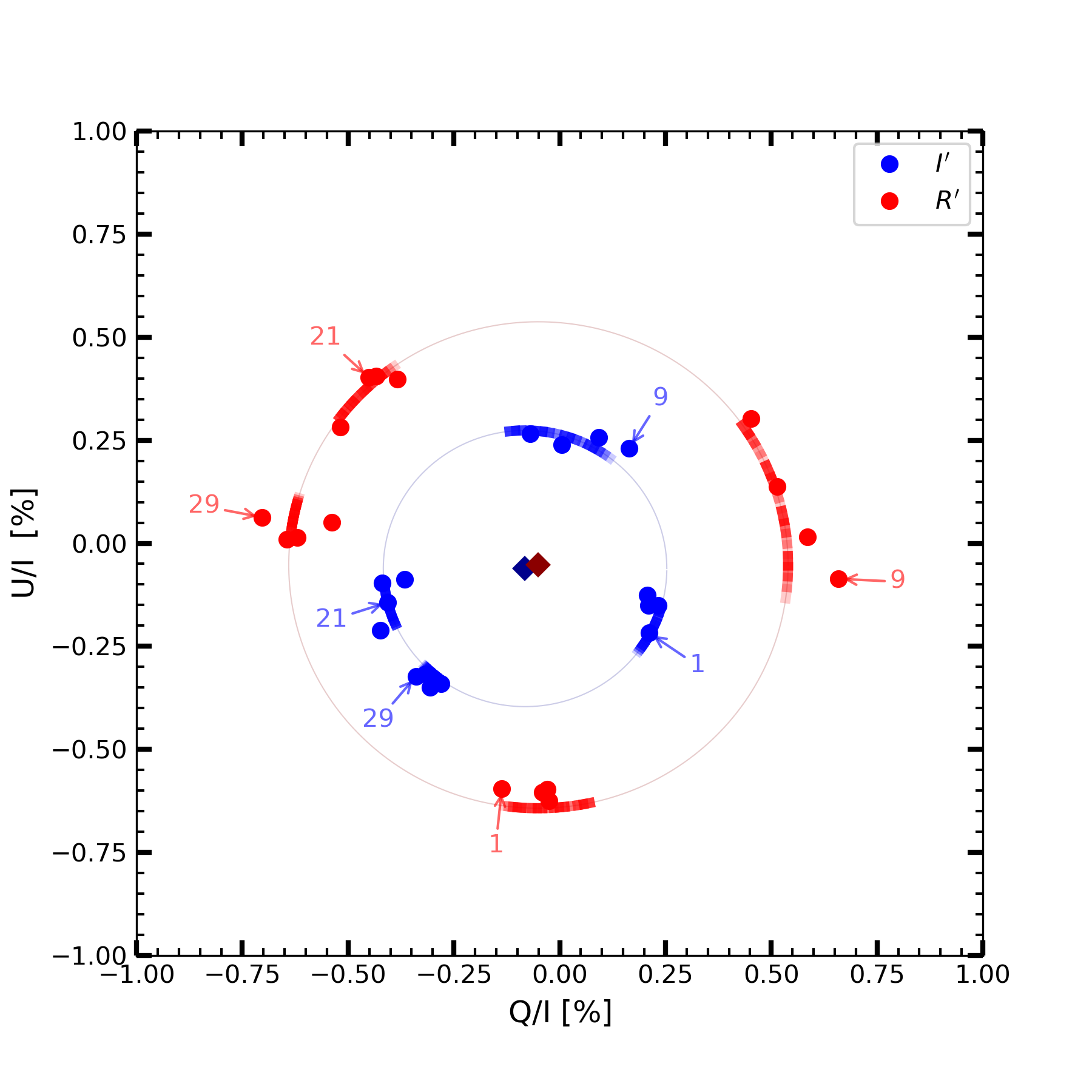}
     \centering
     \caption{Measured I$'$-band (blue dots) and
       R$'$-band (red dots) polarization for HD~169142 for the individual,
       non-corrected, cycle. The best solutions for the corrected
       polarization are shown with diamonds for the two bands.
       The thin circles illustrate the derived
       telescope polarization $p_{\rm tel}$, and the colored lines the derived
       position angle offset $\theta_{\rm tel}$ for the measured cycles. 
       Every fourth cycle number, as defined in Table~\ref{obsSHOW}, is annotated.
      }
\label{TelPolAp6}
\end{figure}
\\
Therefore, the telescope polarization $p_{\rm tel}$ can be derived from the systematic change of the polarization from cycle to cycle along this circle in the $Q/I$-$U/I$ plane. Actually, a more accurate correction should even consider the evolution of $\theta_{\rm para}$ during each polarimetric cycle, which can be very fast for targets near zenith. In this case $\theta_{\rm para}(t)$ differs strongly between the Stokes $Q$ and $U$ measurements introducing a systematic bias effect in the polarization measurement and one should use the following correction formulas for the four measurements $Q^+$, $Q^-$, $U^+$, and $U^-$ of a polarimetric cycle:
\begin{displaymath}
  Q_{\rm cor}^{\pm} = \pm \left( Q - p_{tel}\frac{\int cos(2(\theta_{\rm para}(t)+\delta_{\rm tel})) dt}{nDIT \cdot DIT}\ I \right), 
\end{displaymath}   
\begin{displaymath}
 U_{\rm cor}^{\pm} = \pm \left( U - p_{tel}\frac{\int sin(2(\theta_{\rm para}(t)+\delta_{\rm tel})) dt}{nDIT \cdot DIT}\ I \right). 
\end{displaymath}  
Figure~\ref{TelPolAp0} shows an extreme numerical example for a strong evolution of the parallactic angle by $50^\circ$ during a long polarimetric cycle, resulting in ($q_{\rm c},u_{\rm c}$) position of the cycle in the $Q/I$-$U/I$ plane. The necessary polarization corrections shown in orange (Stokes $Q$) and green (Stokes $U$), calculated from the formulas above, represent a time average $Q/I$ value for the instrument polarization circle sections shown in light blue for the $Q^+$ and $Q^-$ measurements and for the average $U/I$ value for the dark blue $U^+$ and $U^-$ measurements. For this example using the mean parallactic angle of the complete cycle for the U correction would lead to a small correction ($+0.08 \%$) instead of the value ($+0.38 \%$) indicated by the green arrow. Neglecting the temporal evolution effect of $\theta_{\rm para}(t)$ can produce wrong results for the telescope polarization $p_{\rm tel},\theta_{\rm tel}$ and therefore also the ``corrected'' target polarization $(Q_{\rm cor}/I,U_{\rm cor}/I)=(q_{\rm m},u_{\rm m})$. 

We analyzed our fast polarimetry cycles of HD~169142 (all cycles without saturation) using the detailed formula above and searched for the telescope polarization parameters $p_{\rm tel}$ and $\delta_{\rm tel}$, which minimize the scatter for the resulting circle center $(Q_{\rm cor}/I,U_{\rm cor}/I)$. Figure~\ref{TelPolAp6} shows the measured polarization $(Q/I,U/I)$ of each cycle for the I$'$ band and the R$'$ band and the median values for the corrected polarization $(Q_{\rm cor}/I,U_{\rm cor}/I)$ = $( -(0.08\pm0.03) \%, -(0.06\pm0.03)\%$) for the I$'$ band and $( -(0.05\pm0.05)\%, -(0.05\pm0.04)\%$) for the R$'$ band.  Also shown are the circles, which represent the telescope polarization parameters $p_{\rm tel}=0.34 \%$ and $\delta_{\rm tel}=44.4^\circ$ for the I$'$ band and  $p_{\rm tel}=0.59 \%$ and $\delta_{\rm tel}=13.8^\circ$ for the R$'$ band. For our data of HD~169142 the parallactic angle changes during one cycle are less than 6 degrees and the detailed formulae given above yields almost the same result like the approximate correction described in \citet{Schmid18} using just polarization values for the mean parallactic angle $\theta_{\rm para}({\rm mean})$ for each cycle  $Q_{\rm cor}\approx Q - p_{tel}\ \cos(2(\theta_{\rm para}({\rm mean})+\delta_{\rm tel}))\ I $ and $U_{\rm cor} \approx U - p_{tel}\ \sin(2(\theta_{\rm para}({\rm mean})+\delta_{\rm tel}))\ I$.

\section{PSF variations}\label{A2_psf} 

\begin{figure}[h!]
    \includegraphics[width=0.46\textwidth]{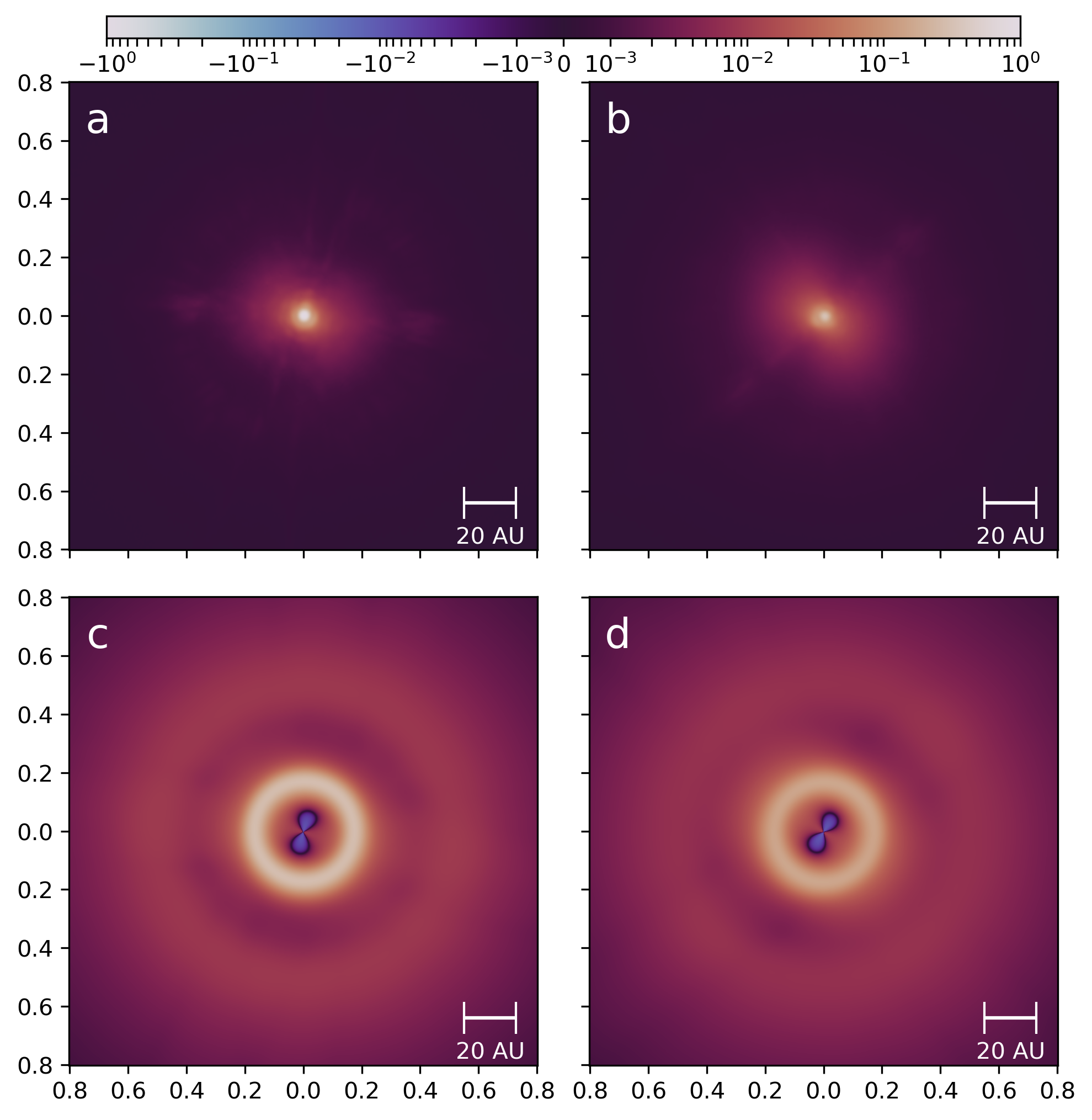}
     \centering
     \caption{I$'$-band PSF variations in the HD~169142 observations on July 9, 2015, and impact on the $Q_\phi$ simulation. The top row shows the best cycle PSF (a) and the worst cycle PSF (b). Both are divided by the maximum of the best cycle PSF. The bottom row illustrates the $Q_\phi$ model (normalized to 1) given in Fig.~\ref{hd24comp_n} (c) convolved with the best PSF (d) and the worst PSF (j).  }
\label{PSFvarAP5}
\end{figure}

\begin{figure}[ht!]
\centering
\begin{subfigure}{0.46\textwidth}
    \includegraphics[width=1\textwidth]{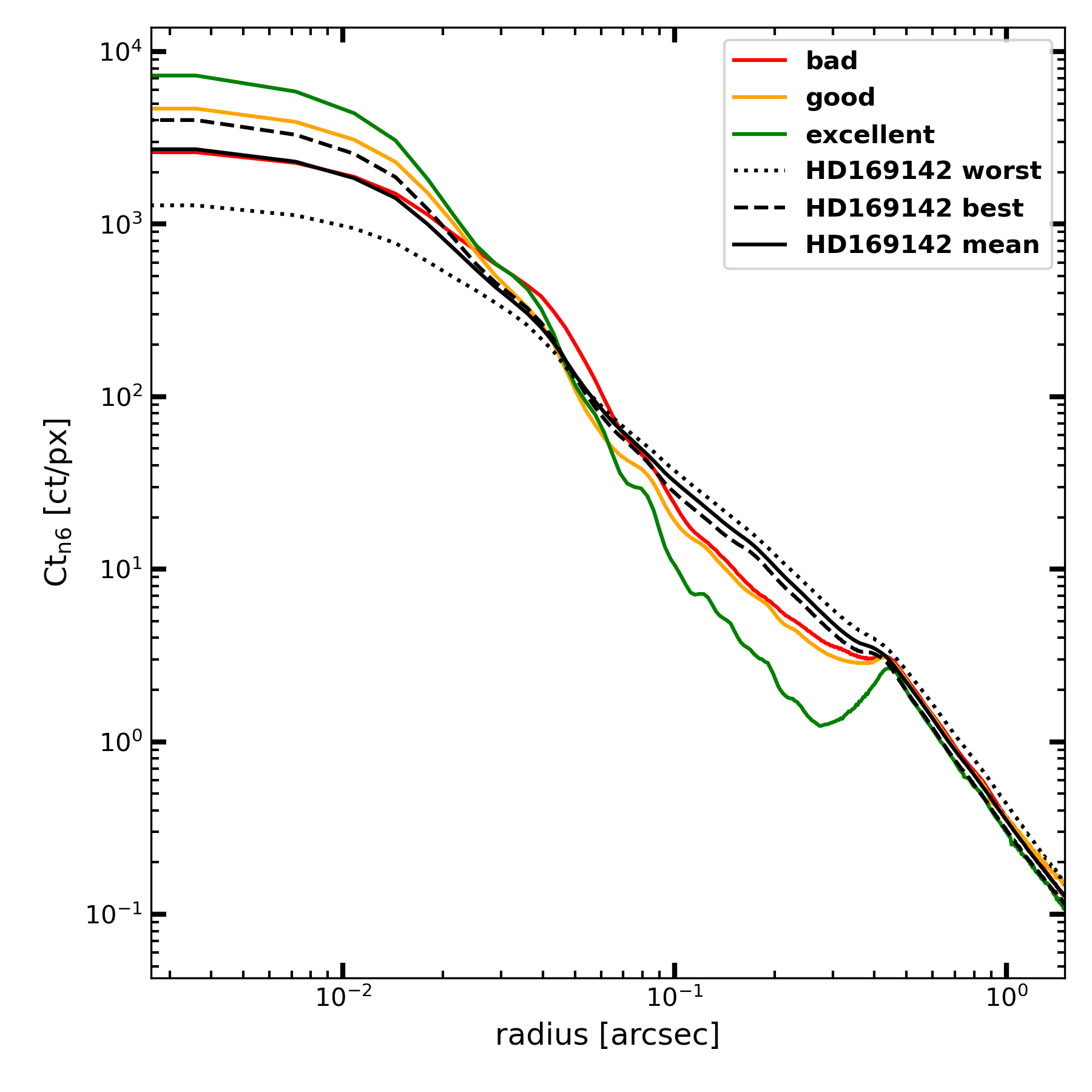}
     \centering
       \caption{Flux normalized radial intensity profiles.}\label{PSFqualAP1}
\end{subfigure}
\begin{subfigure}{0.46\textwidth}
    \includegraphics[width=1\textwidth]{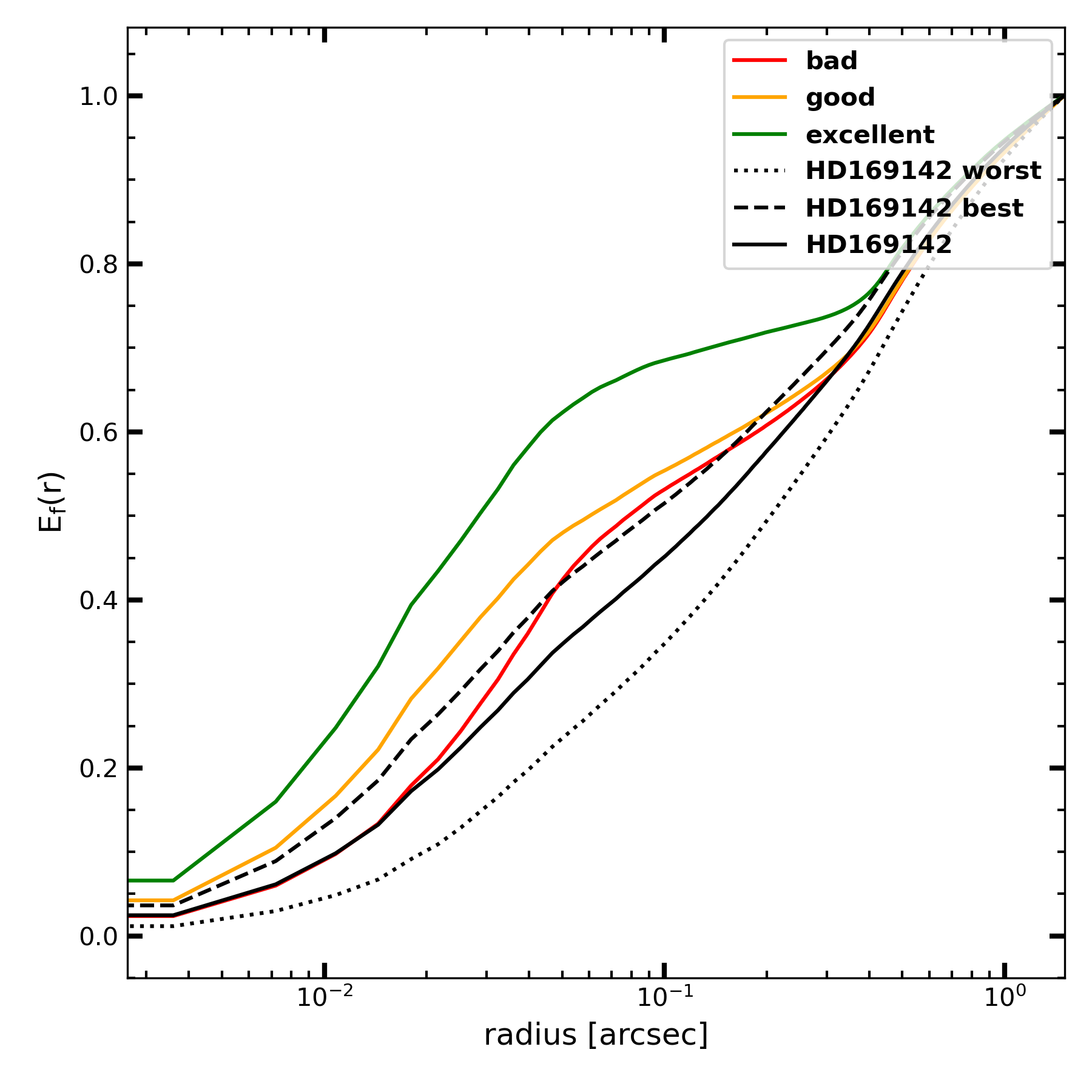}
     \centering
       \caption{Normalized encircled flux.}\label{PSFqualAP2}
\end{subfigure}
\caption{I$'$-band PSF variations in the HD 169142 observations from July 9, 2015, in comparison with other PSF profiles obtained with
SPHERE/ZIMPOL ($\rm N\_I$ filter).}
\label{tbda1_last}
\end{figure}

The shape of the PSF depends strongly on the atmospheric turbulence and the AO performance. As illustrated in Sect.~\ref{ipsfv} this has a strong impact on the observed $Q_\phi$ signal, due to smearing and polarimetric cancelation. In Fig.~\ref{PSFvarAP5} we show the I$'$-band intensity images of HD~169142 with the highest and lowest peak intensity, selected from $I_{\rm peak}^n$ given in Fig.~\ref{atmVARmaxiqphi}. We use these images as proxy for the best and the worst I$'$-band PSF and to assess the effects of the convolution on $Q_\phi$. Figure~\ref{tbda1_last} shows the corresponding best and worst radial intensity profiles for HD~169142, and the mean profile for all unsaturated (fast polarimetry mode) I$'$-band observations from July 7, 2015.   Additionally, we show as comparison the profiles of standard stars for the $\rm N\_I$ filter from \cite{Schmid18} taken under different observing conditions. The profiles are flux normalized and the peak flux can be used as proxy for the Strehl ratio. According to \cite{Schmid18} we derive for the best profiles of HD~169142 an approximate Strehl ratios of about 15~$\%$ for the I$'$ band and 10~$\%$ for the R$'$ band. It should be noted that this kind of Strehl ratio is underestimating the AO-performance, because it includes also stray light effects, detector smearing (e.g., from the frame transfer), and other instrumental effects not related to the AO system. The lower panel in Fig.~\ref{tbda1_last} shows the encircled flux as cumulative radial counts normalized to 1 inside an aperture with r=1.5$''$ aperture as illustration of PSF smearing. However, the logarithmic presentations can be misleading and corresponding profiles on linear scales can be found in \cite{Schmid18}. 

\end{appendix}

\end{document}